\documentclass[aps, prx,twocolumn,longbibliography,superscriptaddress,amsmath,amssymb,floatfix]{revtex4}
\usepackage[latin9]{inputenc}
\usepackage{amssymb}
\usepackage{graphicx}
\usepackage{amsmath}
\usepackage{color}
\usepackage{mathrsfs}
\usepackage{float}
\usepackage{indentfirst}
\usepackage{mathrsfs}
\usepackage{float}
\usepackage{indentfirst}
\usepackage{textcomp}
\usepackage{comment}
\usepackage{mathtools}
\usepackage{natbib,hyperref}
\usepackage{soul}                                       
\usepackage{multirow}

\begin{document}

\title{The ground state of electron-doped $t-t'-J$ model on cylinders}

\author{Yang Shen}
\affiliation{Key Laboratory of Artificial Structures and Quantum Control (Ministry of Education),  School of Physics and Astronomy, Shanghai Jiao Tong University, Shanghai 200240, China}

\author{Xiangjian Qian}
\affiliation{Key Laboratory of Artificial Structures and Quantum Control (Ministry of Education),  School of Physics and Astronomy, Shanghai Jiao Tong University, Shanghai 200240, China}

\author{Mingpu Qin} \thanks{qinmingpu@sjtu.edu.cn}
\affiliation{Key Laboratory of Artificial Structures and Quantum Control (Ministry of Education),  School of Physics and Astronomy, Shanghai Jiao Tong University, Shanghai 200240, China}

\affiliation{Hefei National Laboratory, Hefei 230088, China}

\date{\today}

\begin{abstract}
We perform a comprehensive study of the electron-doped $t-t'-J$ model on cylinders with Density Matrix Renormalization Group (DMRG). We adopt both periodic and anti-periodic boundary conditions along the circumference direction to explore the finite size effect. We study doping levels of $1/6$, $1/8$, and $1/12$ which represent the most interesting region in the phase diagram of electron-doped cuprates. We find that for width-4 and 6 systems, the ground state for fixed doping switches between anti-ferromagnetic Neel state and stripe state under different boundary conditions and with system widths, indicating the presence of large finite size effect in the $t-t'-J$ model. We also have a careful analysis of the $d$-wave pairing correlations which also changes quantitatively with boundary conditions and widths of the system. However, the pairing correlations are enhanced when the system becomes wider for all dopings, suggesting the existence of possible long-ranged superconducting order in the thermodynamic limit. The width-8 results are found to be dependent on the starting state in the DMRG calculation for the kept states we can reach. For width-8 system only Neel (stripe) state can be stabilized in DMRG calculation for $1/12$ ($1/6$) doping, while both stripe and Neel states are stable in the DMRG sweep for $1/8$ doping, regardless of the boundary conditions. These results indicate that $1/8$ doping is likely to lie in the boundary of a phase transition between the Neel phase with lower doping and the stripe phase with higher doping, consistent with the previous study. The sensitivity of ground state on boundary conditions and size observed in this work is similar to that in the $t'$- Hubbard model.    
\end{abstract}

\maketitle

\section{Introduction}

Hubbard model and its descendant $t-J$ model \cite{Hubbard,PhysRevB.37.3759,qin2022hubbard,annurev-conmatphys-031620-102024,PhysRevX.5.041041} play a paradigmatic role in the study of strongly correlated quantum many-body physics \cite{RevModPhys.84.1383,RevModPhys.66.763}.
Though their forms are simple, many exotic quantum states are arguably related to these models. It is believed that they are connected to the high-$T_c$ superconductivity in cuprates \cite{doi:10.1126/science.235.4793.1196,RevModPhys.66.763,PhysRevB.37.3759,nature14165,RevModPhys.84.1383}. Recent state-of-the-art numerical results show that the ground state of pure Hubbard model (with only nearest hoppings) is stripe state \cite{doi:10.1126/science.aam7127,PhysRevResearch.4.013239} without long-ranged superconductivity \cite{PhysRevX.10.031016}, which indicates the necessity to modify the Hamiltonian if we want to pursue superconducting ground state in the single-band framework. 

It was well known that the next nearest neighboring hopping $t'$ term is necessary to account for the particle-hole asymmetry in the phase diagram of cuprates \cite{RevModPhys.84.1383}. Band structure calculations \cite{ANDERSEN19951573,PhysRevB.98.134501} also show the existence of a $t'\approx -0.2$ term in the single band description of cuprates. The effect of $t'$ term on the ground state of Hubbard and $t-J$ models was extensively explored before \cite{PhysRevB.60.R753,PhysRevResearch.2.033073,PhysRevB.95.155116,PhysRevB.102.041106,2018npjQM...3...22H,PhysRevLett.127.097003,PhysRevLett.127.097002,doi:10.1073/pnas.2109978118,lu2023sign,PhysRevLett.132.066002,PhysRevB.106.174507,10.21468/SciPostPhys.12.6.180,PhysRevB.100.195141,2023arXiv231205893C}. The $t'$ term frustrates and weakens the long-range anti-ferromagnetic Neel order in the parent compound. When the system is doped, the stripe order should also be suppressed by $t'$ term which is believed to benefit superconducting pairing. Calculations on $t-t'-J$ model \cite{PhysRevLett.127.097002,PhysRevLett.127.097003,doi:10.1073/pnas.2109978118} show the ground state turns to a superconducting one in the electron doped side. Superconductivity was also found in the $t'$- Hubbard model \cite{xu2023coexistence} and matches quantitatively with the phase diagram of cuprates, in which superconductivity is stronger in the hole doped side than the electron doped side. 

One of the important finding in \cite{xu2023coexistence} is that the ground state of the $t'$- Hubbard model becomes very sensitive to the system width and boundary conditions, which indicates the presence of large finite size effect in the model. For this reason, the authors in \cite{xu2023coexistence} utilized twist averaged boundary conditions and careful finite size scaling to obtain physical quantity in the thermodynamic limit. For the $t-t'-J$ model, the most recent study \cite{PhysRevLett.132.066002,doi:10.1073/pnas.2109978118} on cylinders with finite width show the superconductivity is stronger in the electron doped side than the hole doped side if exists, contradictory to the phase diagram of cuprates and the recent results of $t'$-Hubbard model \cite{xu2023coexistence}. Given the presence of very large finite size effect in the $t'$-Hubbard model, it is natural to ask whether $t-t'-J$ model also has large finite size effect.

In this work, we have a comprehensive study of the $t-t'-J$ model with Density Matrix Renormalization Group (DMRG) \cite{PhysRevLett.69.2863,PhysRevB.48.10345,SCHOLLWOCK201196}. We adopt both periodic and anti-periodic boundary conditions along the circumference direction and study systems with different widths to explore the finite size effect \cite{xu2023coexistence}.  We only study the electron doped side in this work, which was believed to be simper than the hole doped side, without the competition of different states \cite{RevModPhys.84.1383}. We study doping levels of $1/6$, $1/8$, and $1/12$ which represent the most interesting region in the phase diagram of electron-doped cuprates.    

Our results show that the ground state of $t-t'-J$ model is also very sensitive to the boundary conditions and width of the systems, similar to the $t'$-Hubbard model \cite{xu2023coexistence}. We find that for width-4 and 6 systems, the ground state for fixed doping switches between anti-ferromagnetic Neel state and stripe state under different boundary conditions and with system widths. We find the pairing correlation also changes quantitatively with boundary conditions and widths. But for all dopings, the pairing correlations are enhanced when the system becomes wider, suggesting the existance of possible long-ranged superconducting pairing in the thermodynamic limit. The width-8 results are found to be dependent on the starting state in the DMRG calculation for the kept states we can reach. For width-8 system only Neel (stripe) state can be stabilized in DMRG calculation for $1/12$ ($1/6$) doping, while both stripe and Neel states are stable in the DMRG sweep for $1/8$ doping. Although a reliable finite width scaling of the DMRG results is infeasible at present, these results indicate that $1/8$ doping is likely to lie in the boundary of the phase transition between the Neel state with lower doping and stripe state with higher doping, consistent with the previous study \cite{doi:10.1073/pnas.2109978118}.

The rest of this paper is organized as follows. In Sec.~\ref{Model and methods} we discuss the model and the method we employ. Our results are presented in Sec.~\ref{results}: starting with the spin and charge
densities, followed by an analysis of the pairing correlations. In Sec.~\ref{discussions}, we have a discussion of the relationship between superconductivity and stripe. We conclude in
Sec.~\ref{conclusion and perspective}.

\begin{table*}[]
	\centering
	\resizebox{\textwidth}{!}{%
		\begin{tabular}{|cc|ll|ll|ll|}
			\hline
			\multicolumn{2}{|l|}{\multirow{2}{*}{}}                                       & \multicolumn{2}{c|}{\textbf{width-4}}                                         & \multicolumn{2}{c|}{\textbf{width-6}}                                                & \multicolumn{2}{c|}{\textbf{width-8}}                                               \\ \cline{3-8} 
			\multicolumn{2}{|l|}{}                                                        & \multicolumn{1}{c|}{\textbf{PBC}}        & \multicolumn{1}{c|}{\textbf{APBC}} & \multicolumn{1}{c|}{\textbf{PBC}}               & \multicolumn{1}{c|}{\textbf{APBC}} & \multicolumn{1}{c|}{\quad \quad \quad \textbf{PBC} \quad \quad \quad}              & \multicolumn{1}{c|}{\textbf{APBC}} \\ \hline
						\multicolumn{1}{|c|}{\multirow{3}{*}{\textbf{1/12}}} & \textbf{spin}   & \multicolumn{1}{l|}{Neel, short-ranged}   & filled stripe                      & \multicolumn{1}{l|}{Neel, short-ranged}          & Neel, (quasi) long-ranged           & \multicolumn{2}{l|}{\multirow{2}{*}{only Neel stablized}}                         \\ \cline{2-6} 
			\multicolumn{1}{|c|}{}                                      & \textbf{charge} & \multicolumn{1}{l|}{half-filled stripe}  & filled stripe                      & \multicolumn{1}{l|}{1/3 filled stripe}          & 1/3 filled stripe                  & \multicolumn{2}{l|}{}               \\ \cline{2-8}                         
			\multicolumn{1}{|c|}{}                                      & \textbf{pair}   & \multicolumn{1}{l|}{power-law, $K_{sc} \approx 1.1(0.98)$}                    & exponential, $\xi_{sc} \approx 9.82(9.9)$                                   & \multicolumn{1}{l|}{power-law, $K_{sc} \approx 0.49(0.49)$}                           & power-law, $K_{sc} \approx 0.48(0.41)$                                    & \multicolumn{1}{l|}{}                          &                                    \\ \hline
			\multicolumn{1}{|c|}{\multirow{3}{*}{\textbf{1/8}}}  & \textbf{spin}   & \multicolumn{1}{l|}{Neel, short-ranged}   & filled stripe                      & \multicolumn{1}{l|}{filled stripe, short-ranged} & Neel, (quasi) long-ranged         & \multicolumn{2}{l|}{\multirow{2}{*}{both filled stripe $\&$ Neel stablized}}                   \\ \cline{2-6} 
			\multicolumn{1}{|c|}{}                                      & \textbf{charge} & \multicolumn{1}{l|}{half-filled stripe}  & filled stripe                      & \multicolumn{1}{l|}{filled stripe, short-ranged} & uniform                           & \multicolumn{2}{l|}{}               \\ \cline{2-8} 
			\multicolumn{1}{|c|}{}                                      & \textbf{pair}   & \multicolumn{1}{l|}{power-law, $K_{sc} \approx 1.3(1.38)$}                    & exponential, $\xi_{sc} \approx 13.3(13.4)$                                   & \multicolumn{1}{l|}{power-law, $K_{sc} \approx 0.60(0.82)$}                           &  power-law, $K_{sc} \approx 0.59(0.69)$                                  & \multicolumn{1}{l|}{}                          &                                    \\ \hline
			\multicolumn{1}{|c|}{\multirow{3}{*}{\textbf{1/6}}}  & \textbf{spin}   & \multicolumn{1}{l|}{stripe, short-ranged} & filled stripe                      & \multicolumn{1}{l|}{filled stripe, short-ranged}              & Neel,(quasi) long-ranged           & \multicolumn{2}{l|}{\multirow{2}{*}{only filled stripe stablized}}                       \\ \cline{2-6} 
			\multicolumn{1}{|c|}{}                                      & \textbf{charge} & \multicolumn{1}{l|}{modulated stripe}             & filled stripe                      & \multicolumn{1}{l|}{filled stripe, short-ranged}              & uniform                            & \multicolumn{2}{l|}{}                         \\ \cline{2-8} 
			\multicolumn{1}{|c|}{}                                      & \textbf{pair}   & \multicolumn{1}{l|}{power-law, $K_{sc} \approx 1.5(1.8)$}                    & exponential, $\xi_{sc} \approx 14.4(14.3)$                                    & \multicolumn{1}{l|}{power-law, $K_{sc} \approx 1.0(1.2)$}                           & power-law, $K_{sc} \approx 0.76(0.83)$                                    & \multicolumn{1}{l|}{}                          &                                    \\ \hline
		\end{tabular}%
	}
	\caption{Summary of all the results, including the spin and charge densities, as well as the singlet pair-pair correlations. All systems have $d$-wave pairing symmetry except the $1/6$ doping case on width-4 cylinder under PBC, for which the pairing symmetry is plaquette $d$-wave \cite{PhysRevB.102.041106}. $K_{sc}$ and $\xi_{sc}$ are obtained from the fit of pair-pair correlation with the reference bond set as the 8th (4th for the values in the parentheses) vertical bond from the left edge.}
\label{table_sum}
\end{table*}

\begin{figure*}[t]
	\includegraphics[width=0.23\textwidth]{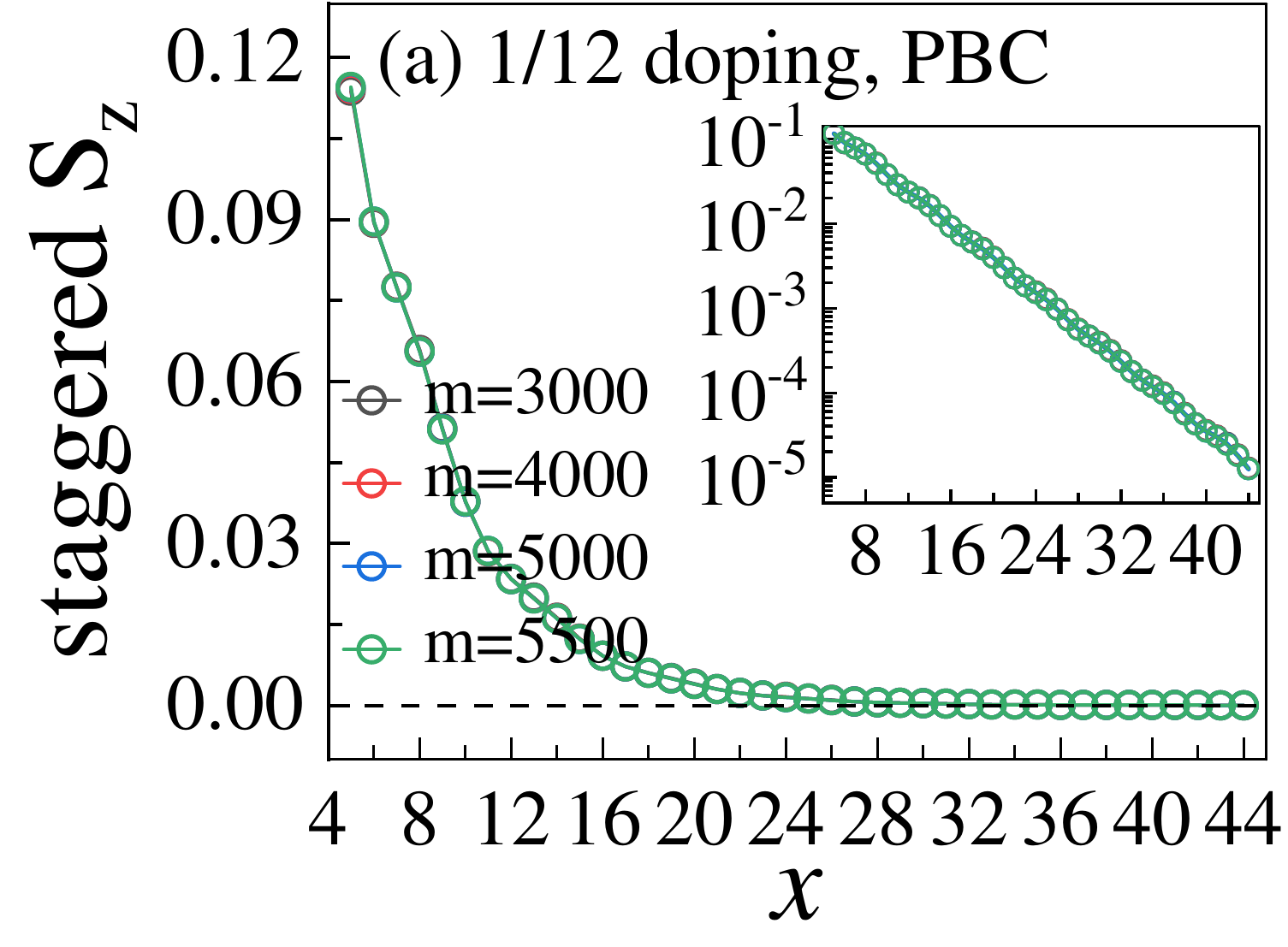}
	\includegraphics[width=0.23\textwidth]{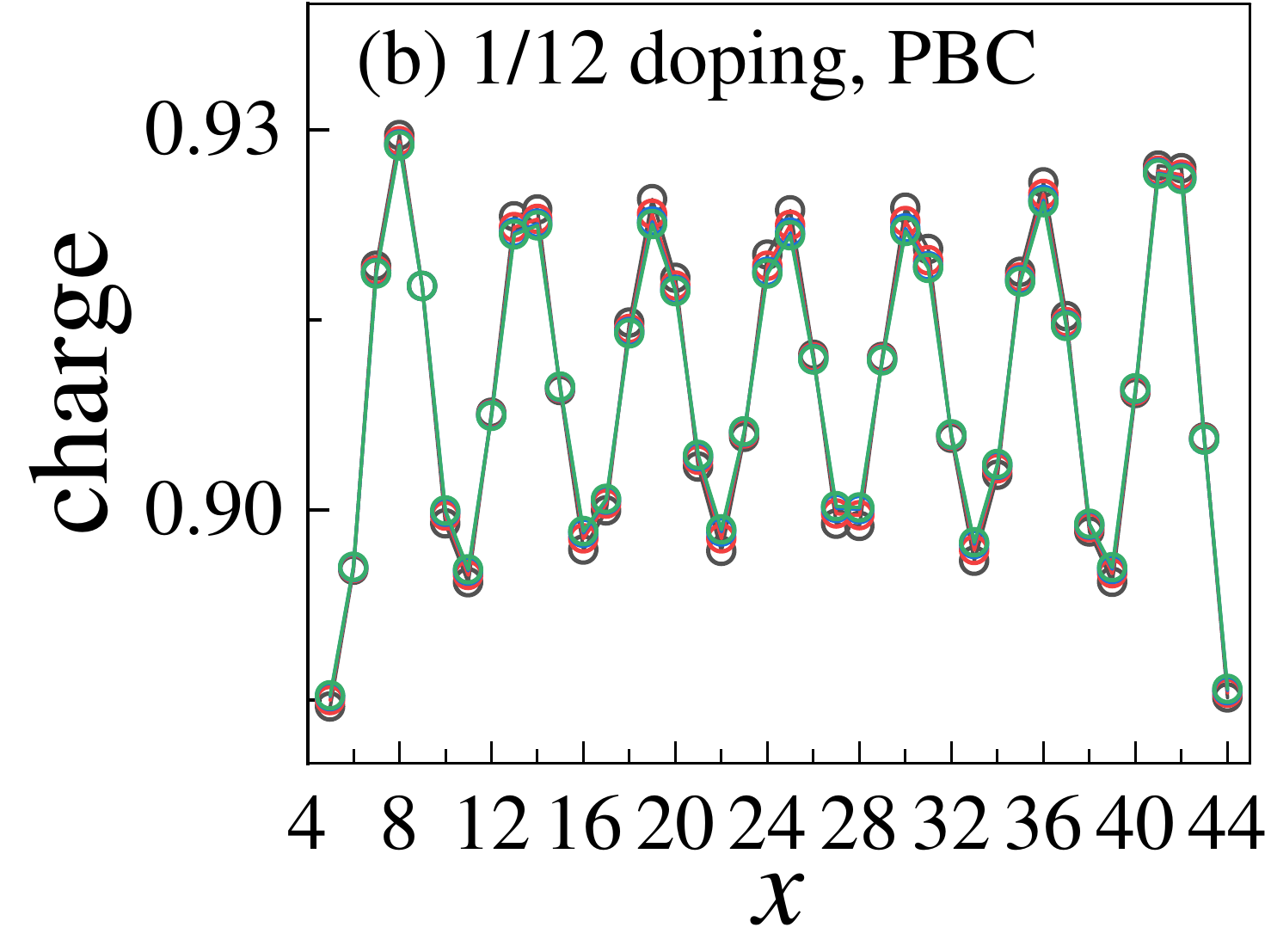}
	\includegraphics[width=0.23\textwidth]{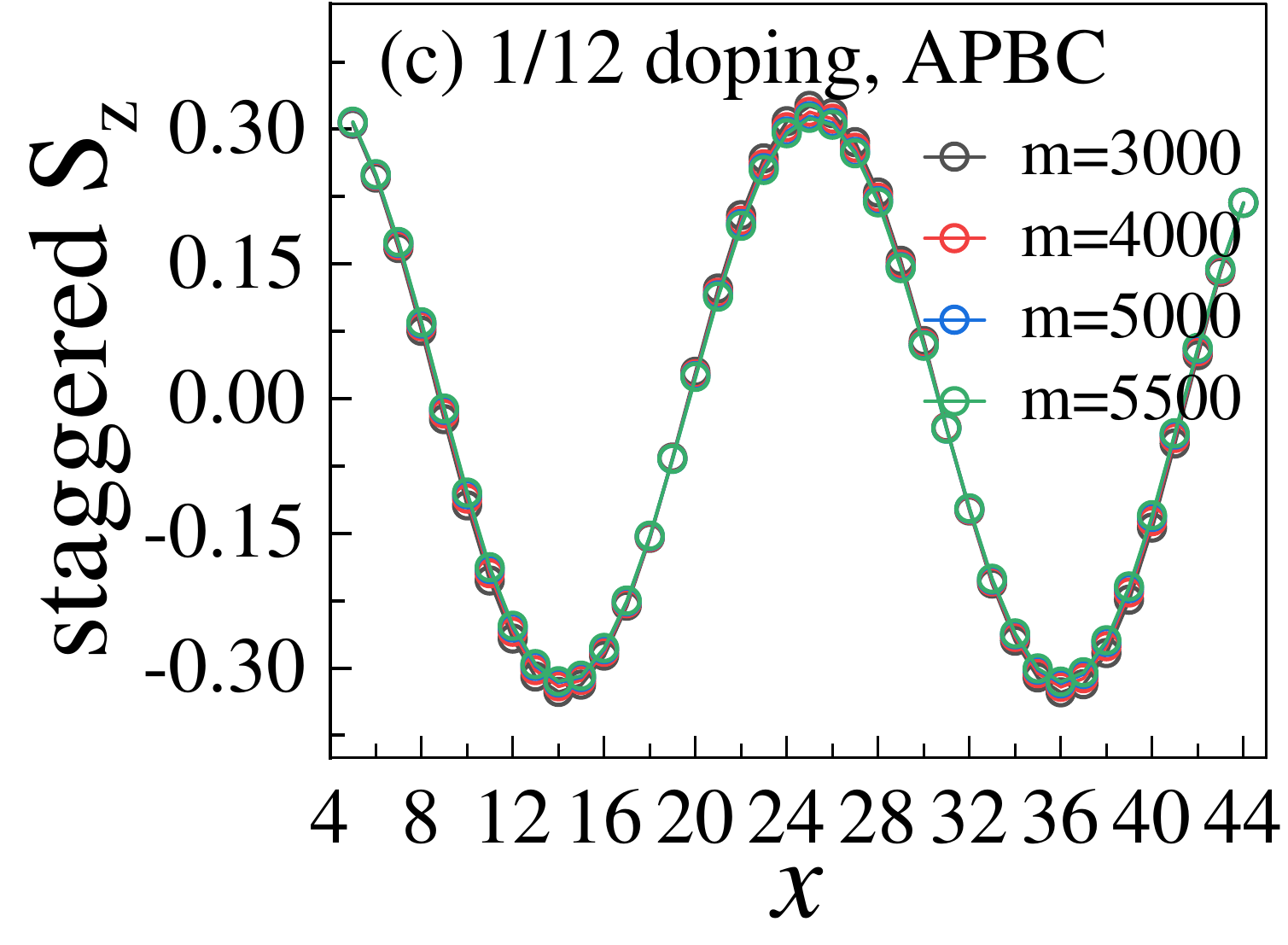}
	\includegraphics[width=0.23\textwidth]{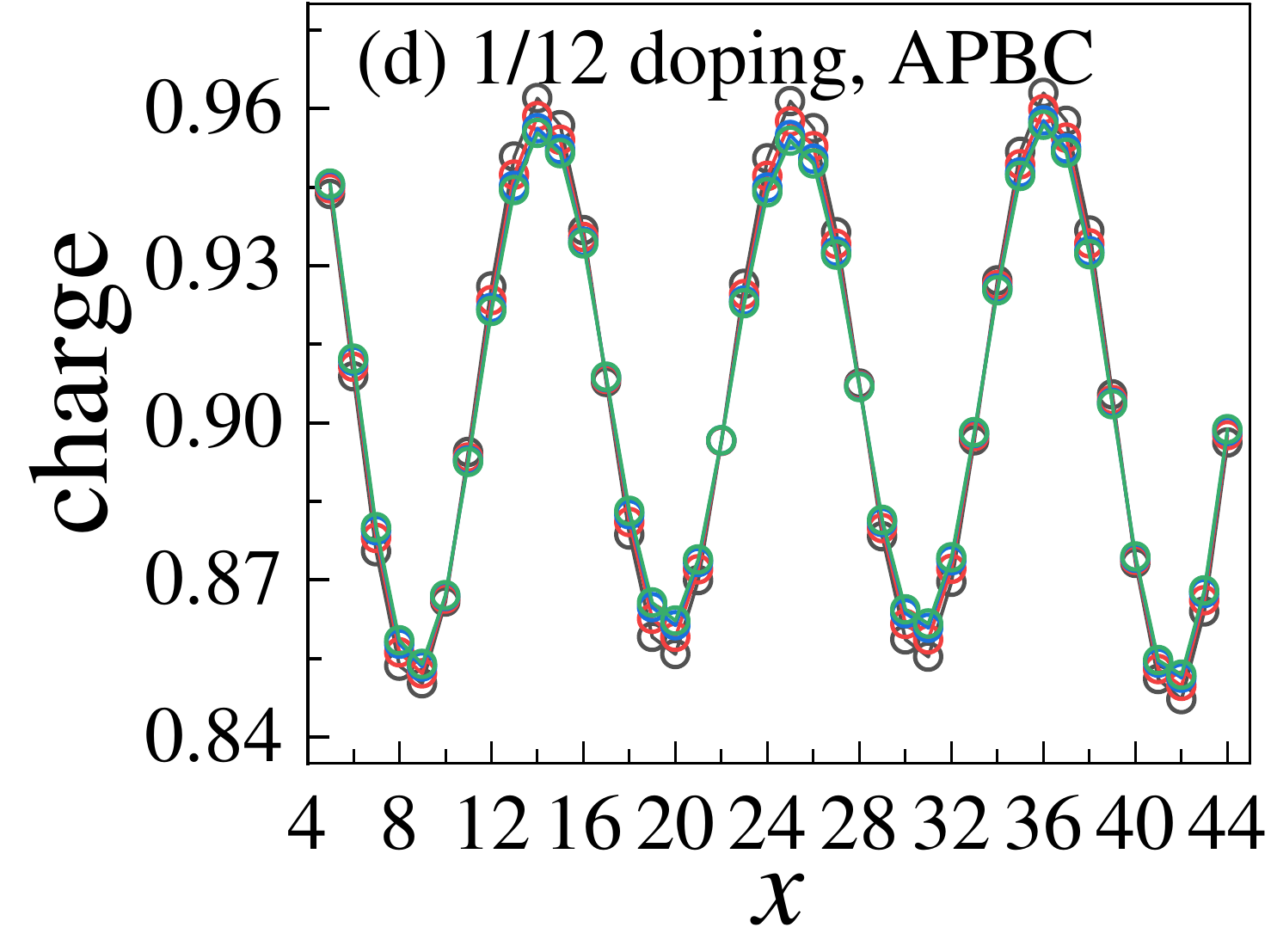}

	\includegraphics[width=0.23\textwidth]{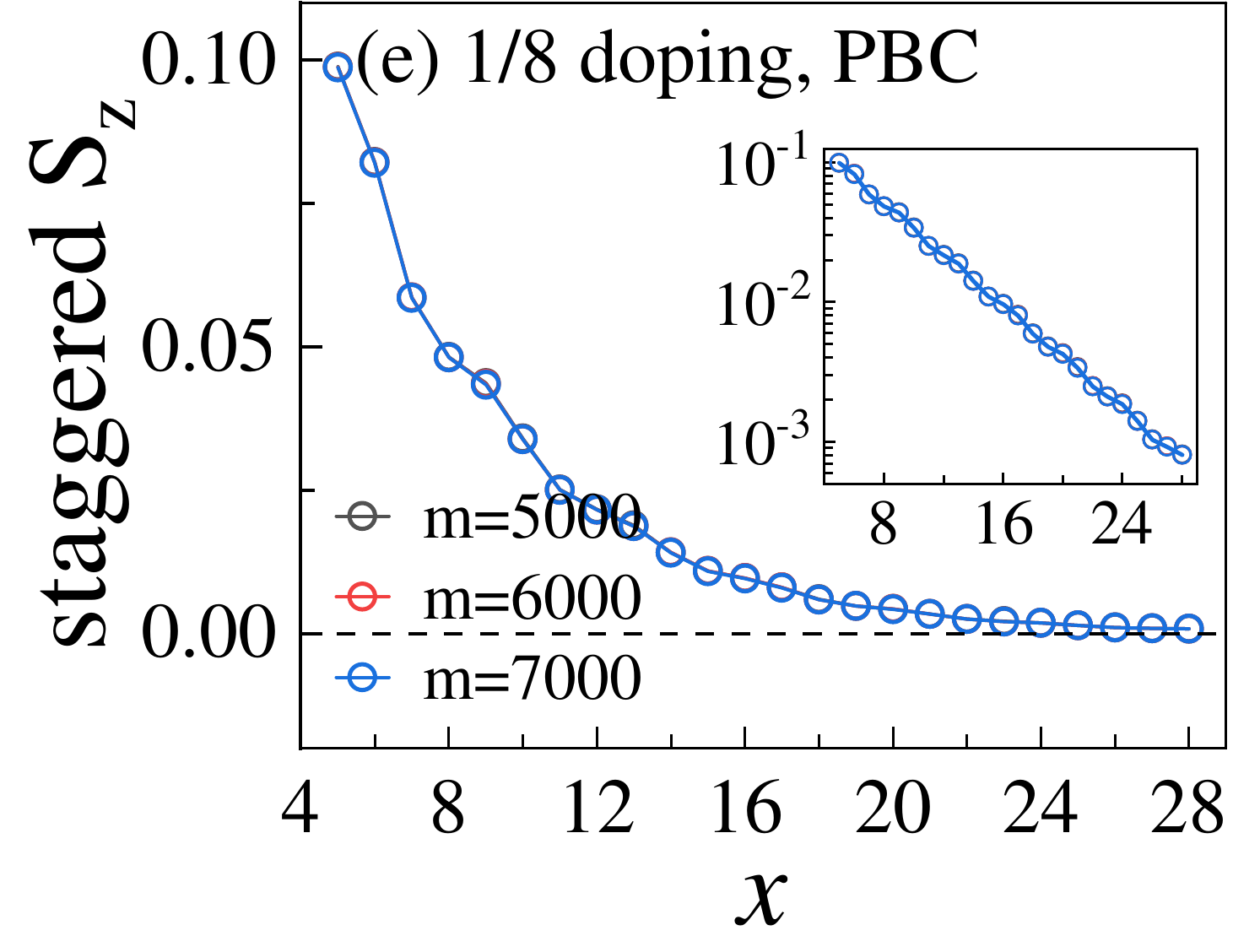}
	\includegraphics[width=0.23\textwidth]{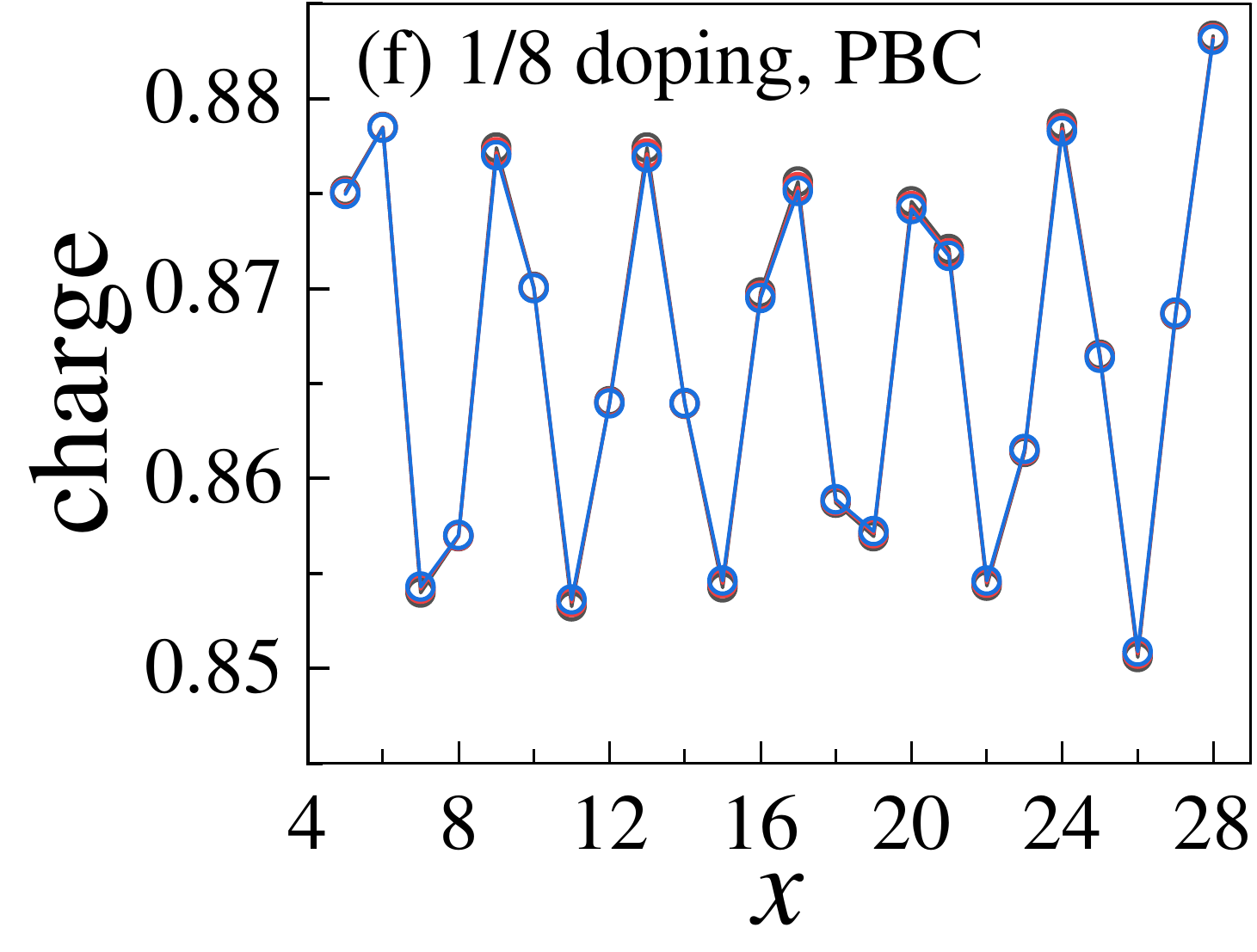}
	\includegraphics[width=0.23\textwidth]{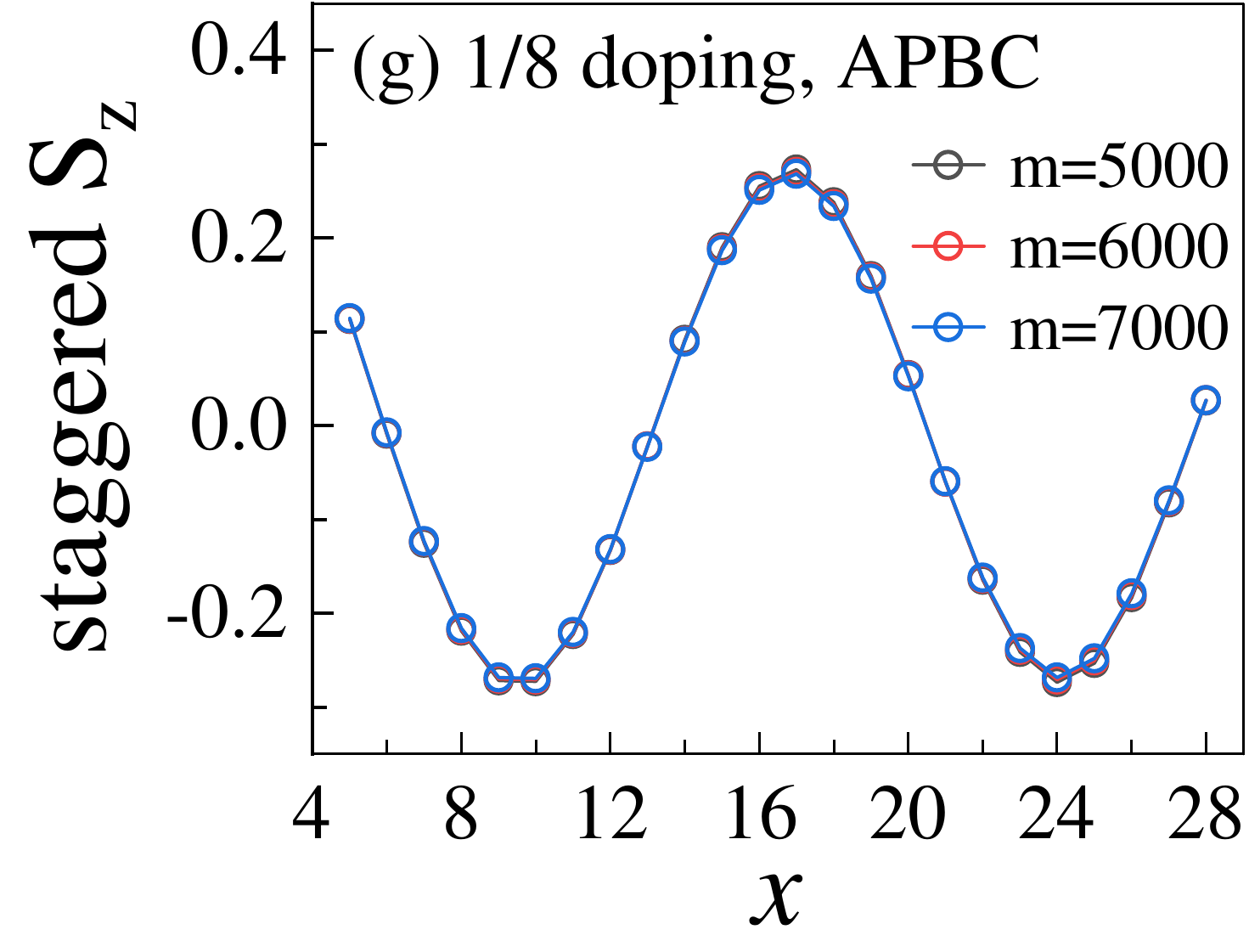}
	\includegraphics[width=0.23\textwidth]{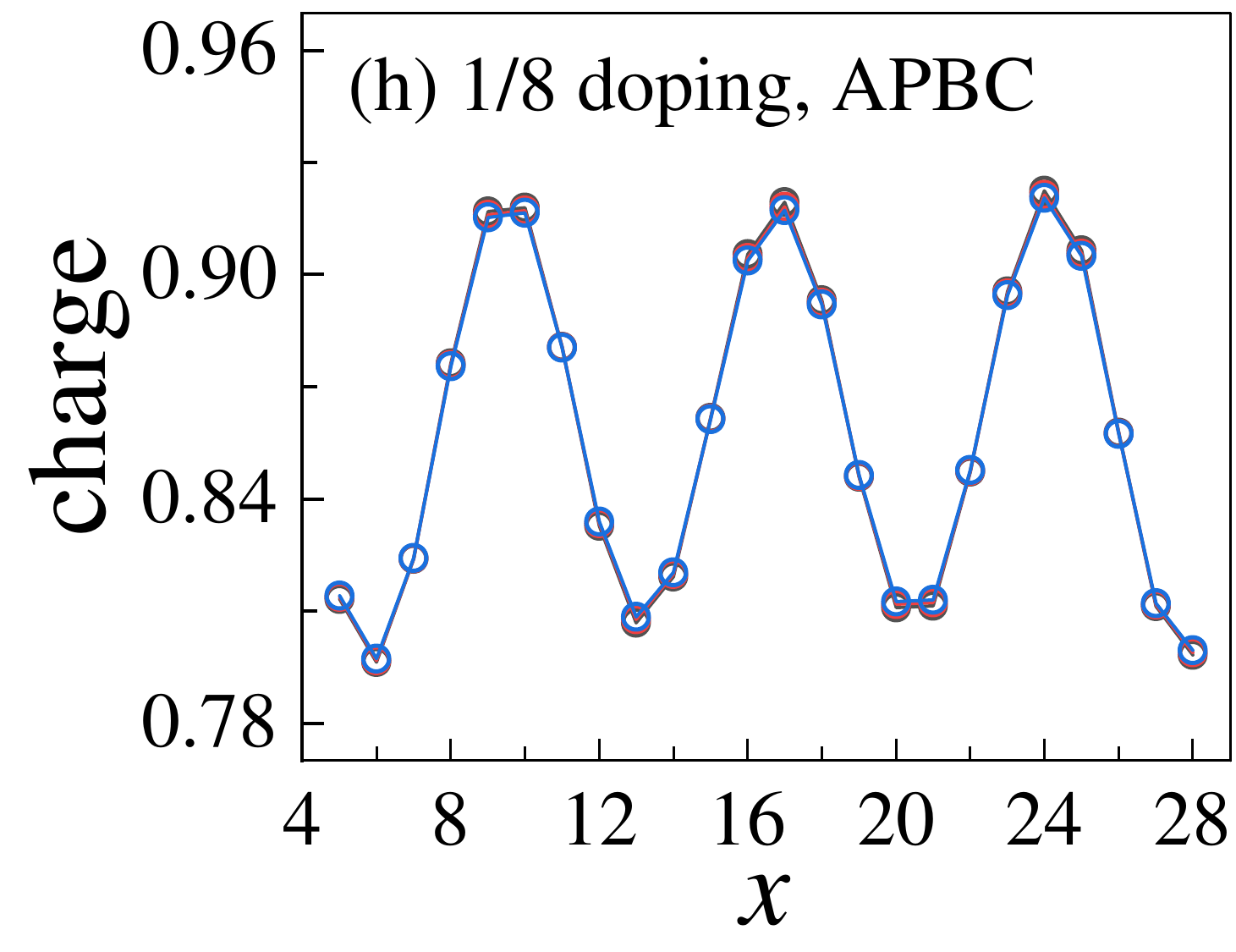}

	\includegraphics[width=0.23\textwidth]{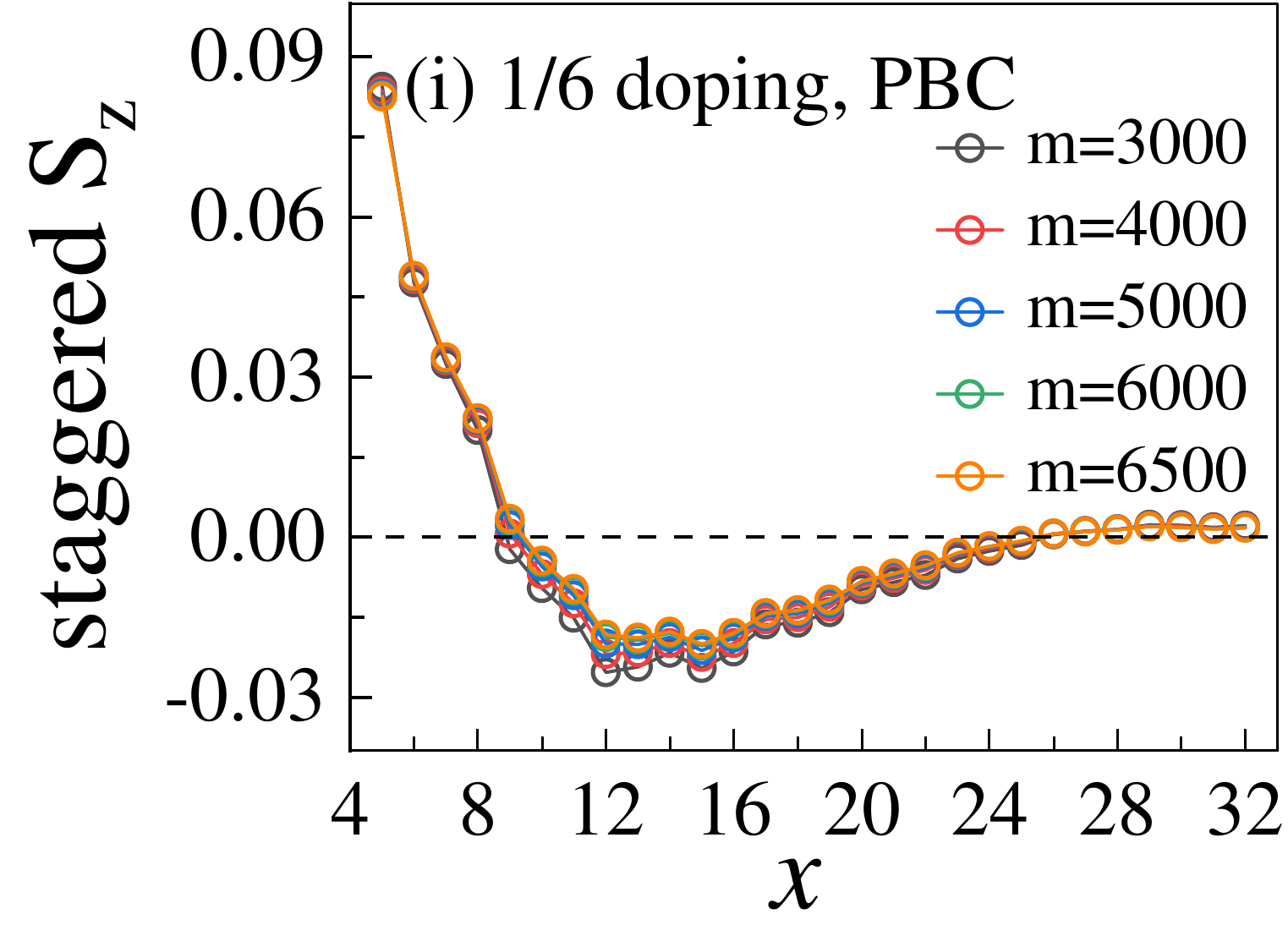}
	\includegraphics[width=0.23\textwidth]{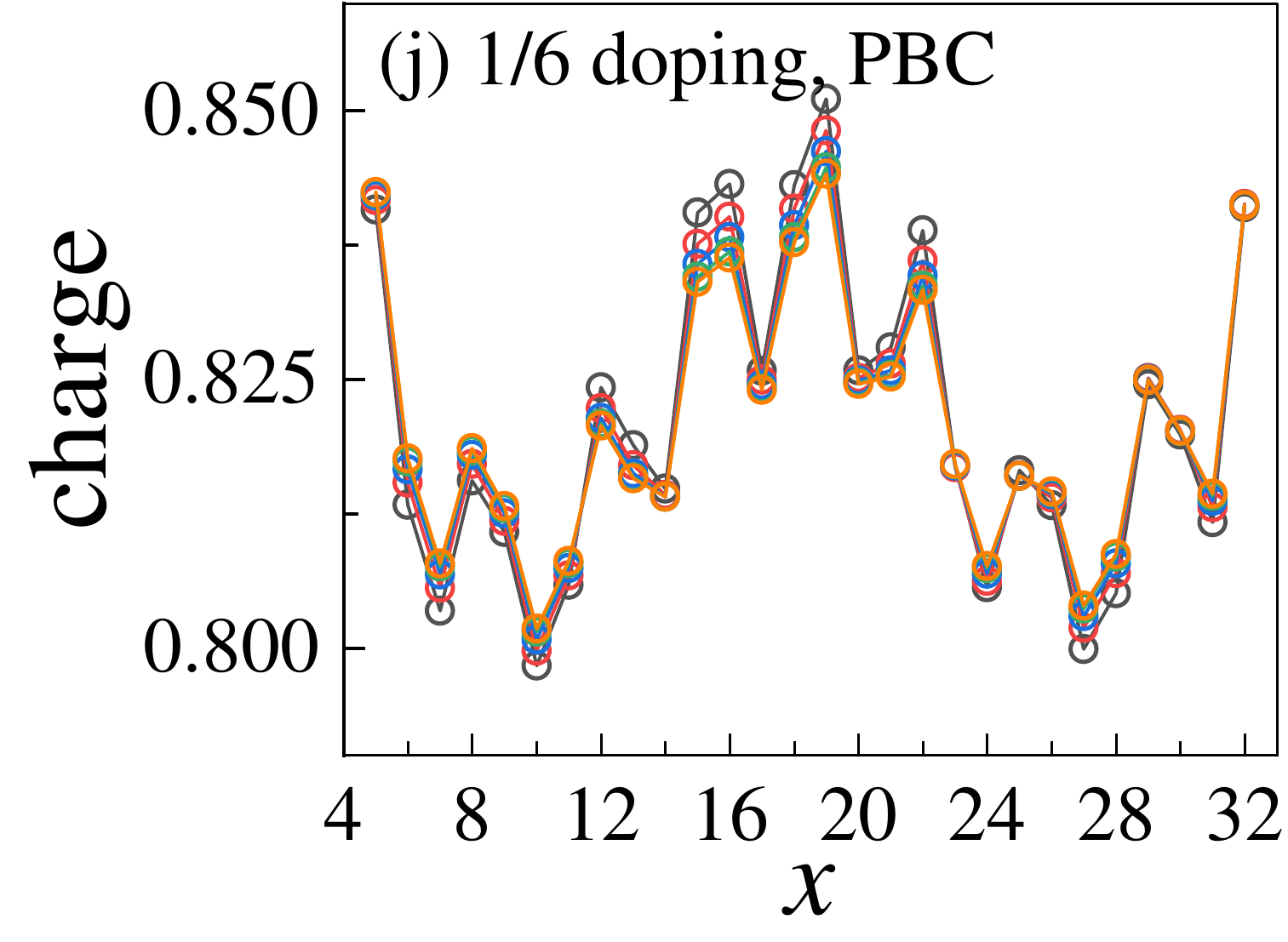}
	\includegraphics[width=0.23\textwidth]{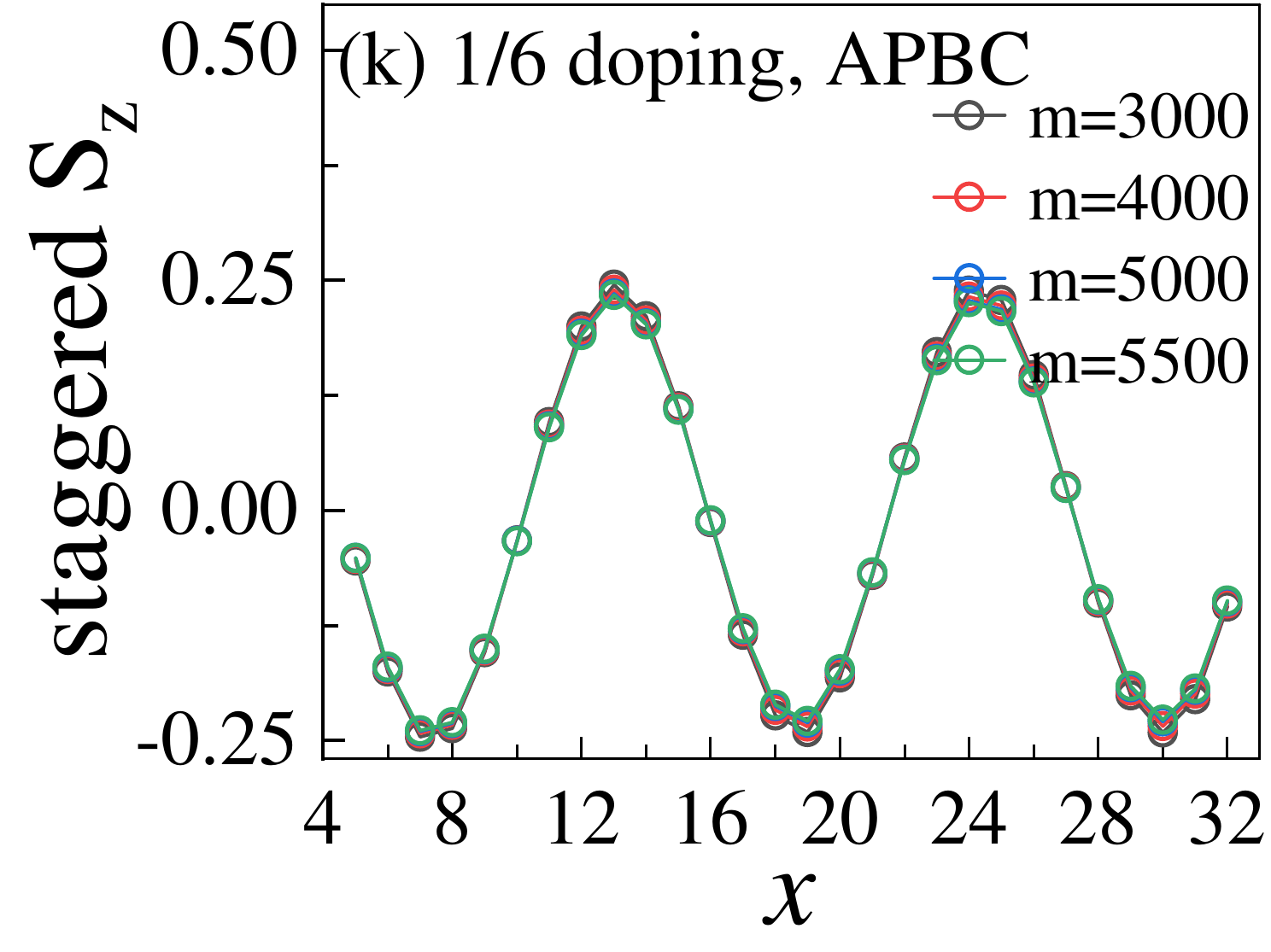}
	\includegraphics[width=0.23\textwidth]{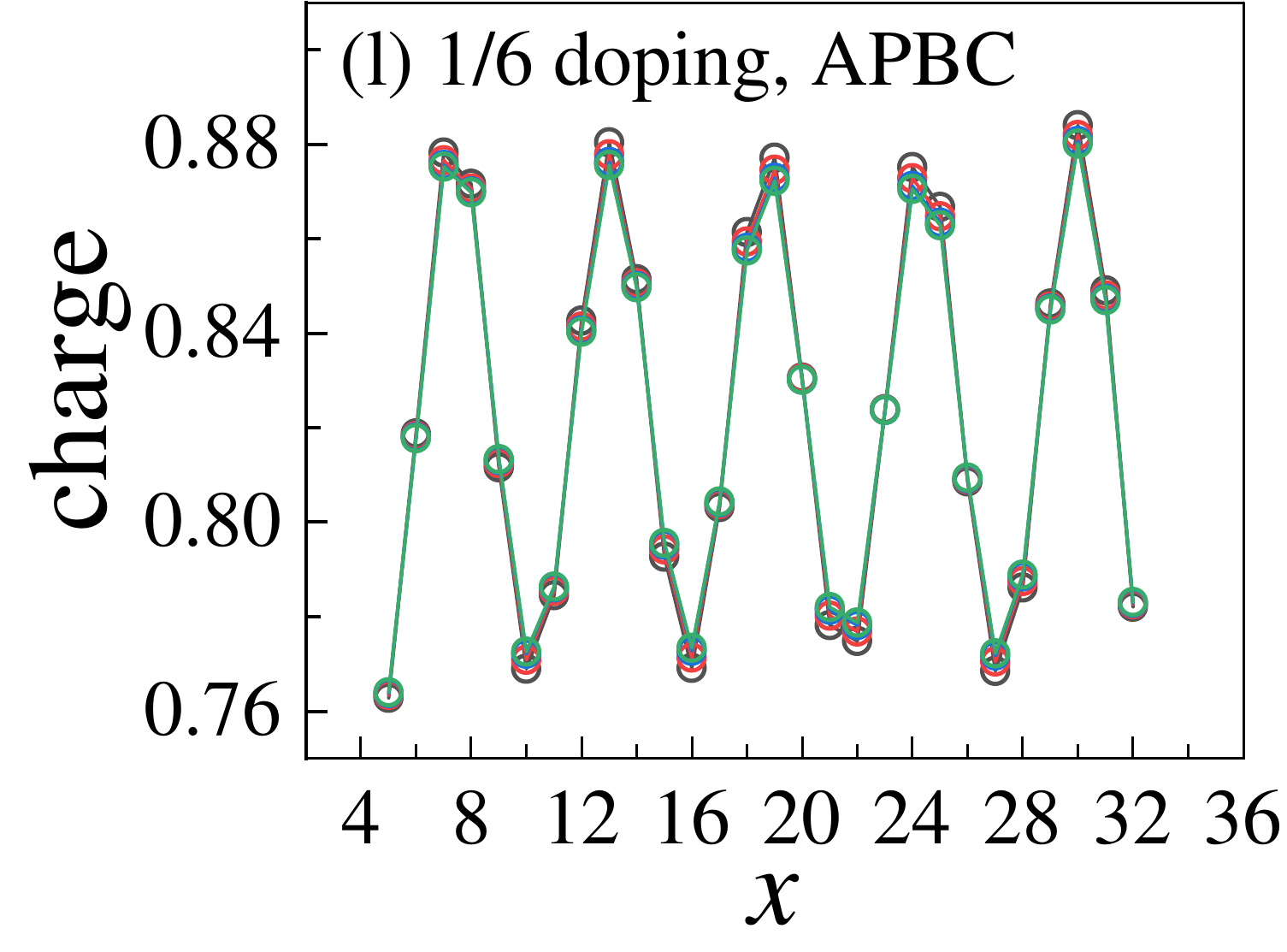}
	
	\caption{Staggered spin density and electron density for $1/12$ (top), $1/8$ (middle) and $1/6$ (bottom) dopings on width-4 cylinders (system sizes are $4 \times 48$, $4 \times 32$ and $4 \times 36$ for $1/12$, $1/8$ and $1/6$ dopings, respectively) under PBC (left two columns) and APBC (right two columns). Results of finite kept states are plotted to show the convergence of the DMRG calculations. The insets in (a) and (e) are plotted in semi-logarithmic scale to show the exponential decay of the staggered spin density for $1/12$ and $1/8$ dopings under PBC. The dashed horizontal lines represent zero for the staggered spin density.}
	\label{width-4}
\end{figure*}

\begin{figure*}[t]
	\includegraphics[width=0.23\textwidth]{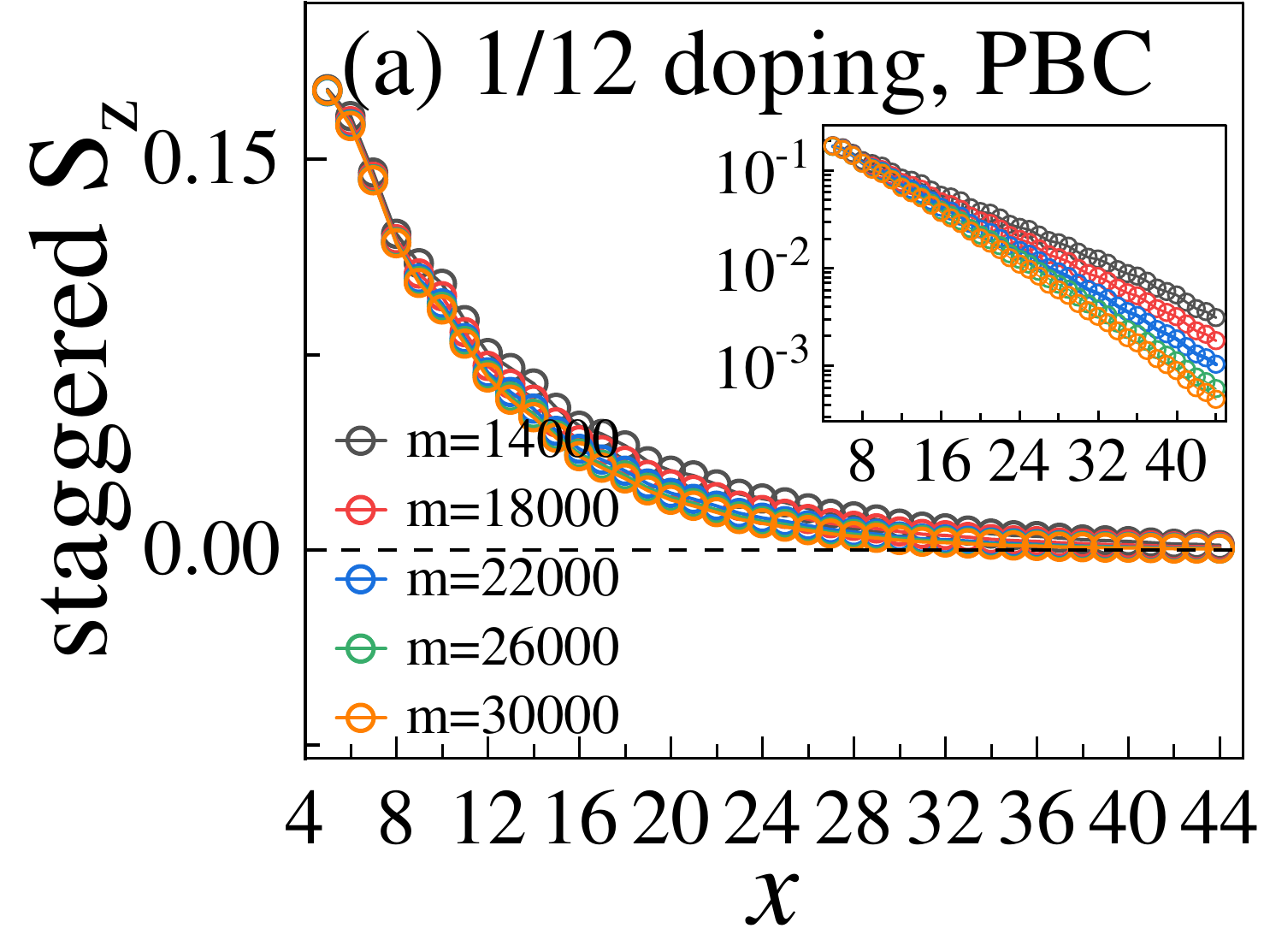}
	\includegraphics[width=0.23\textwidth]{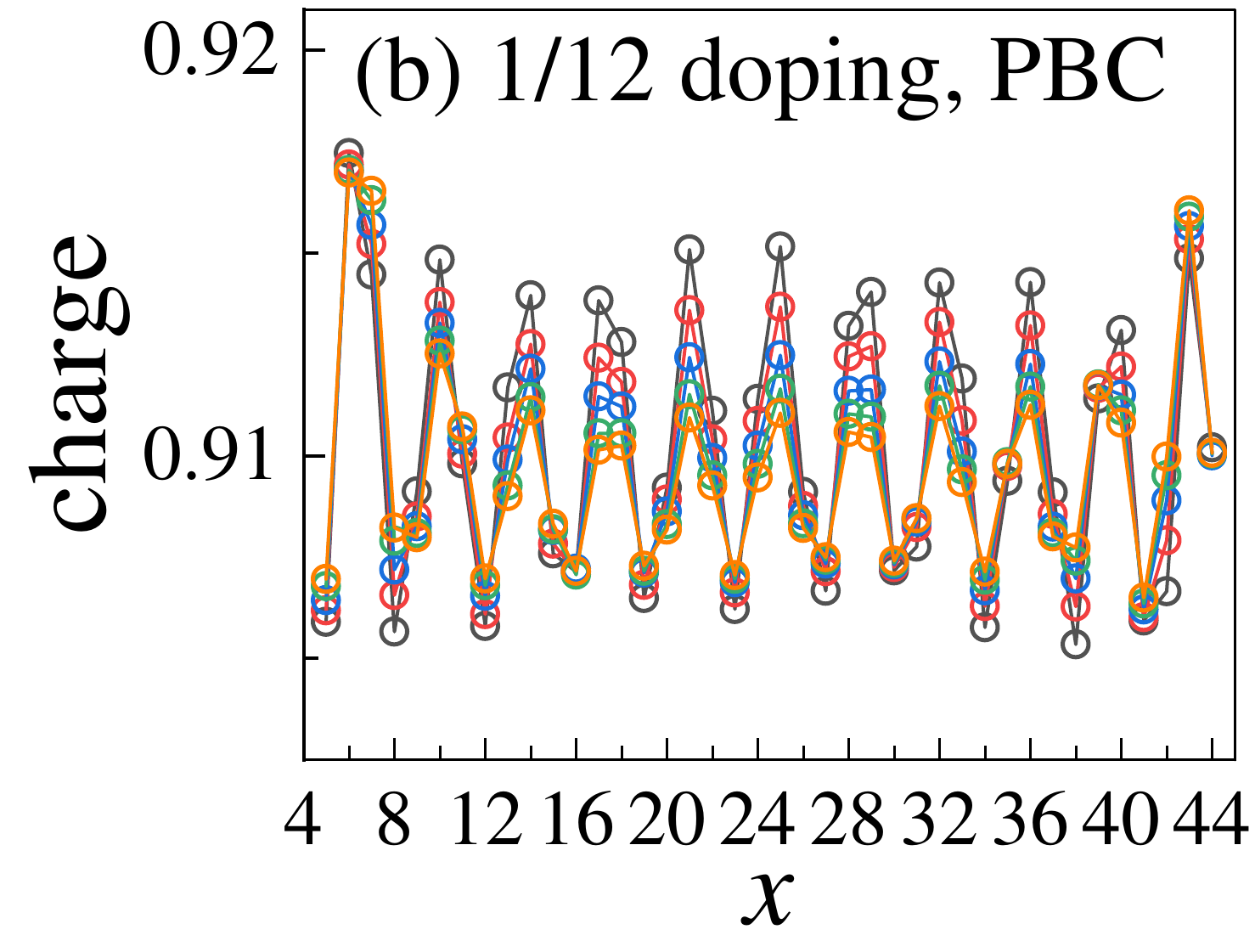}
	\includegraphics[width=0.23\textwidth]{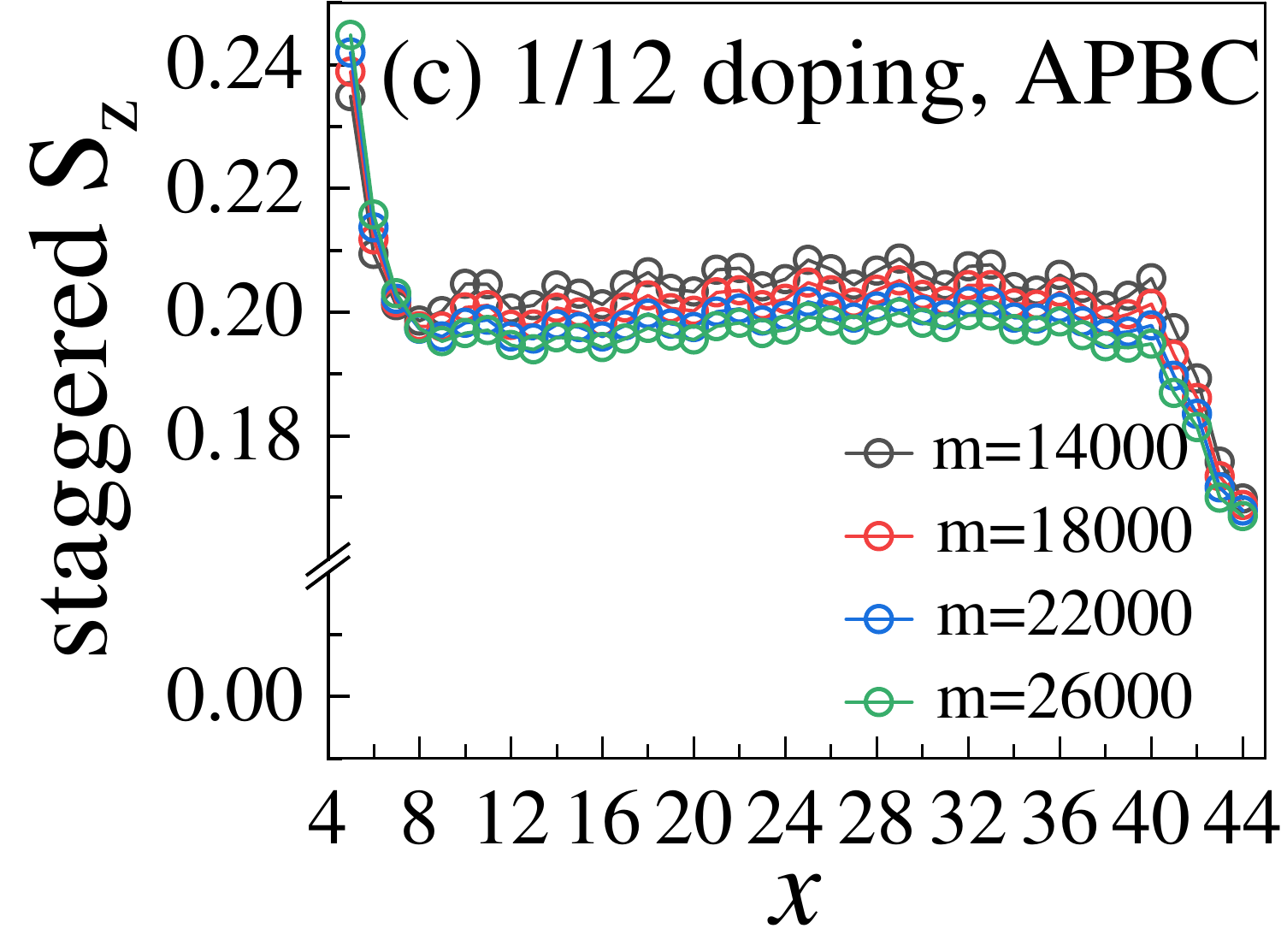}
	\includegraphics[width=0.23\textwidth]{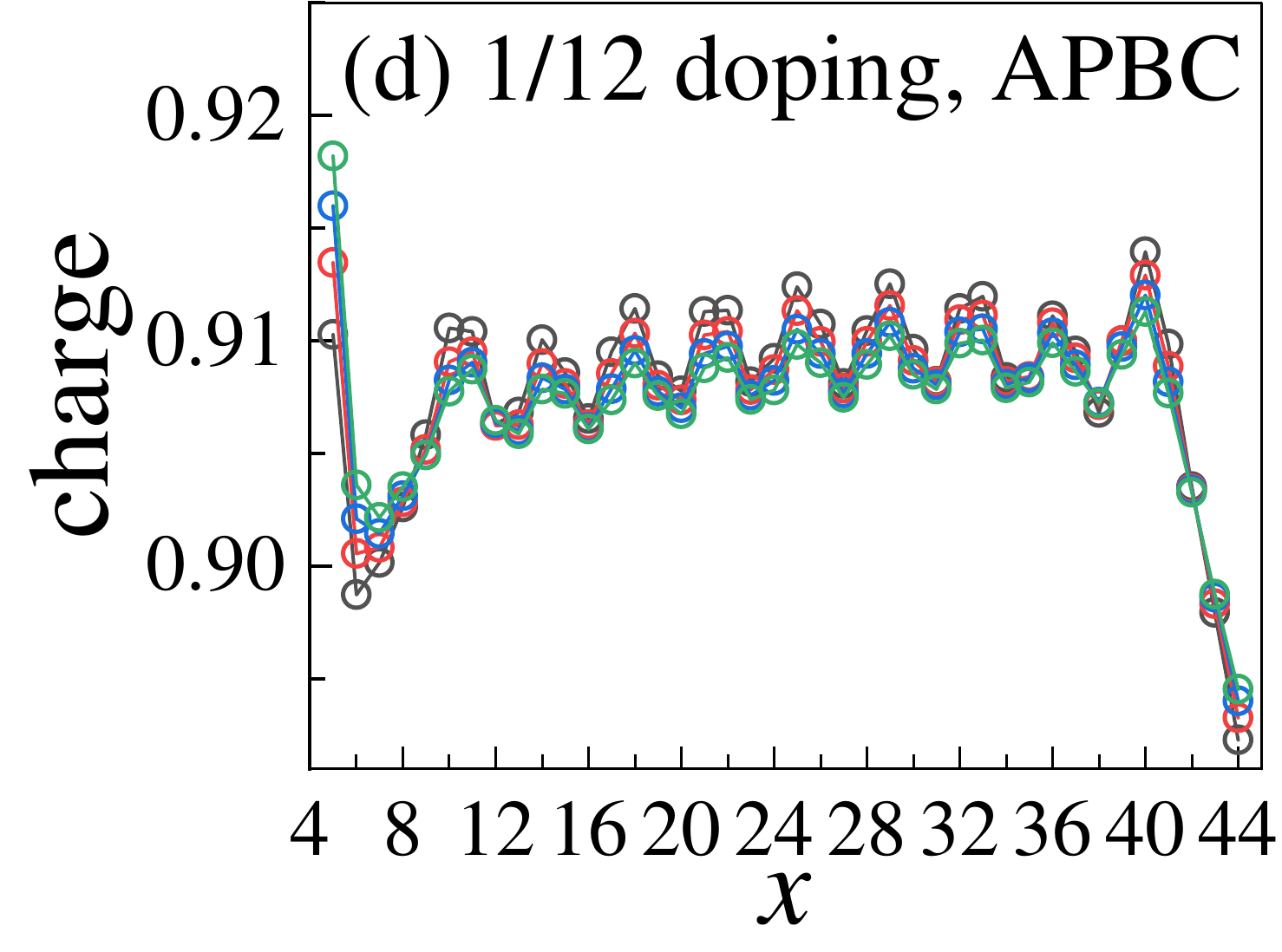}
	
	\includegraphics[width=0.23\textwidth]{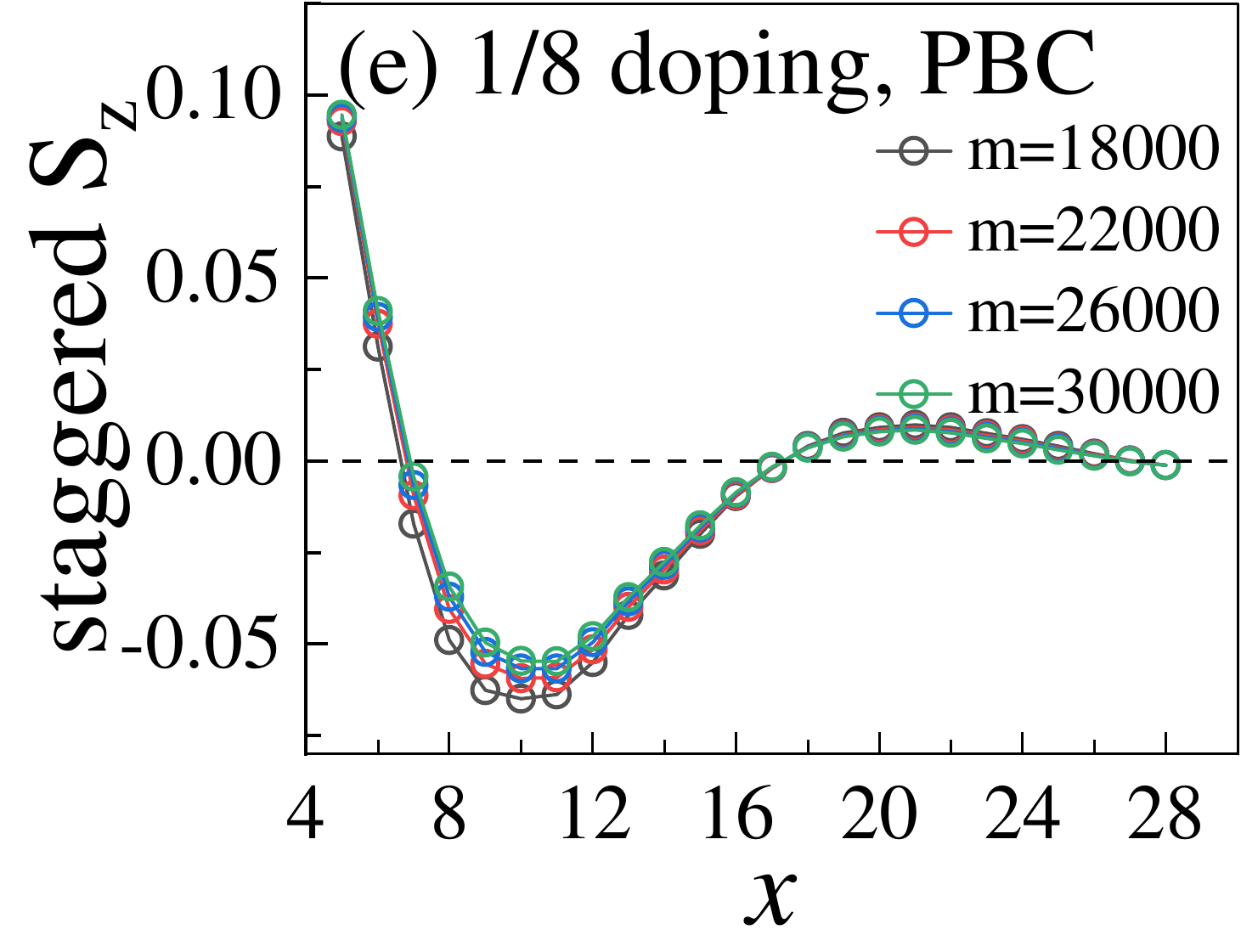}
	\includegraphics[width=0.23\textwidth]{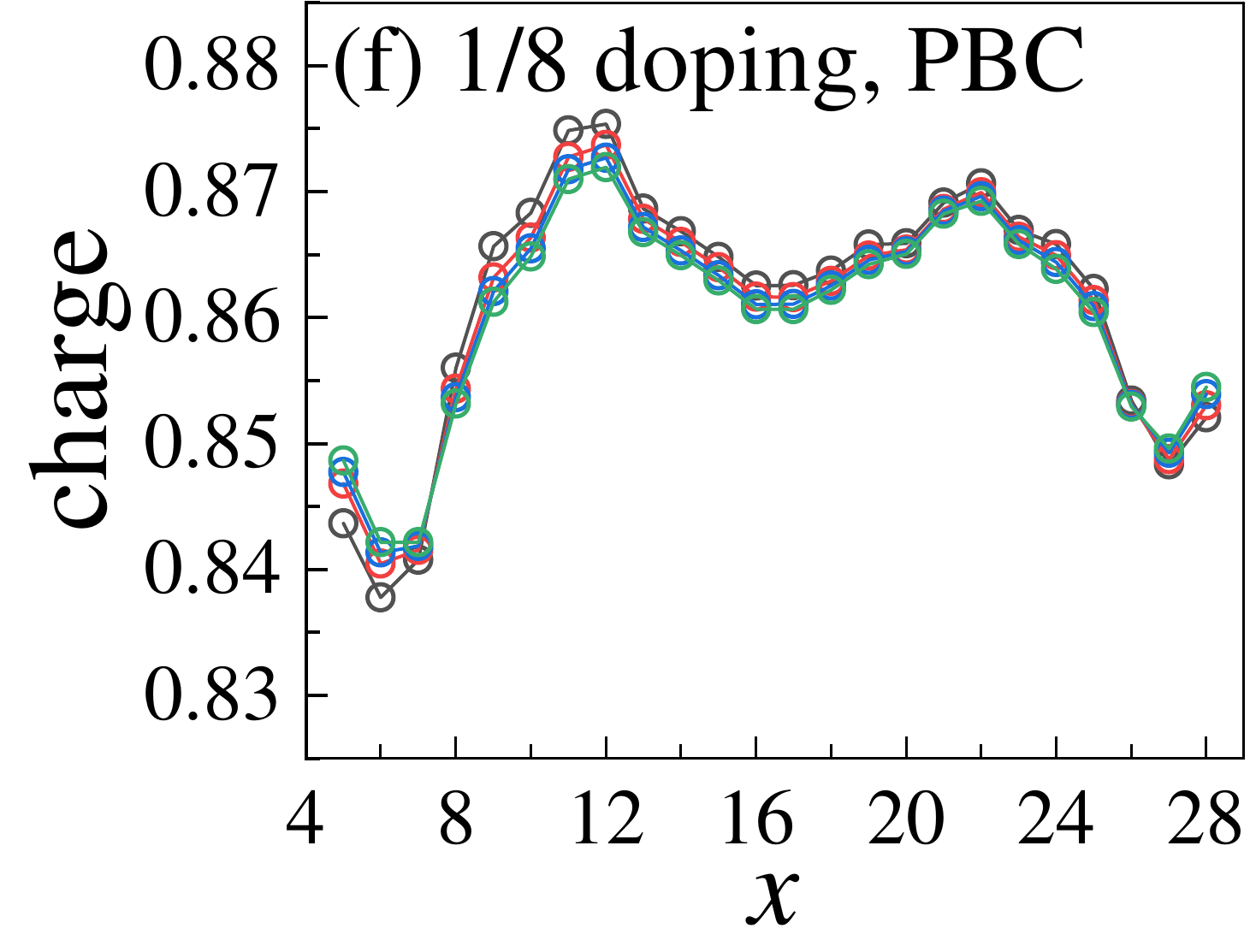}
	\includegraphics[width=0.23\textwidth]{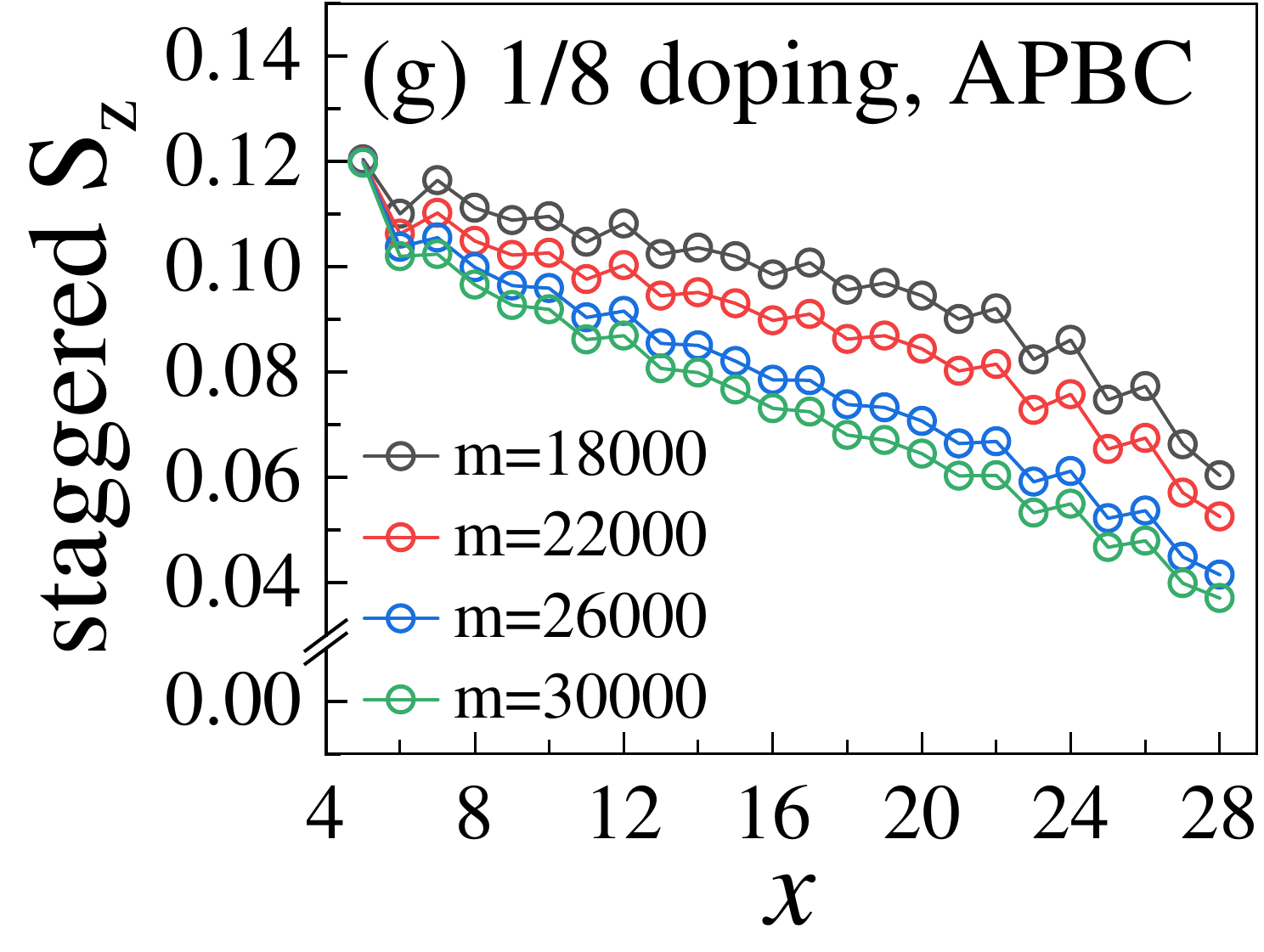}
	\includegraphics[width=0.23\textwidth]{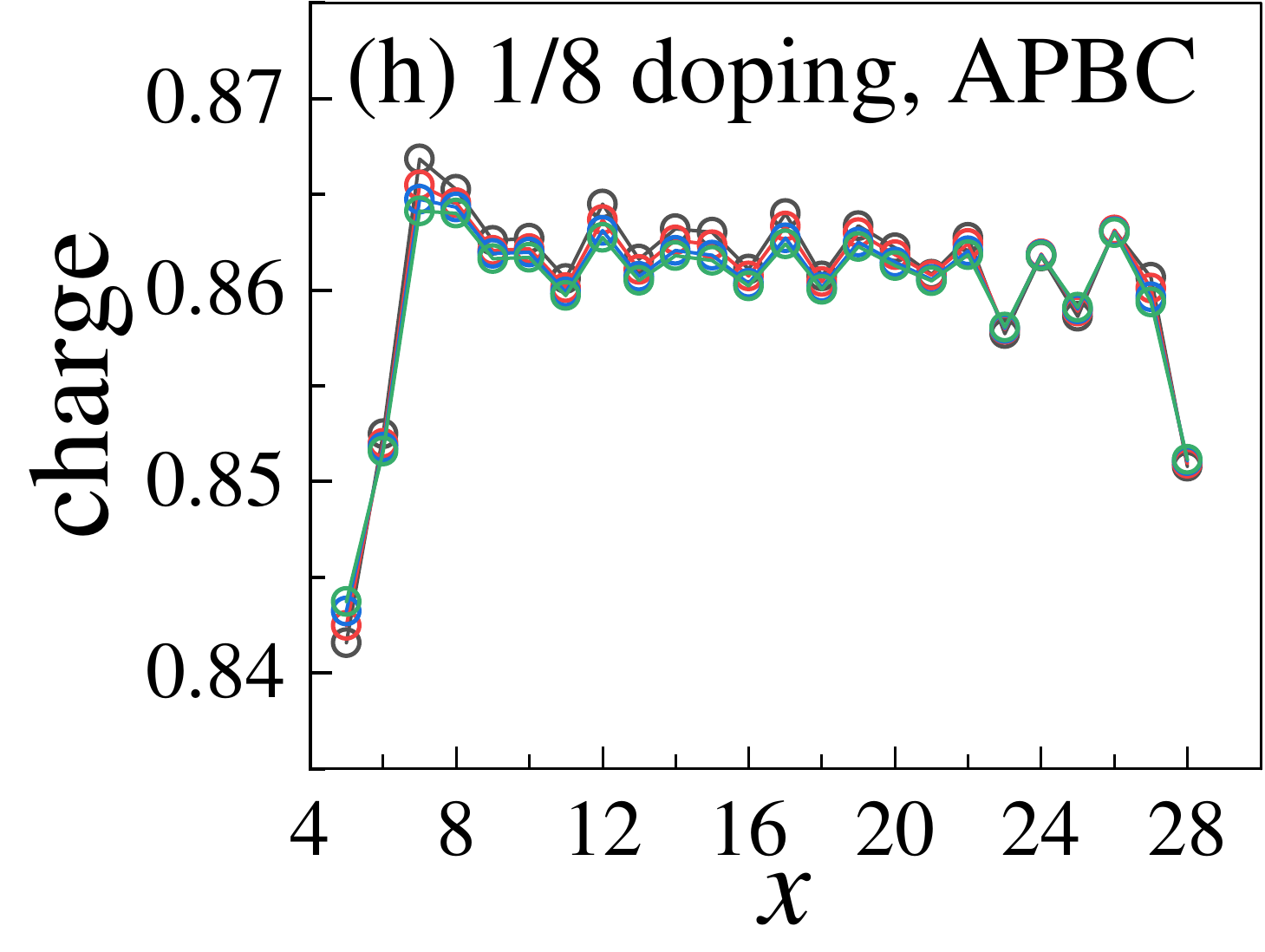}
	
	\includegraphics[width=0.23\textwidth]{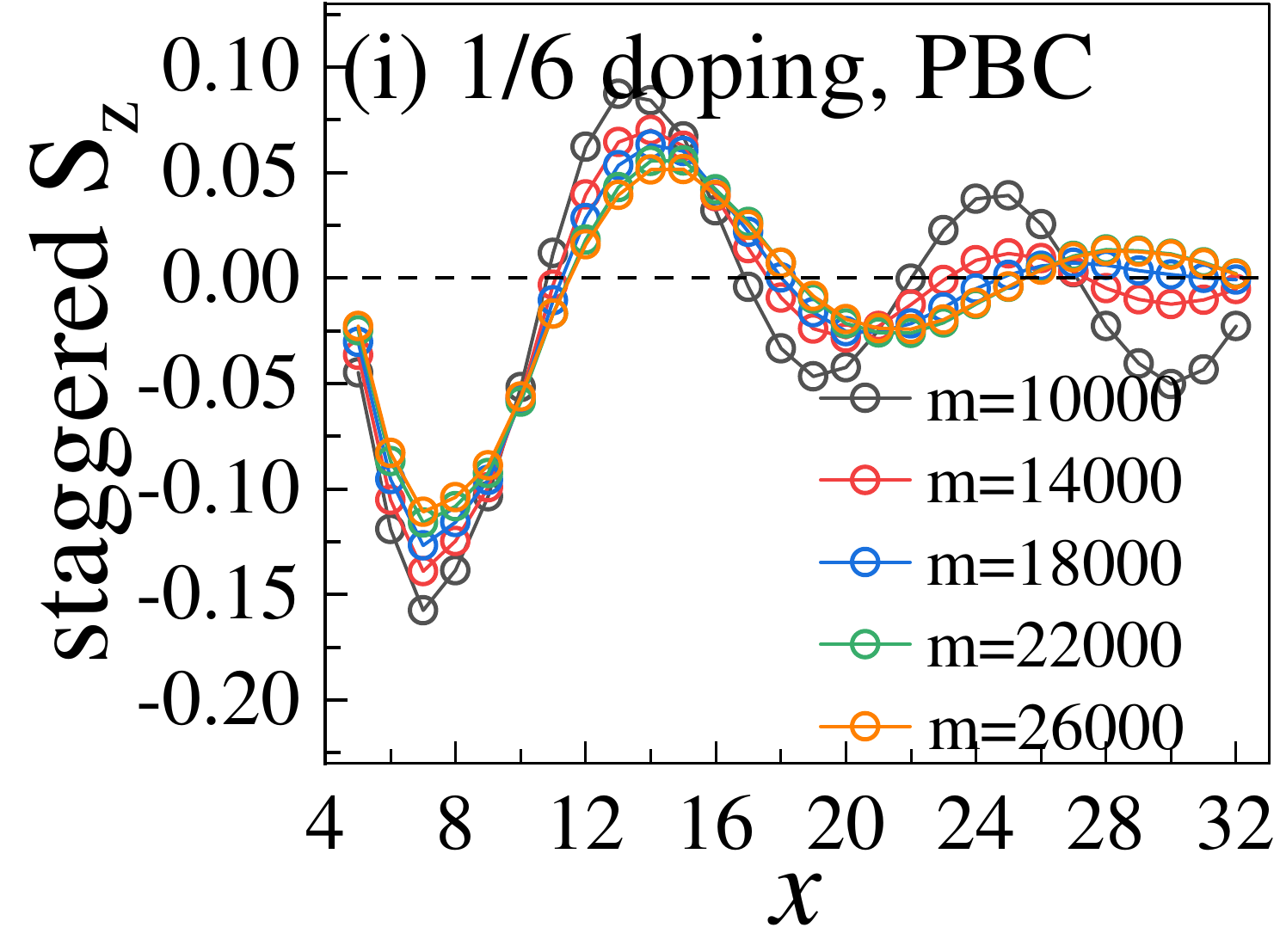}
	\includegraphics[width=0.23\textwidth]{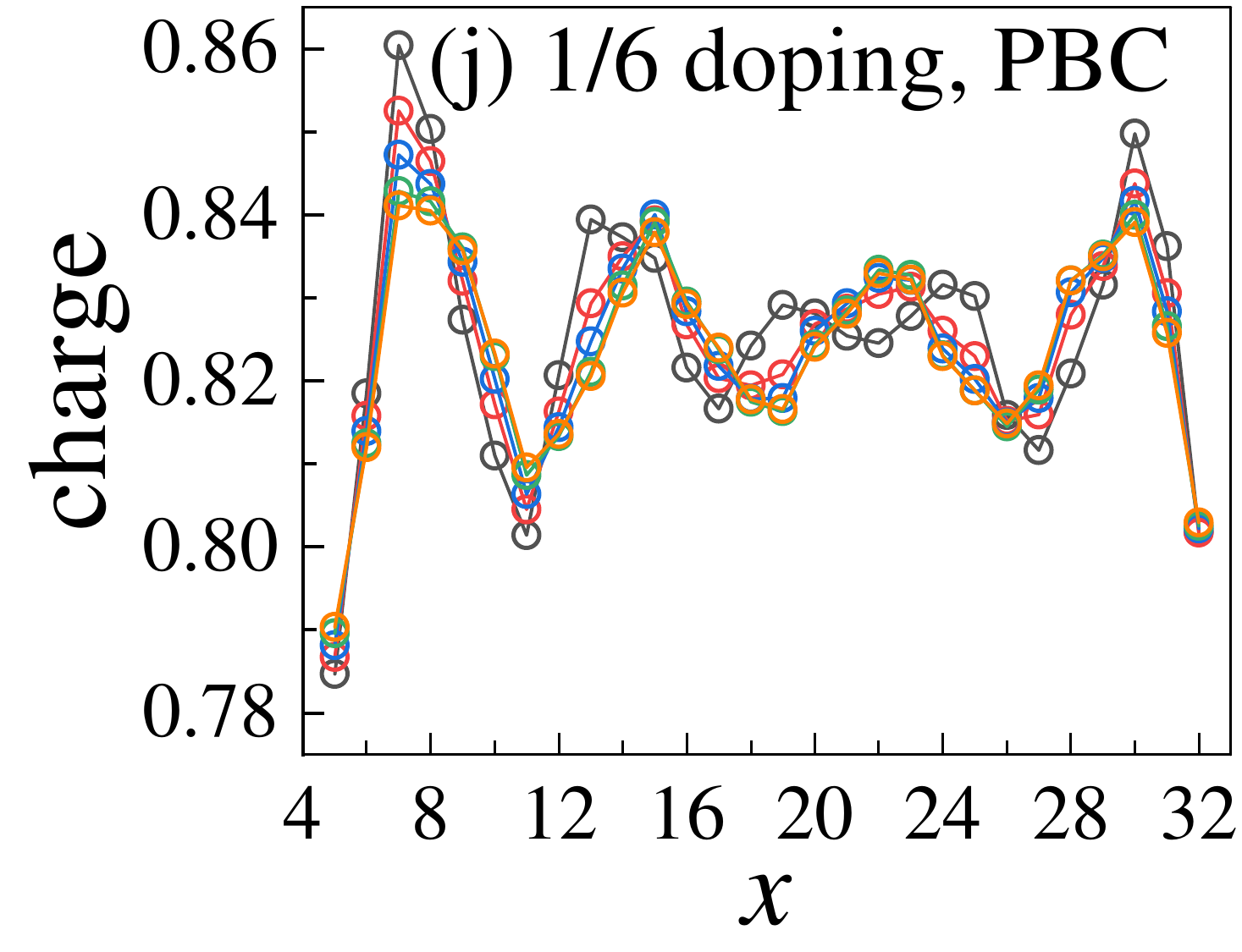}
	\includegraphics[width=0.23\textwidth]{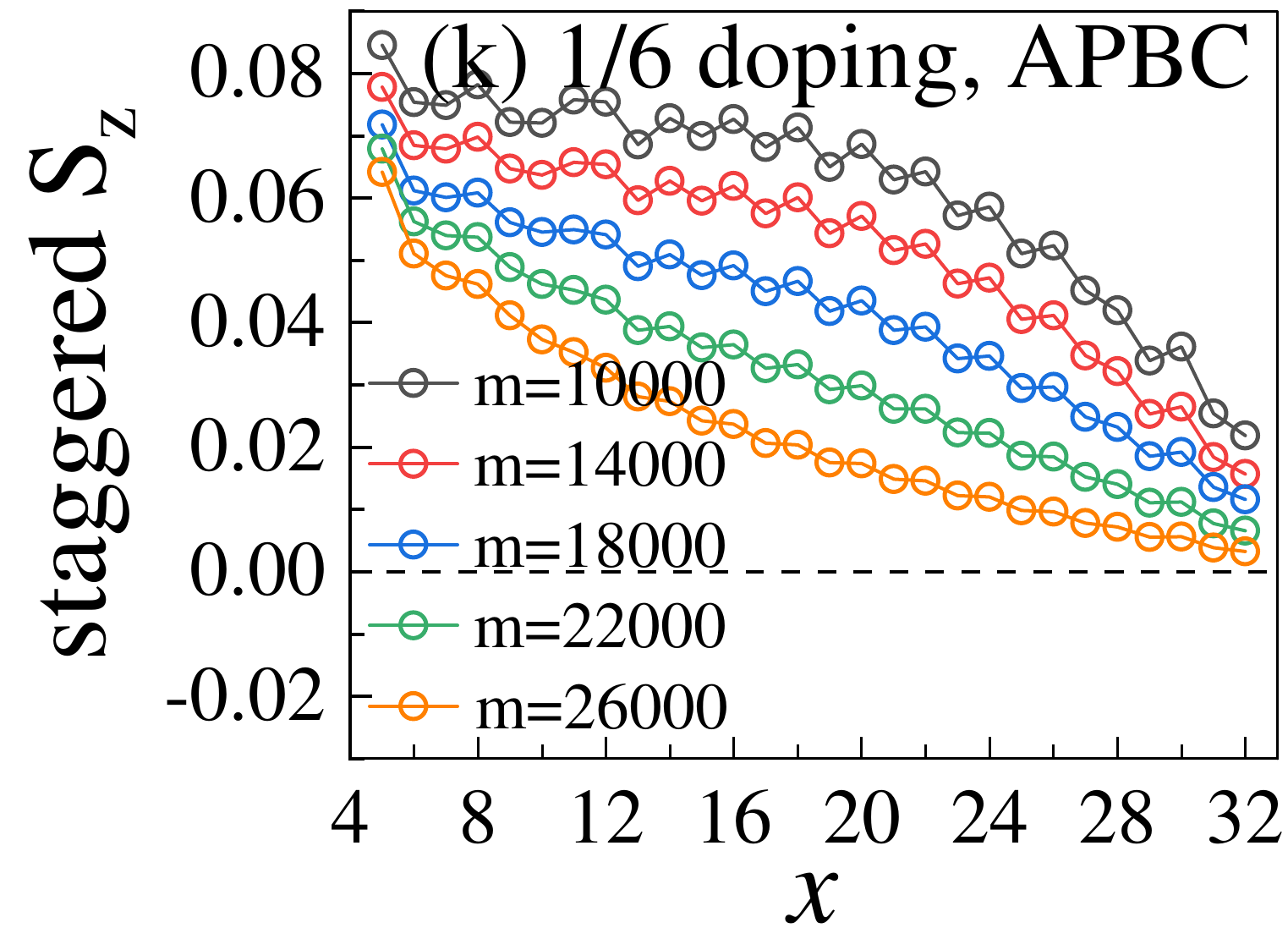}
	\includegraphics[width=0.23\textwidth]{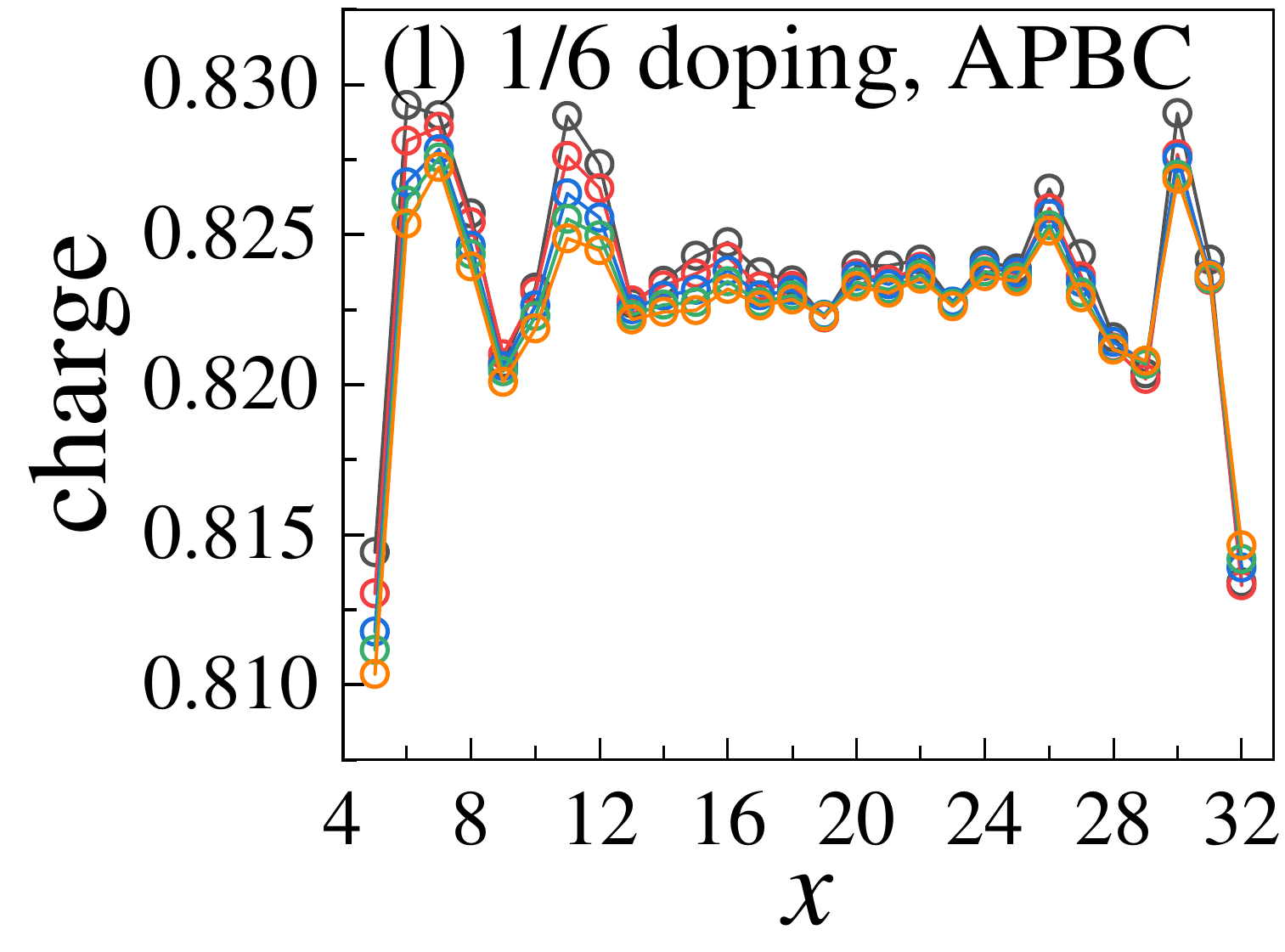}
	\caption{Similar as in Fig.~\ref{width-4} but for systems with width 6. Staggered spin density and electron density for $1/12$ (top), $1/8$ (middle) and $1/6$ (bottom) dopings on width-6 cylinders (system sizes are $6 \times 48$, $6 \times 32$ and $6 \times 36$ for $1/12$, $1/8$ and $1/6$ dopings, respectively) under PBC (left two columns) and APBC (right two columns). Results of finite kept states are plotted to show the convergence of the DMRG calculations. The inset in (a) is plotted in semi-logarithmic scale to show the exponential decay of the staggered spin density for $1/12$ doping under PBC. The dashed horizontal lines represent zero for the staggered spin density.}
	\label{width-6}
\end{figure*}

\begin{figure*}[t]
	\includegraphics[width=0.23\textwidth]{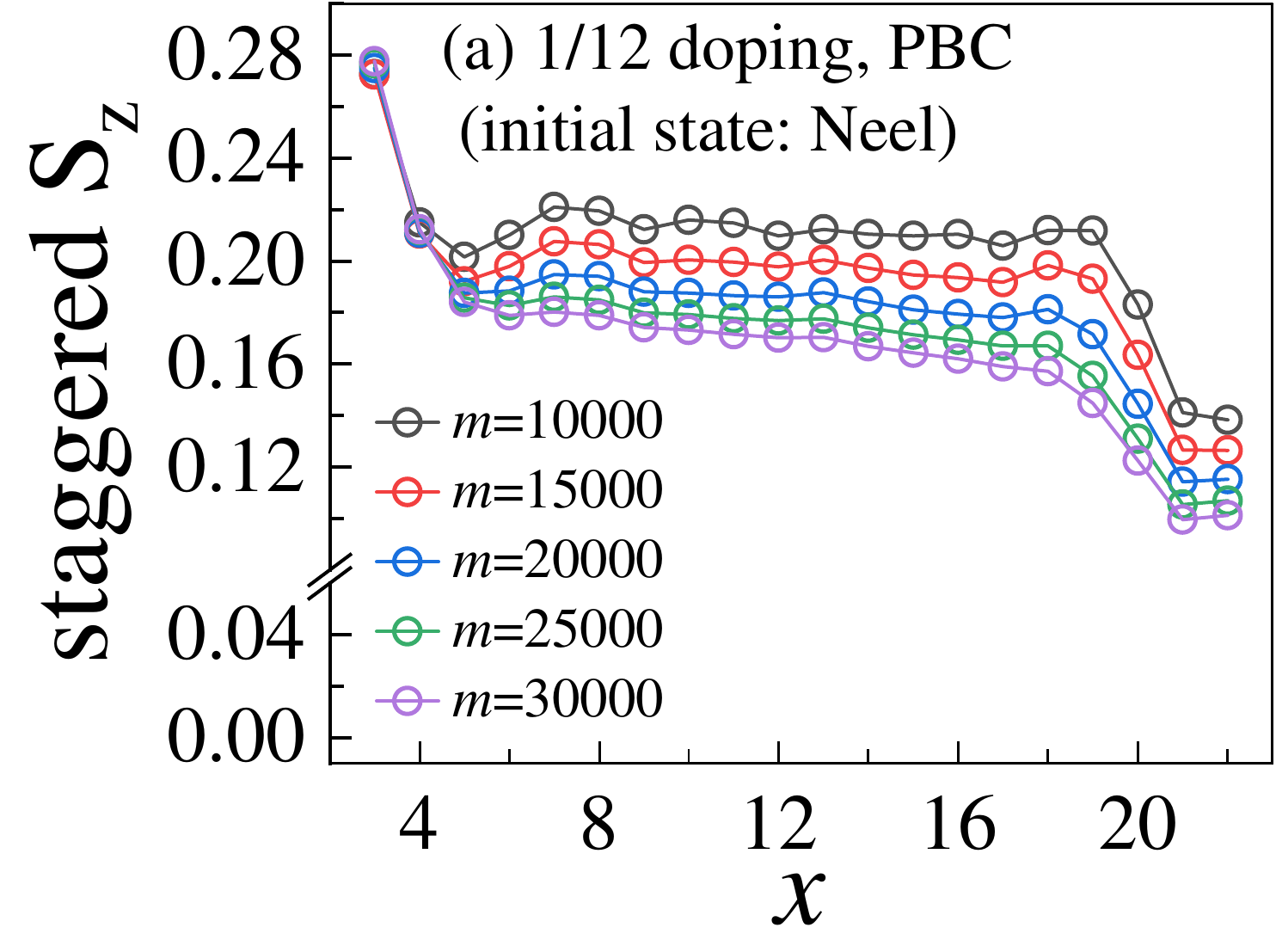}
	\includegraphics[width=0.23\textwidth]{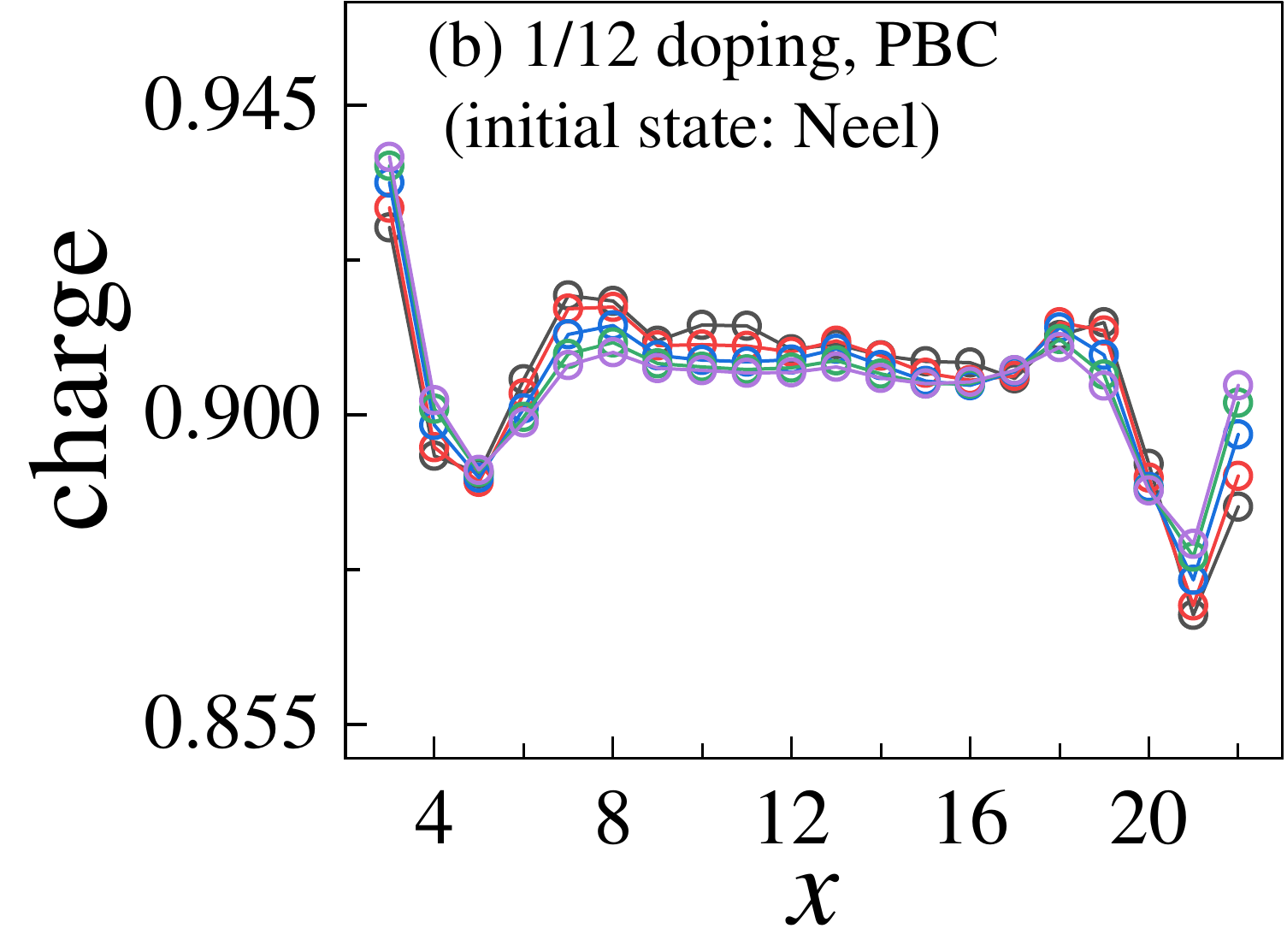}
	\includegraphics[width=0.23\textwidth]{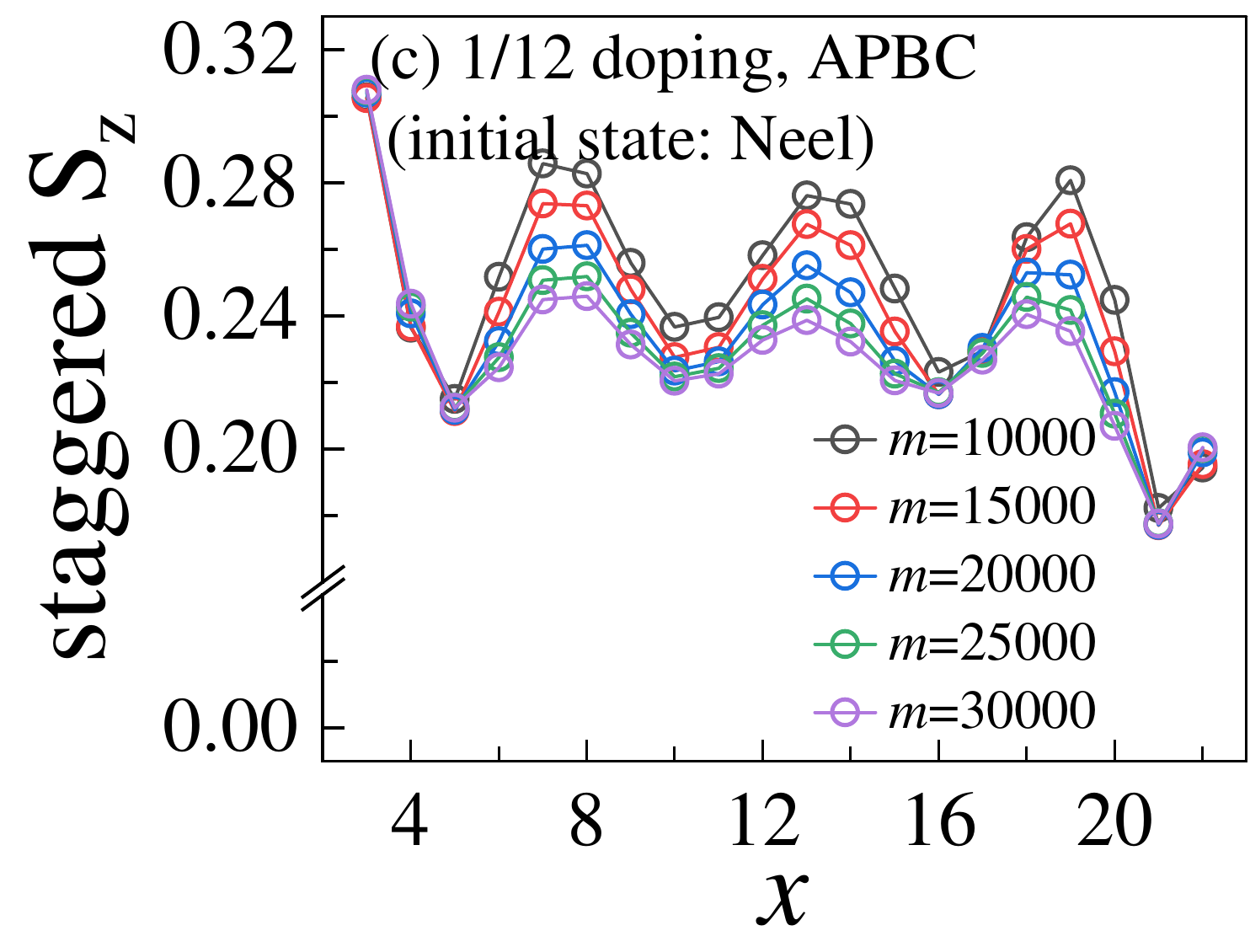}
	\includegraphics[width=0.23\textwidth]{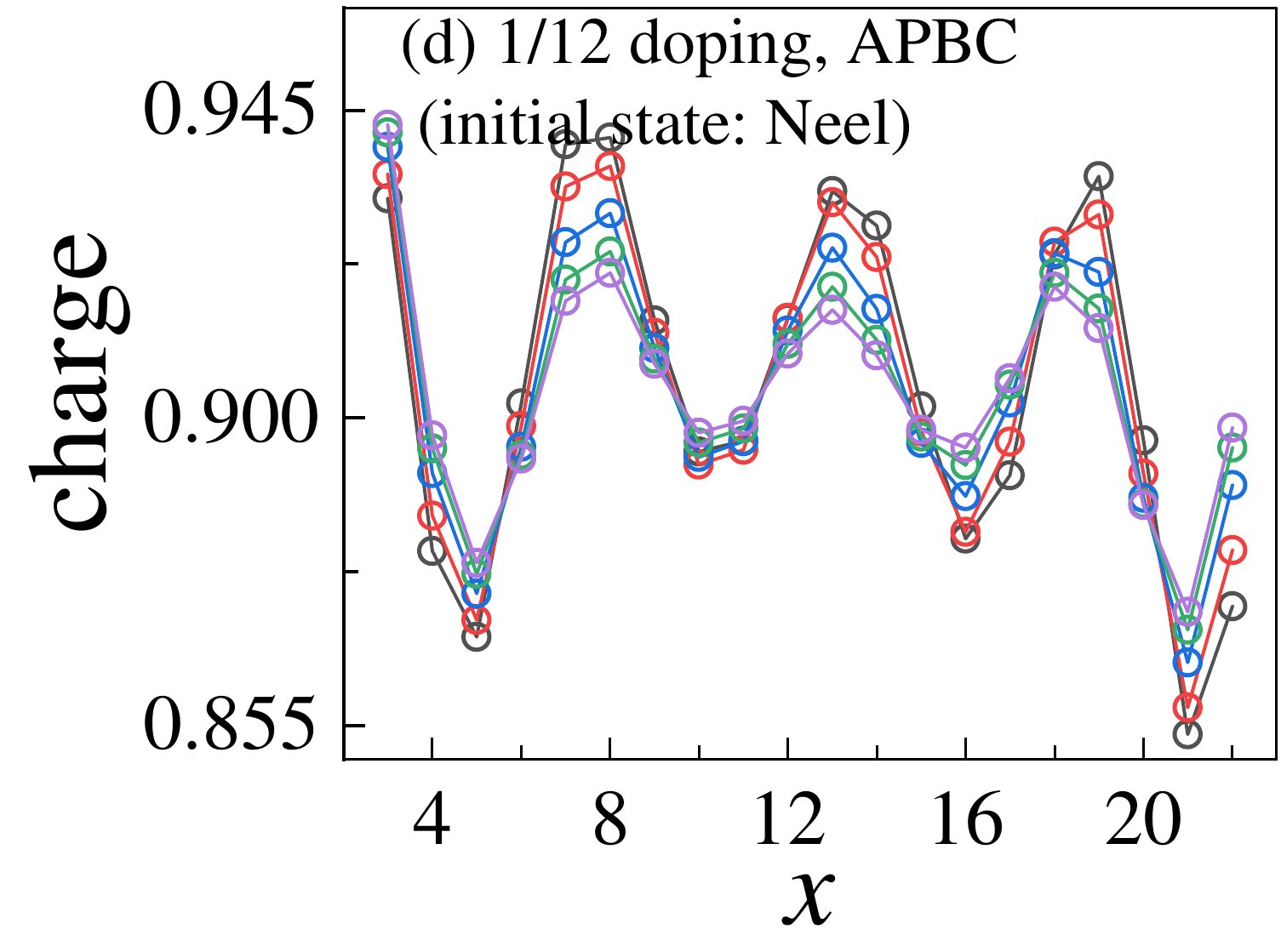}
	
	\includegraphics[width=0.23\textwidth]{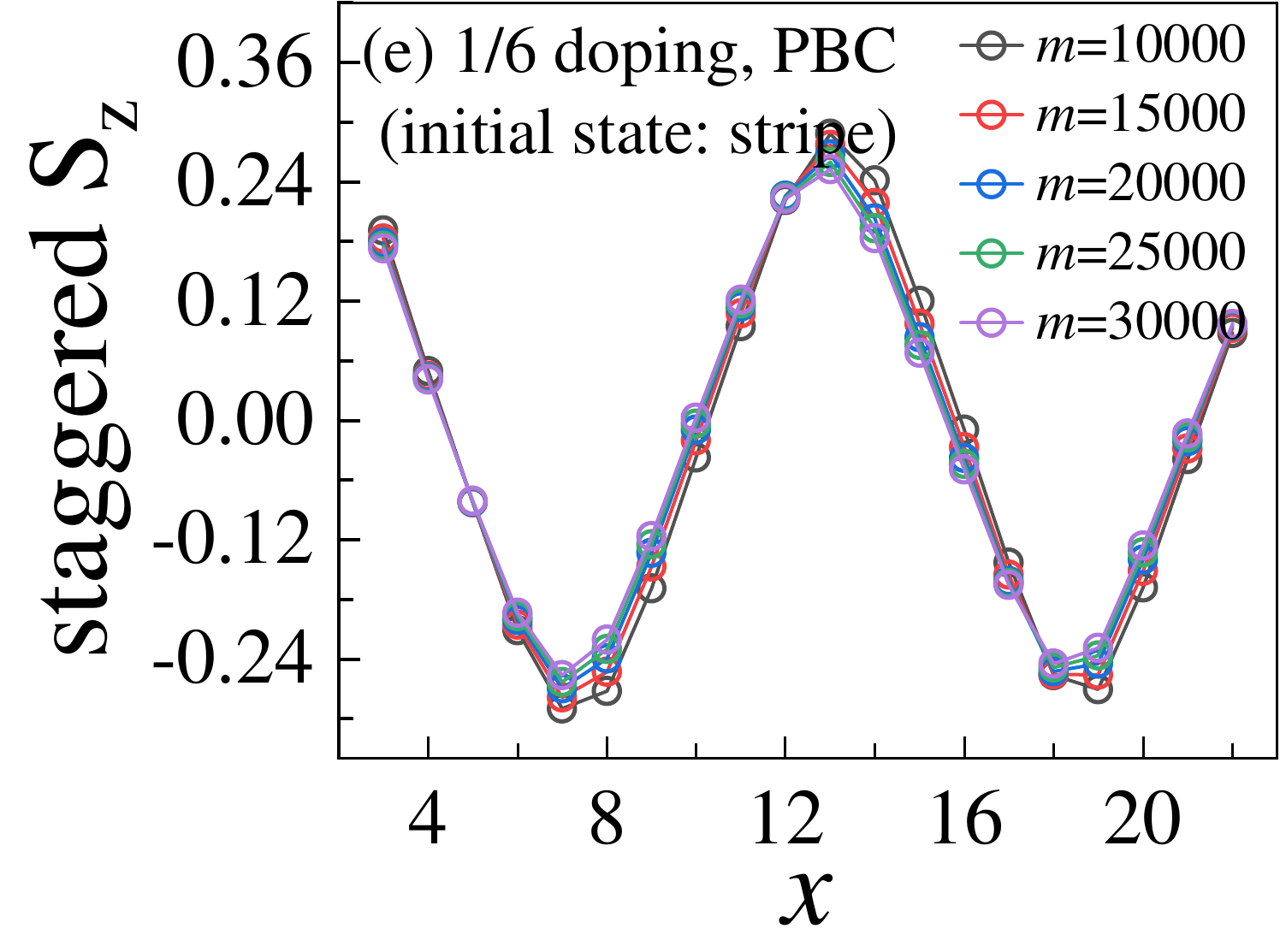}
	\includegraphics[width=0.23\textwidth]{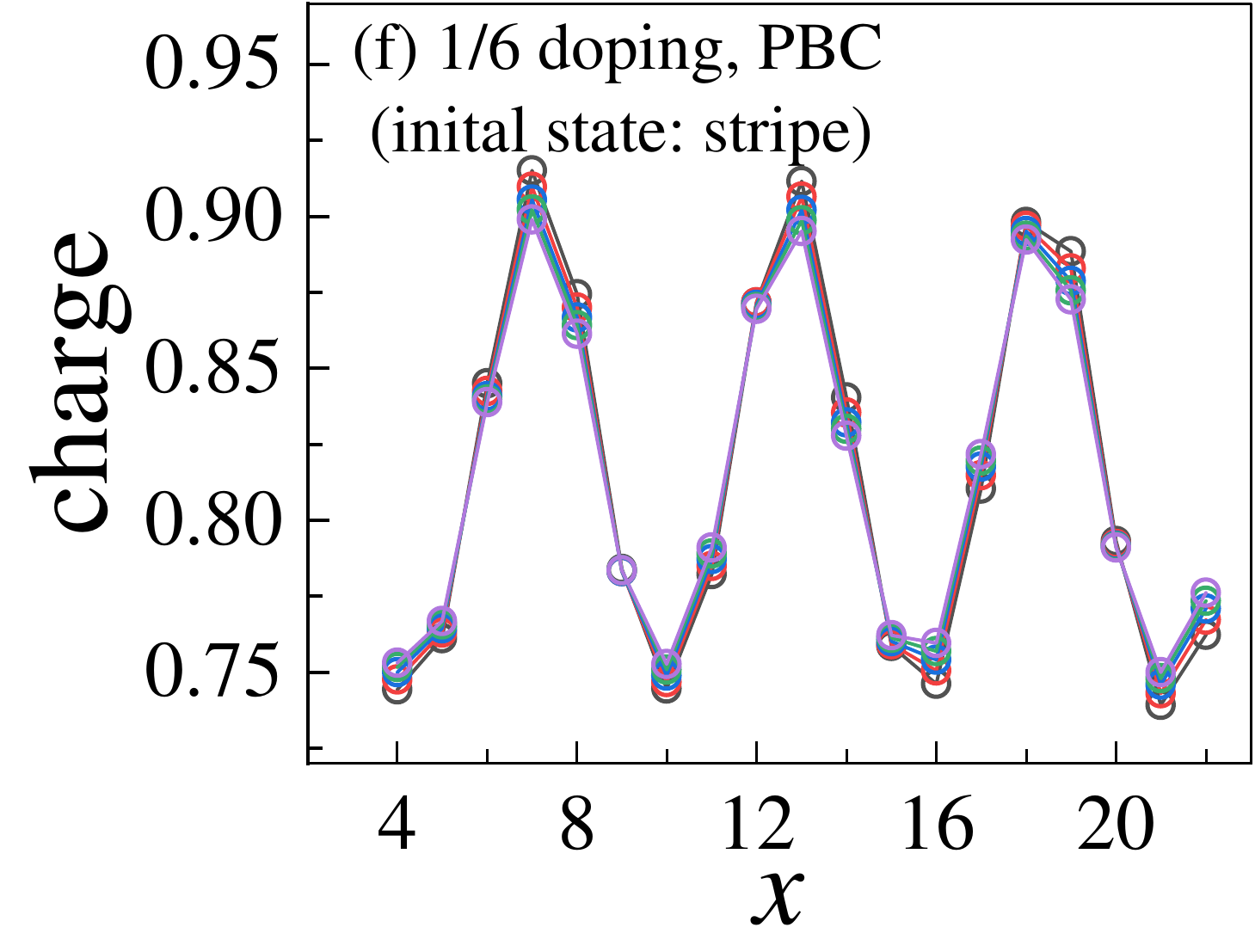}
	\includegraphics[width=0.23\textwidth]{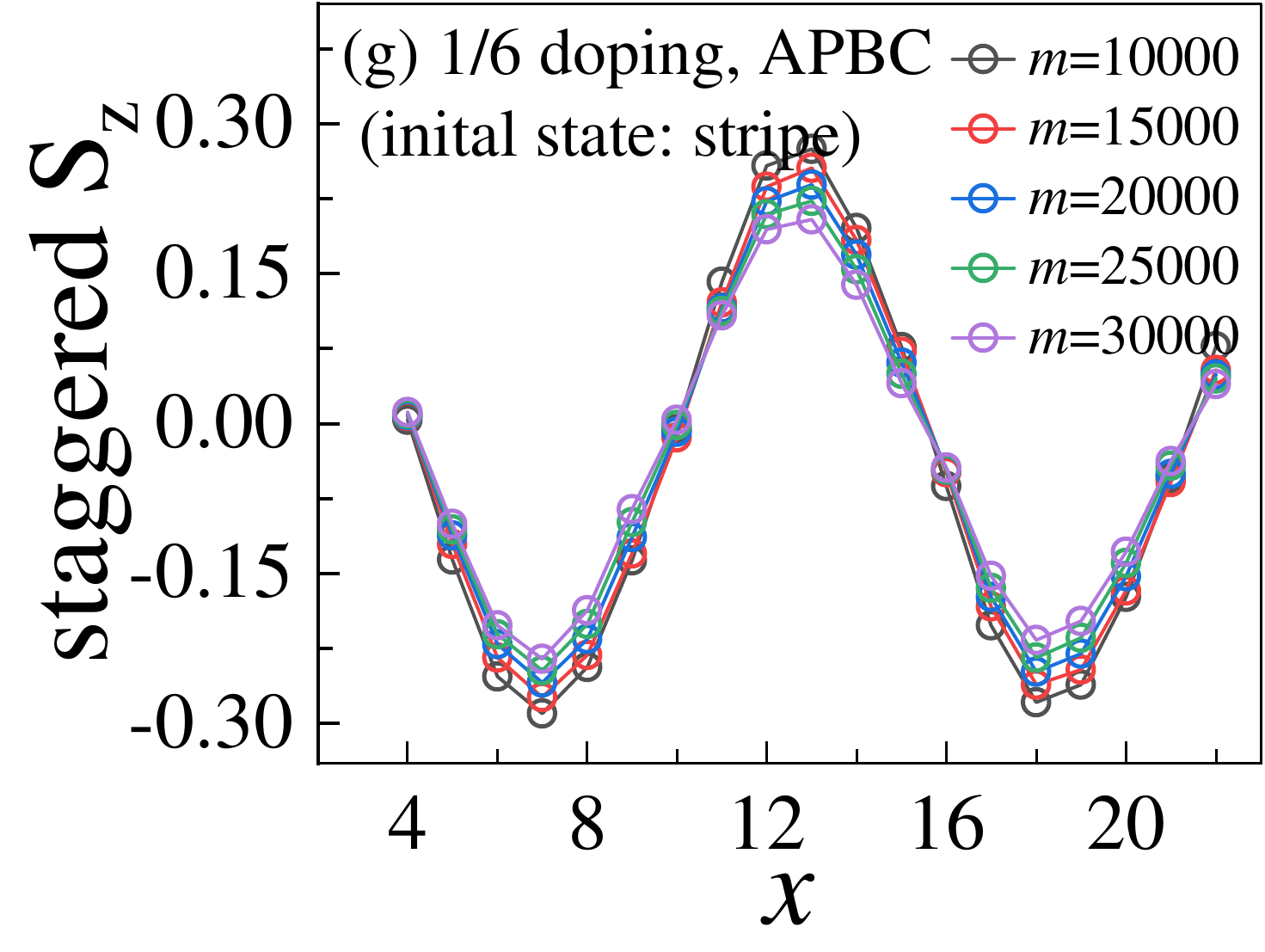}
	\includegraphics[width=0.23\textwidth]{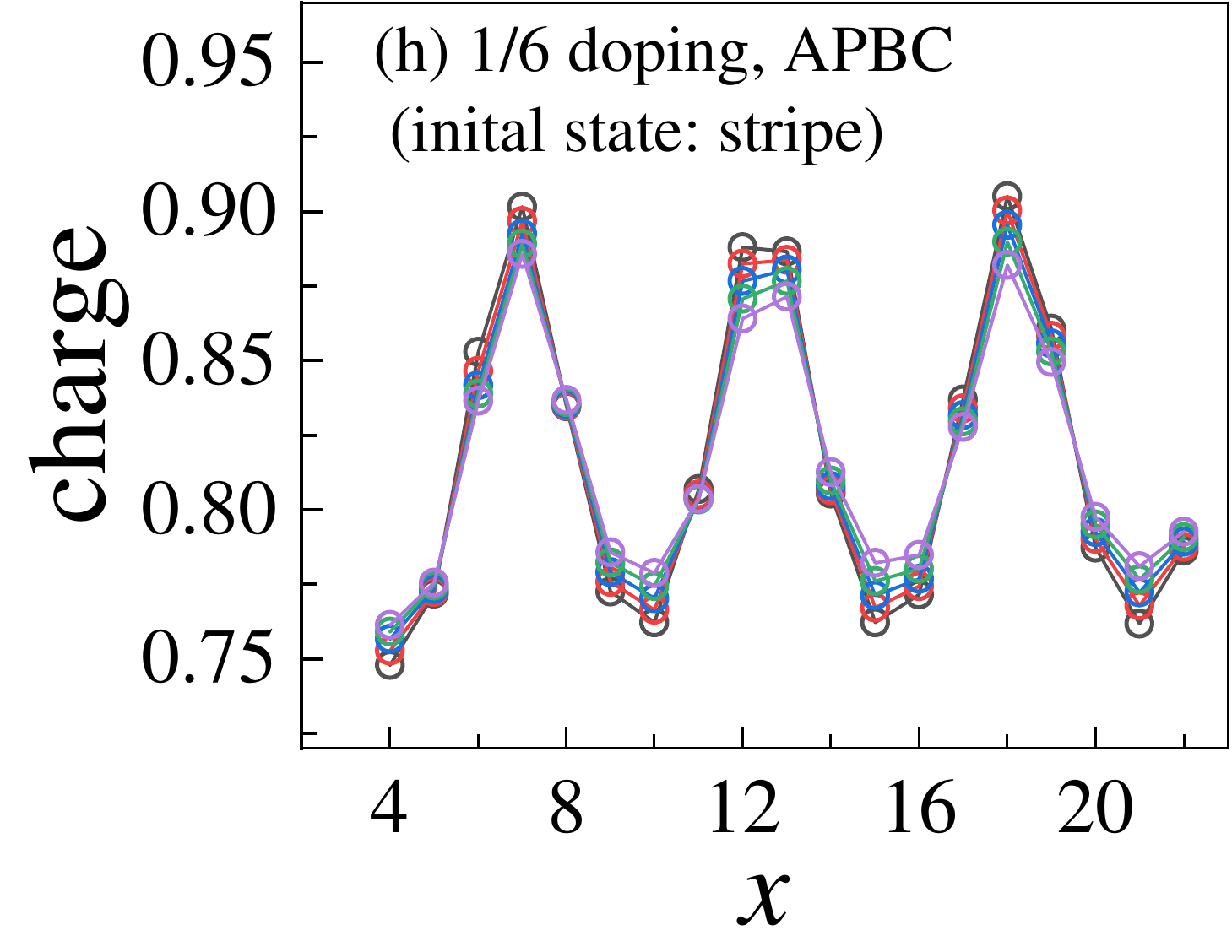}
	
	\caption{Staggered spin density and electron density for $1/12$ (top)and $1/6$ (bottom) dopings (system sizes for both dopings are $8 \times 24$) on width-8 cylinders with PBC (left two columns) and APBC (right two columns). We try both Neel and filled stripe state as the starting states in the DMRG calculations but only Neel (stripe) state can be stabilized in the DMRG sweeps for $1/12$ {($1/6$)} doping. We only show the numerically stabilized states.}
	\label{width-8-1}
\end{figure*}

\begin{figure*}[t]
	\includegraphics[width=0.23\textwidth]{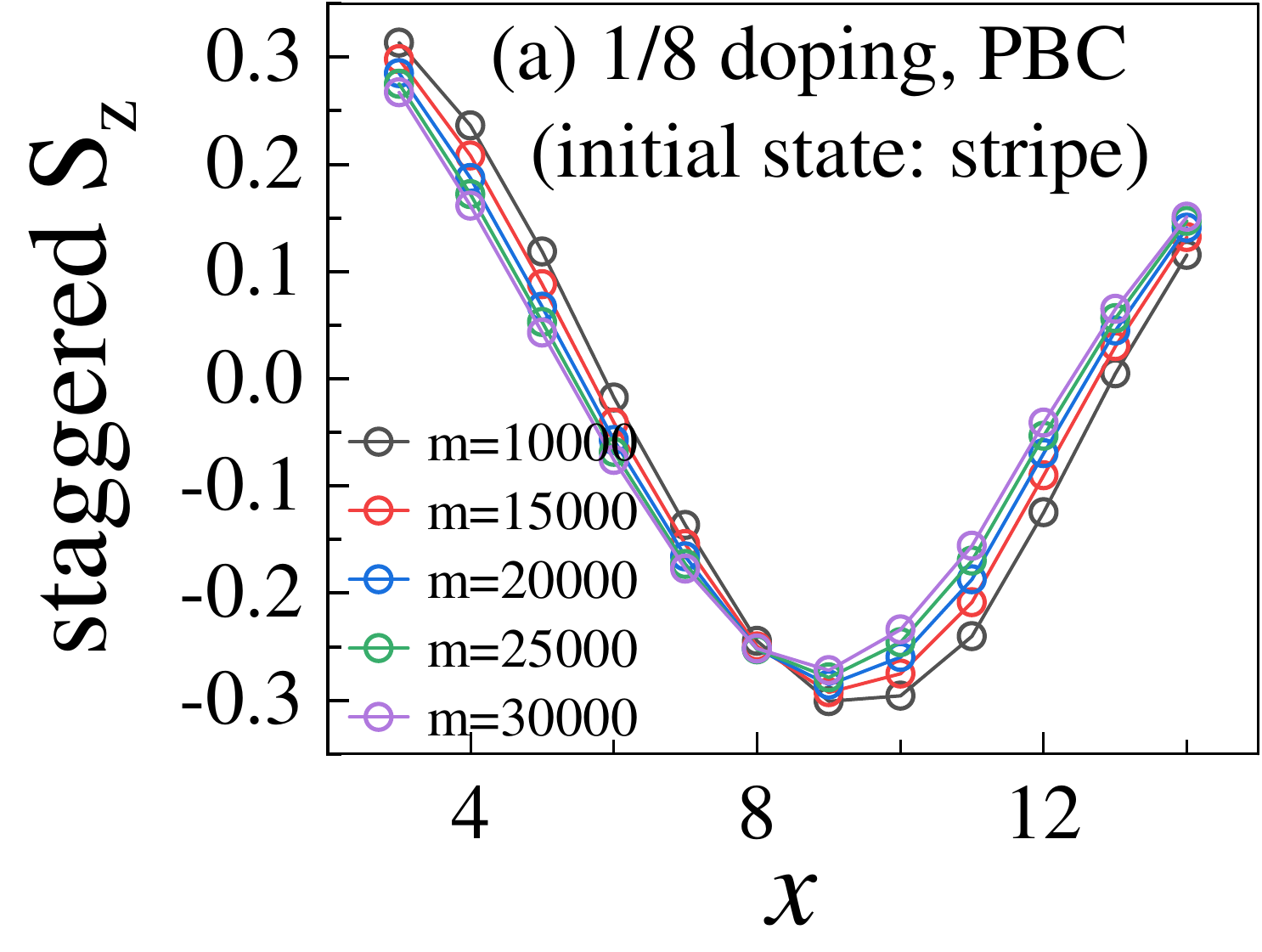}
	\includegraphics[width=0.23\textwidth]{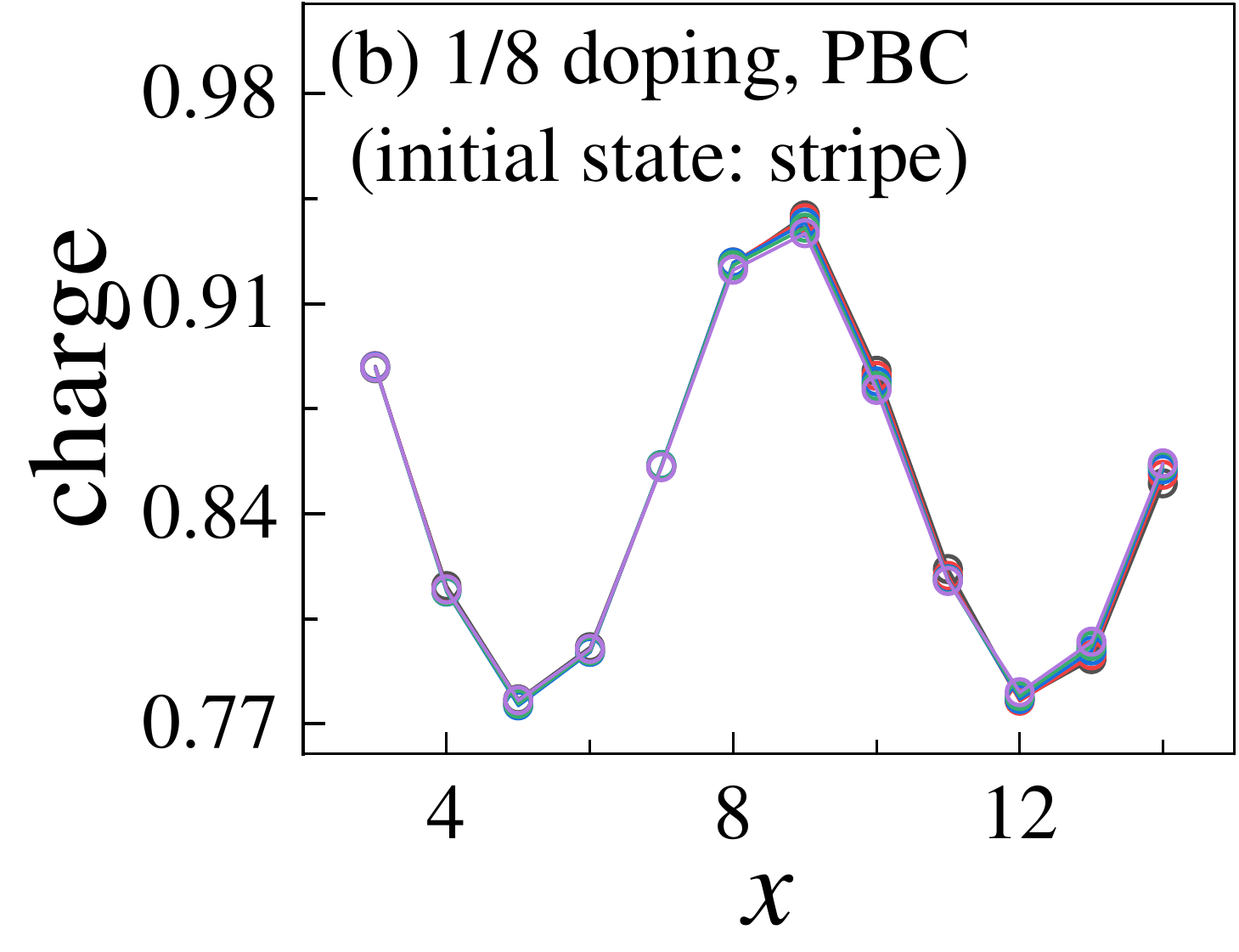}
	\includegraphics[width=0.23\textwidth]{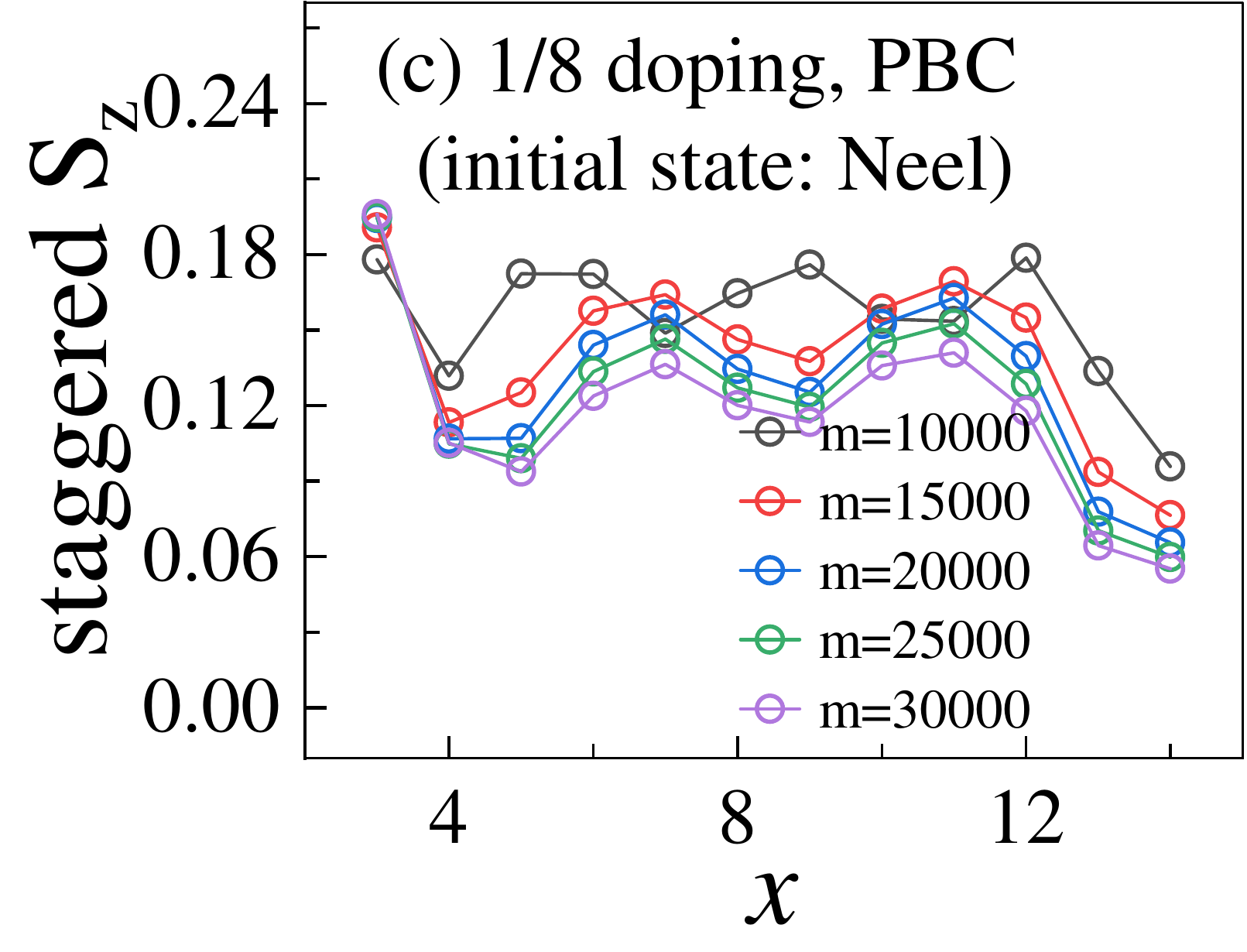}
	\includegraphics[width=0.23\textwidth]{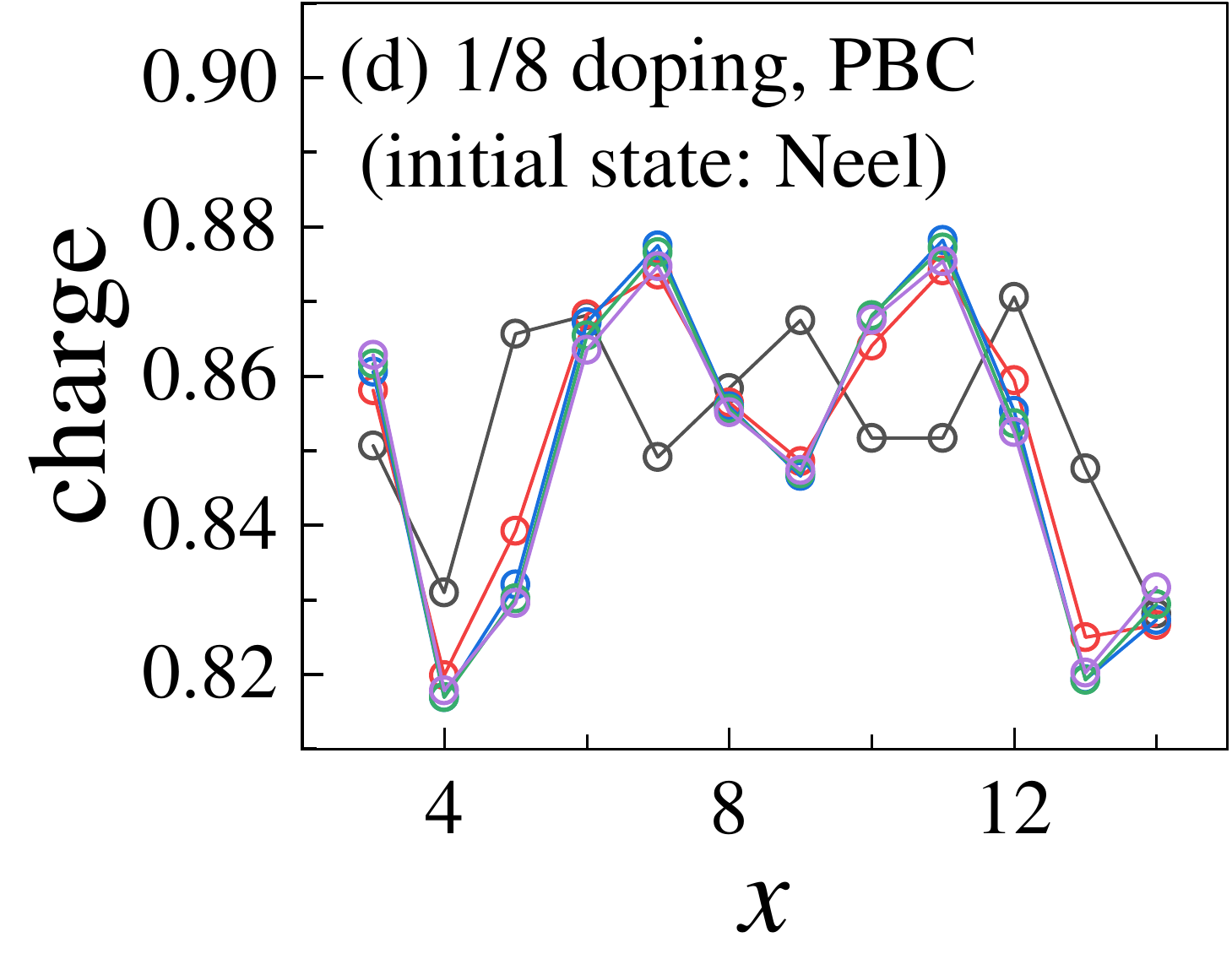}
	\includegraphics[width=0.23\textwidth]{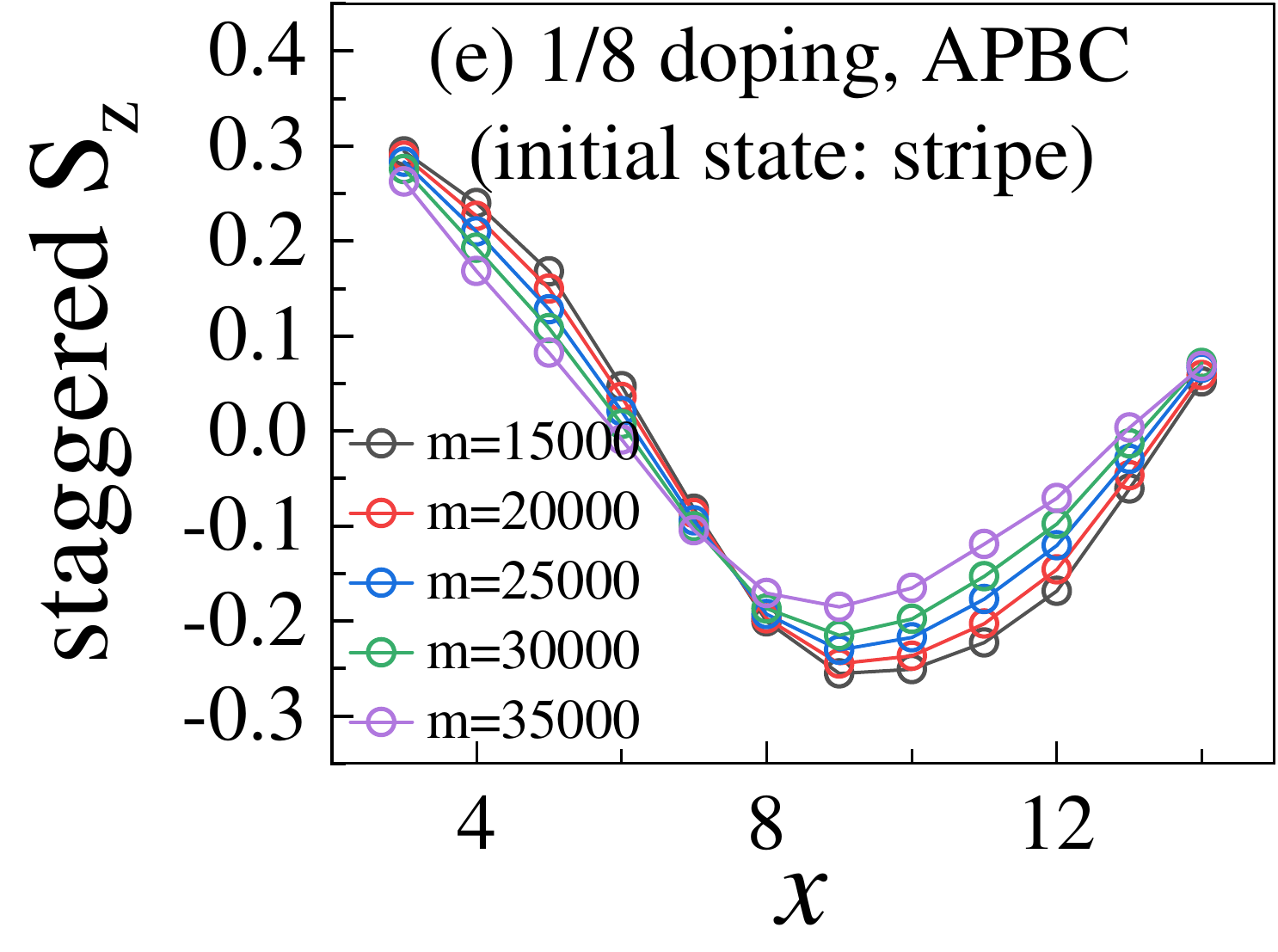}
	\includegraphics[width=0.23\textwidth]{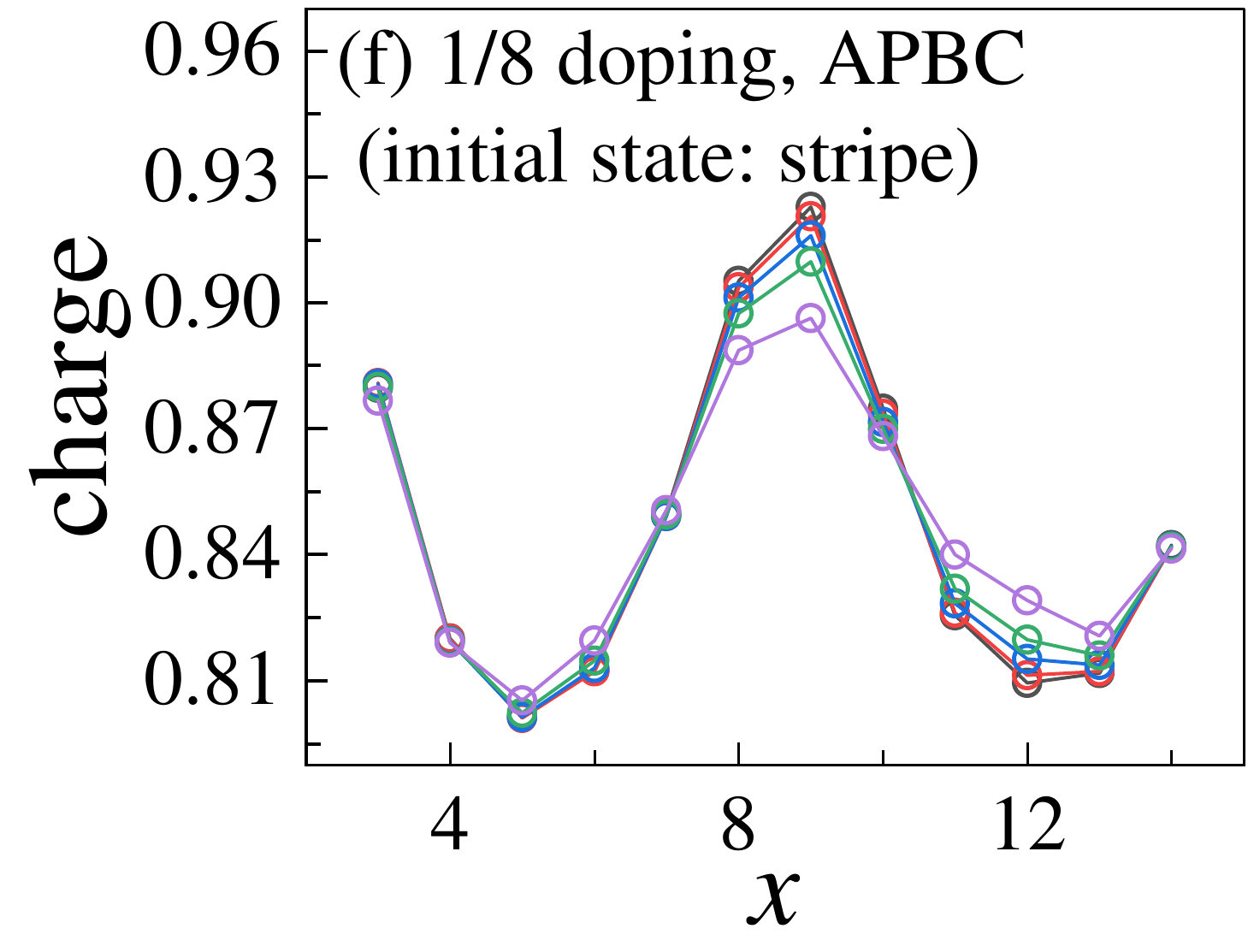}
	\includegraphics[width=0.23\textwidth]{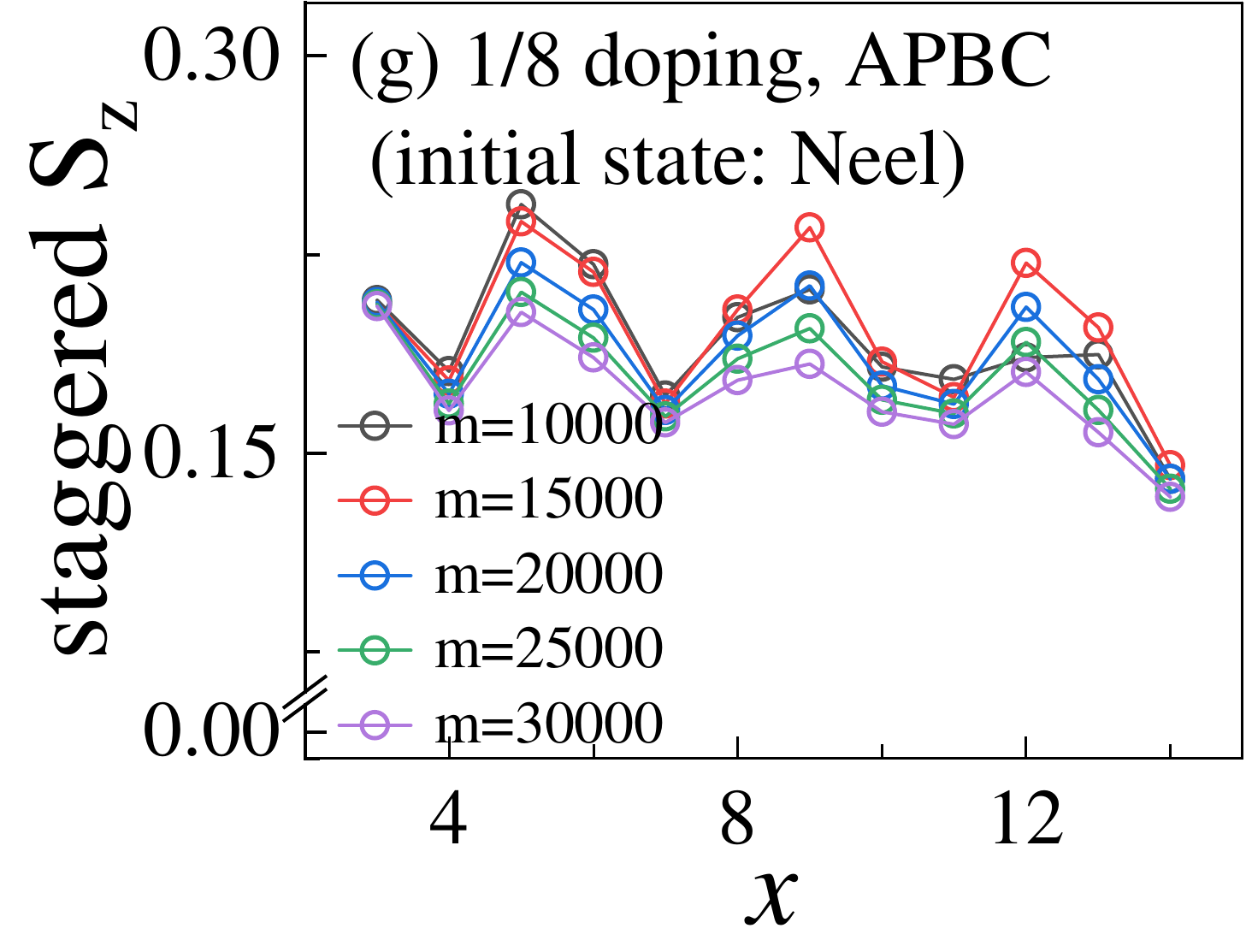}
	\includegraphics[width=0.23\textwidth]{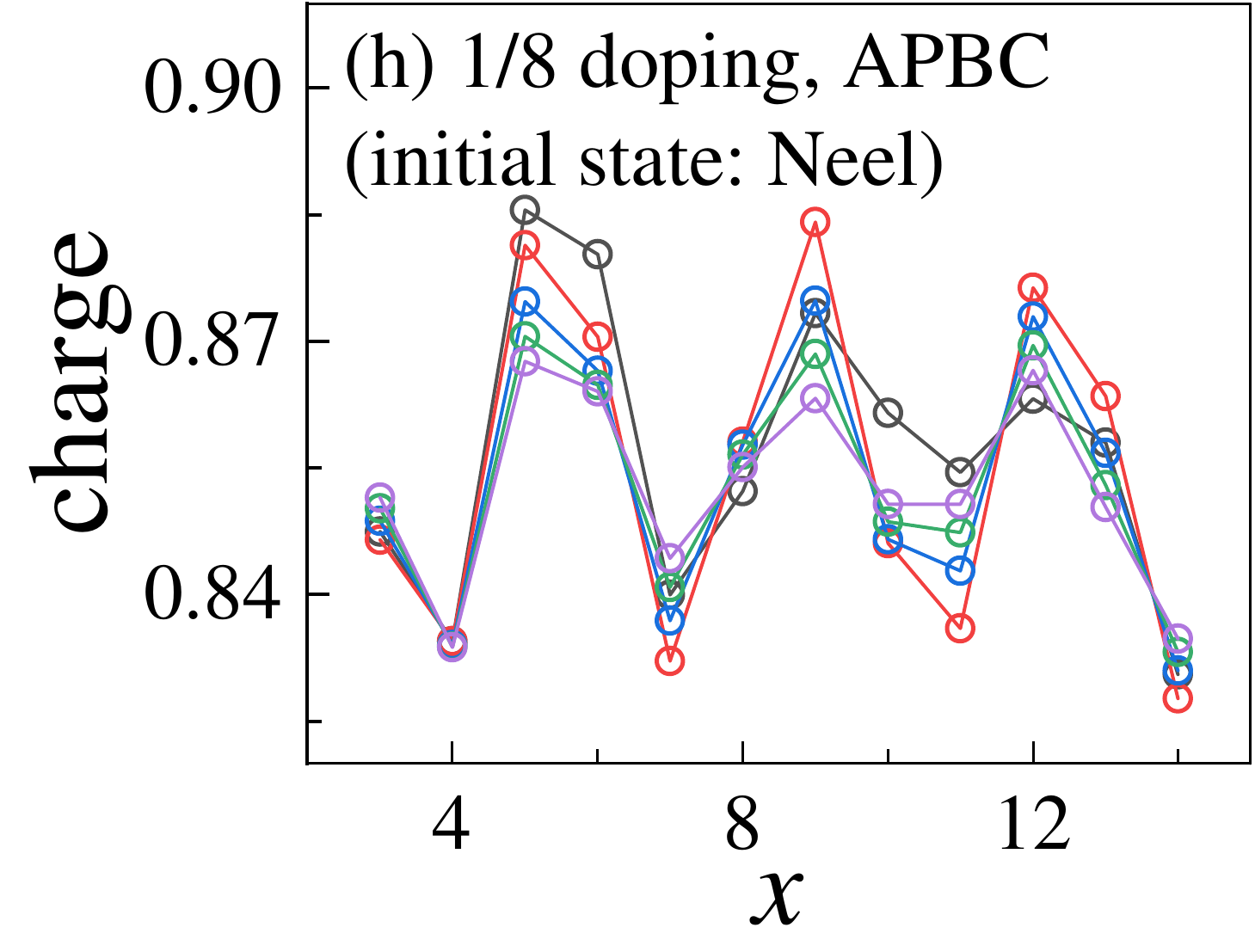}
	
	\caption{Staggered spin density and electron density for $1/8$ doping levels on width-8 cylinders ($8 \times 16$) with PBC (top) and APBC (bottem). In the DMRG sweep, both Neel and filled stripe states can be stabilized.}
	\label{width-8-2}
\end{figure*}

\section{Model and methods} 
\label{Model and methods} 
The Hamiltonian for $t-t'-J$ model is defined as
\begin{equation}
H=-t\sum_{\langle i,j\rangle,s}c_{i,s}^{\dagger}c_{j,s}-t^{\prime}\sum_{\langle\langle i,j\rangle\rangle,s}c_{i,s}^{\dagger}c_{j,s}+J\sum_{\langle i,j\rangle}S_{i}.S_{j}
\end{equation}
where $\hat{c}_{i,s}^{\dagger}(\hat{c}_{i,s})$ creates (annihilates) an electron on site $i=(x_i,y_i)$ with spin $s$ and $\hat{S}_i = 1/2 \times \hat{c}^\dagger_{i,s} \hat{\sigma} \hat{c}_{i,s'}$ ($\hat{\sigma}$ is the Pauli matrix). The doubly occupied states are explicitly excluded in the Hilbert space. We consider electron hopping terms up to next-nearest neighbors ($t'$) and set the nearest neighboring hopping $t$ as energy unit. We choose $J = 1/3$ which is relevant to cuprates. We intend to study the physics related to the electron doped cuprates in this work. But in principle, only hole doping is allowed in $t-t'-J$ model because double occupancy is excluded in the local Hilbert space. However, by connecting the $t-t'-J$ model as the large $U$ limit of the $t'$-Hubbard model and through a particle-hole transformation, we can investigate the electron doped cuprates by studying the hole doped $t-t'-J$ model with the sign of $t'$ reversed from $-0.2$ in cuprates to $0.2$ \cite{ANDERSEN19951573,PhysRevB.98.134501}.  We study cylinders with size $L_x \times L_y$ and impose periodic or anti-periodic boundary conditions (PBC or APBC) along the closed circumference direction. We apply antiferromagnetic pinning fields with strength $h_m =0.5$ on the left open edges in the calculation, which allows us to calculate spin density instead of the more demanding correlation function, and make the DMRG calculation easier to converge. The averaged hole concentration away from half-filling is defined as $\delta=N_h/N$ with $N_h=\sum_i(1-n_i)$ ($\hat{n}_i=\sum_s\hat{c}_{i,s}^{\dagger}\hat{c}_{i,s}$ and $N=L_x \times L_y$) . We study dopings with $\delta = 1/12$, $1/8$, and $1/6$ which represents the most interesting region in the phase diagram of cuprates.

The method we employ in this work is DMRG. In the calculations, the total particle number of each spin species and total $S_z$ are conserved. For width-4 and 6 systems, we push the number of the kept states to ensure the calculation is converged (with truncation error in the order of $10^{-6}$ for width-4 systems) or in the converged region (with truncation error in the order of $10^{-5}$ for width-6 systems). For width-8 systems, within the kept states ($D = 30000$) we can reach, the results depend on the starting state in the DMRG sweep. So we try different states (Neel and stripe) to start the DMRG calculation and try to determine which one is stable during the sweep.

\section{results}
\label{results} 
In Table~\ref{table_sum}, we summarize the results for spin and charge density as well as pair-pair correlation for the $t-t'-J$ model on cylinders with different boundary conditions and widths for dopings $1/12$, $1/8$, and $1/6$. For the pair-pair correlation, $K_{sc}$ ($\xi_{sc}$) implies it has a power (exponential)-law decay. The values of $K_{sc}$ and $\xi_{sc}$ are obtained from the fit of pair-pair correlation with the reference bond set as the 8th vertical bond from the left edge, while the values in the parentheses are results by setting the reference bond as the 4th vertical bond from the left edge.  In Table~\ref{table_sum}, we can find the ground state switches between stripe and antiferromagnetic Neel state with different boundary conditions and widths. For all dopings, the pair-pair correlations are enhanced when the cylinder becomes wider from $4$ to $6$. The DMRG calculations for width 8 systems are not well converged for the largest kept states we can reach ($D = 30000$). But we try different starting states in the DMRG calculation and find that only Neel (stripe) state can be stabilized in DMRG calculation for $1/12$ ($1/6$) doping, while both stripe and Neel states are stable in the DMRG sweep for $1/8$ doping, regardless of the boundary conditions. These results indicate that $1/8$ doping is likely to lie in the boundary of the phase transition between the Neel and stripe phase, consistent with previous study \cite{doi:10.1073/pnas.2109978118}. In the following, we discuss the details of the results.   

\subsection{spin and charge density}
The staggered spin density
is defined as $(-1)^{{i_x + i_y}} \times \langle \hat{S}_i^z \rangle$ with $\hat{S}_i^z=1/2 \times (n_{i\uparrow}-n_{i\downarrow})$ and charge density is defined as the averaged density along the circumference direction
$n_i=\sum_{y=1}^{L_{y}}\left\langle {n}(i_x,i_y)\right\rangle / L_{y}$.

\subsubsection{Width-4 results}

Fig.~\ref{width-4} shows the staggered spin density and electron density for $1/12$ (top), $1/8$ (middle) and $1/6$ (bottom) dopings on width-4 cylinders under PBC (left two columns) and APBC (right two columns). Results for different kept states in DMRG calculation are plotted and we can find the good convergence of the results with kept states.

We first discuss results under PBC. For $1/12$ and $1/8$ dopings, the staggered spin density has the same sign and decays exponentially (see the insets of Fig.~\ref{width-4} (a) and (e)) with the distance to the antiferromagnetic pinning fields on the left open edge, indicating the systems have short-ranged Neel correlation, while the charge density shows a half-filled stripe structure with periodic modulation with $\lambda=1/(2\delta)$. For $1/6$ doping, the spin density displays short-ranged stripe structure and the charge density has a small fluctuation without clear pattern. 

The ground states are different under APBC. For all the three dopings, we find that the staggered spin density displays a spatial modulation with a wavelength ($\lambda=2/\delta$) twice that of the charge density and has a $\pi$ phase flip at the hole-concentrated sites, indicating the systems all have a filled stripe ground state.

\subsubsection{Width-6 results}
The sensitivity of ground state to boundary conditions in width-4 system reflects the finize size effect in the model,
so we further study width-6 systems. We show the staggered spin density and electron density for $1/12$ (top), $1/8$ (middle) and $1/6$ (bottom) dopings on width-6 cylinders under PBC (left two columns) and APBC (right two columns) in Fig.~\ref{width-6}. We plot results with different kept states to show the convergence process. 

For $1/12$ doping under both PBC and APBC, we find the spin density is Neel-type (no sign change in staggered spin density) and charge has $1/3$ filled stripe structure, though the Neel order in spin is short-ranged under PBC but (quasi) long-ranged under APBC. 

For both $1/8$ and $1/6$ dopings under PBC, spin and charge density results indicate the systems have short-ranged filled stripe order.
But under APBC, they have quasi-long-ranged Neel order in spin with nearly uniform charge density. 

It seems that the antiferromagnetic Neel correlation in spin density at $1/12$ doping is well established on the width-6 cylinders. But for $1/8$ and $1/6$ dopings, the ground state changes under different boundary conditions. We also notice that the ground states are different between width-4 and width-6 systems with same doping and boundary conditions, indicating the large finite size effect in the model.

\subsubsection{Width-8 results}
As expected, the DMRG calculation for width-8 systems is challenging. For the largest kept states we can reach (D = 30000), the truncation error in DMRG calculation is still in the order of $10^{-4}$, which indicates the calculation is not well converged. Nevertheless, we perform DMRG calculations with different starting states (antiferromagnetic Neel and filled stripe states) and try to determine whether they are stable in the DMRG sweep. We find that for width-8 system, only Neel (stripe) state can be stabilized in DMRG calculation for $1/12$ ($1/6$) doping, while both stripe and Neel states are stable in the DMRG sweep for $1/8$ doping, regardless of the boundary condtitions. We show the stable states in Fig.~\ref{width-8-1} for $1/12$ and $1/6$ dopings and in Fig.~\ref{width-8-2} for $1/8$ doping. We notice that the Neel state at $1/12$ doping persists in the width-8 system from width-6 system and the charge is more uniform in width-8 system. The fact that both filled stripe and Neel states are stable in the DMRG sweep for $1/8$ doping indicates that $1/8$ doping is likely to lie in the boundary of the phase transition between the Neel state at lower doping and the stripe state at higher doping.

In the electron doped side of the phase diagram of cuprates, the antiferromagnetic Neel phase in the parent compound stretches to electron doping larger than $0.1$. In our calculation, even though a reliable extrapolation with system width is infeasible, it is highly likely that the ground state for $1/12$ doping has antiferromagnetic Neel order. In this sense, the $t-t'-J$ model is in accordance with cuprates.      

\begin{figure}[t]
	\includegraphics[width=0.23\textwidth]{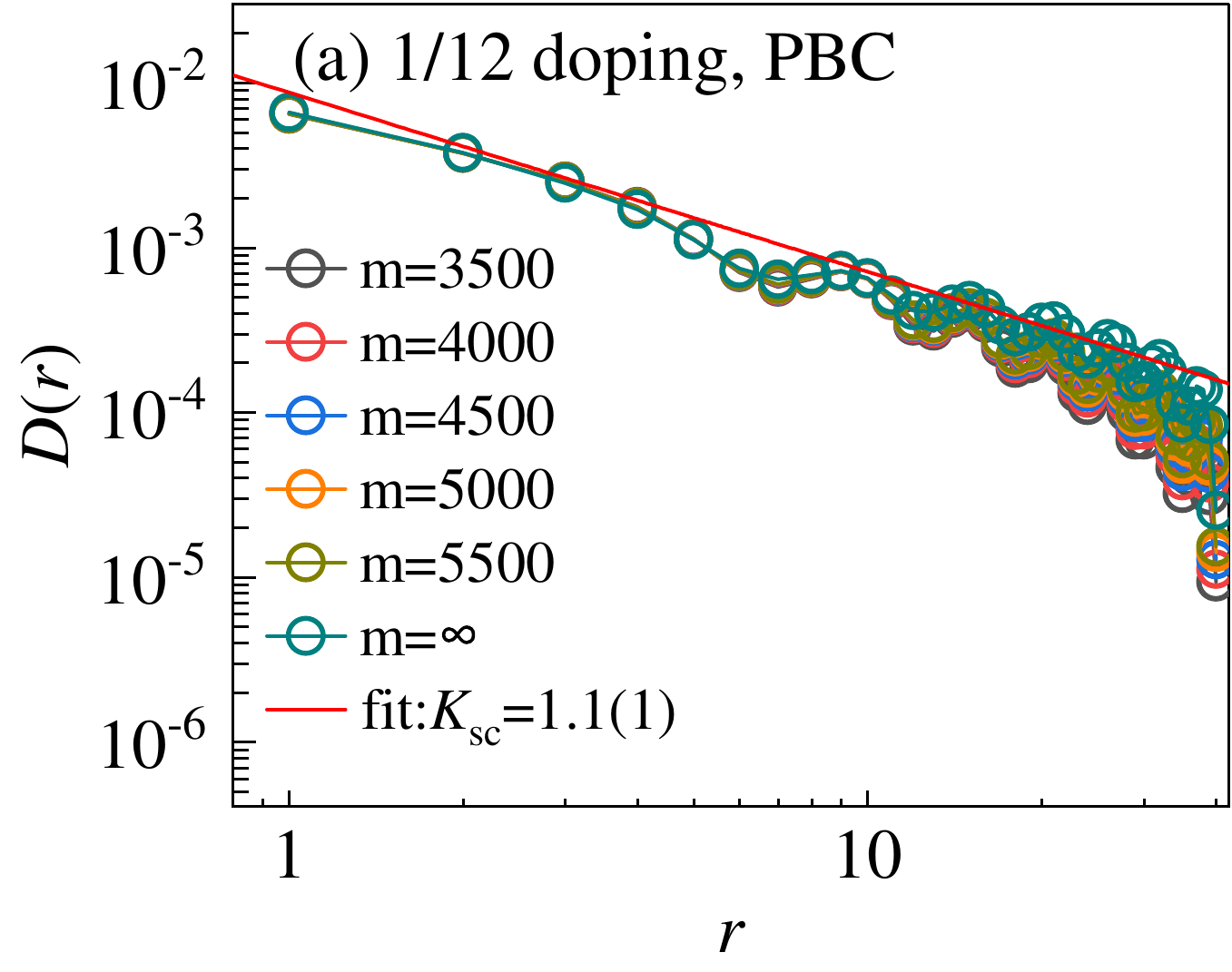}
	\includegraphics[width=0.23\textwidth]{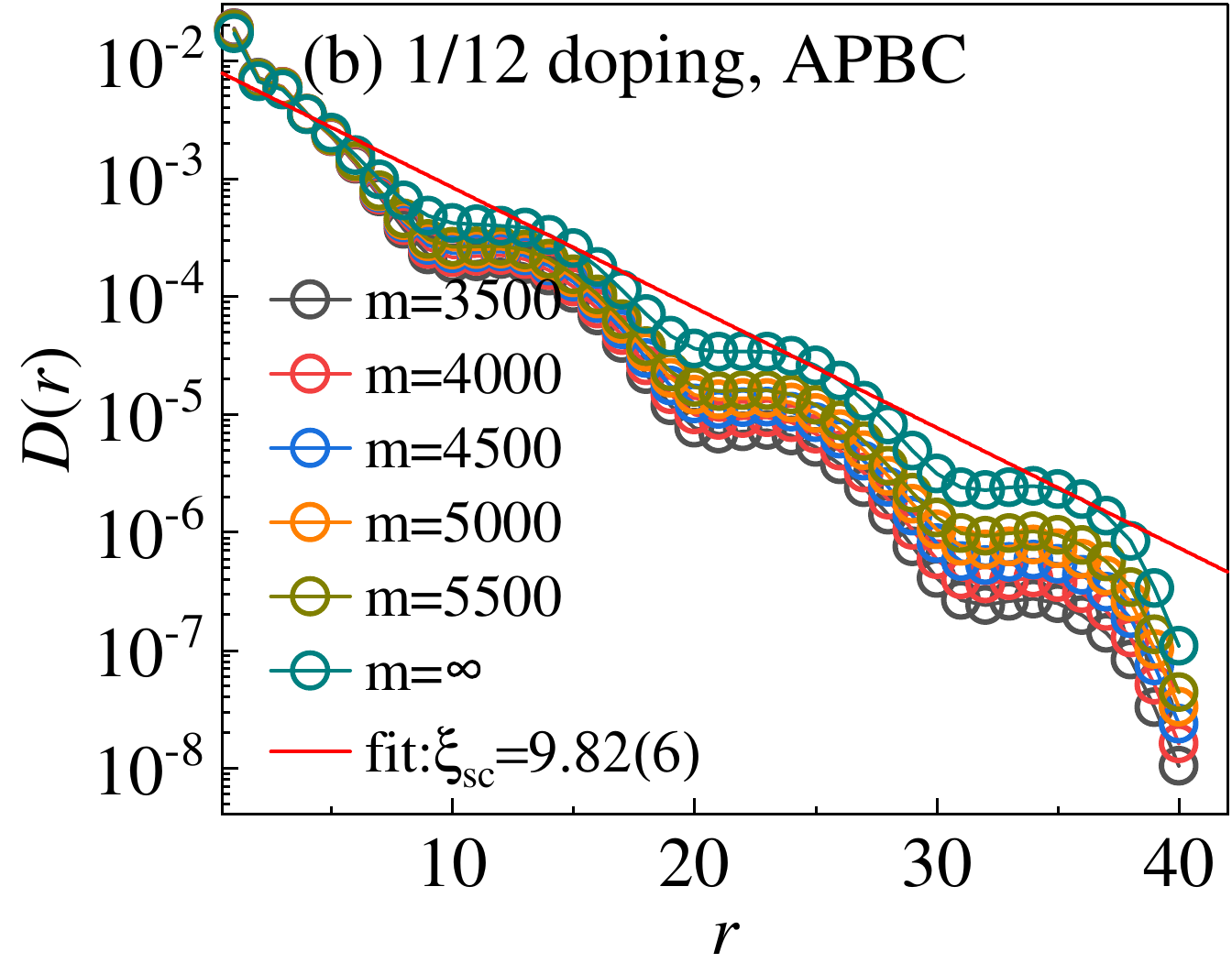}
	\includegraphics[width=0.23\textwidth]{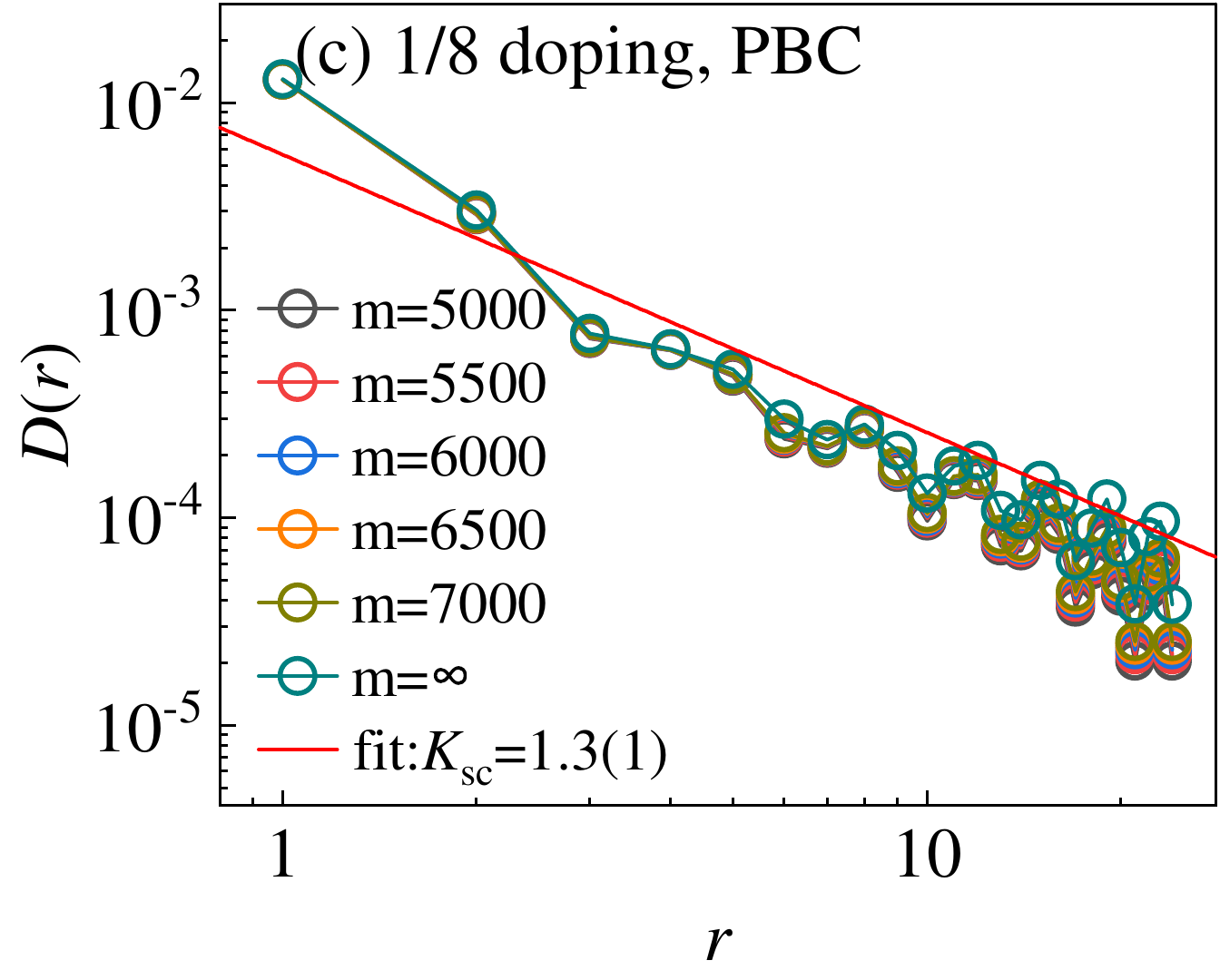}
	\includegraphics[width=0.23\textwidth]{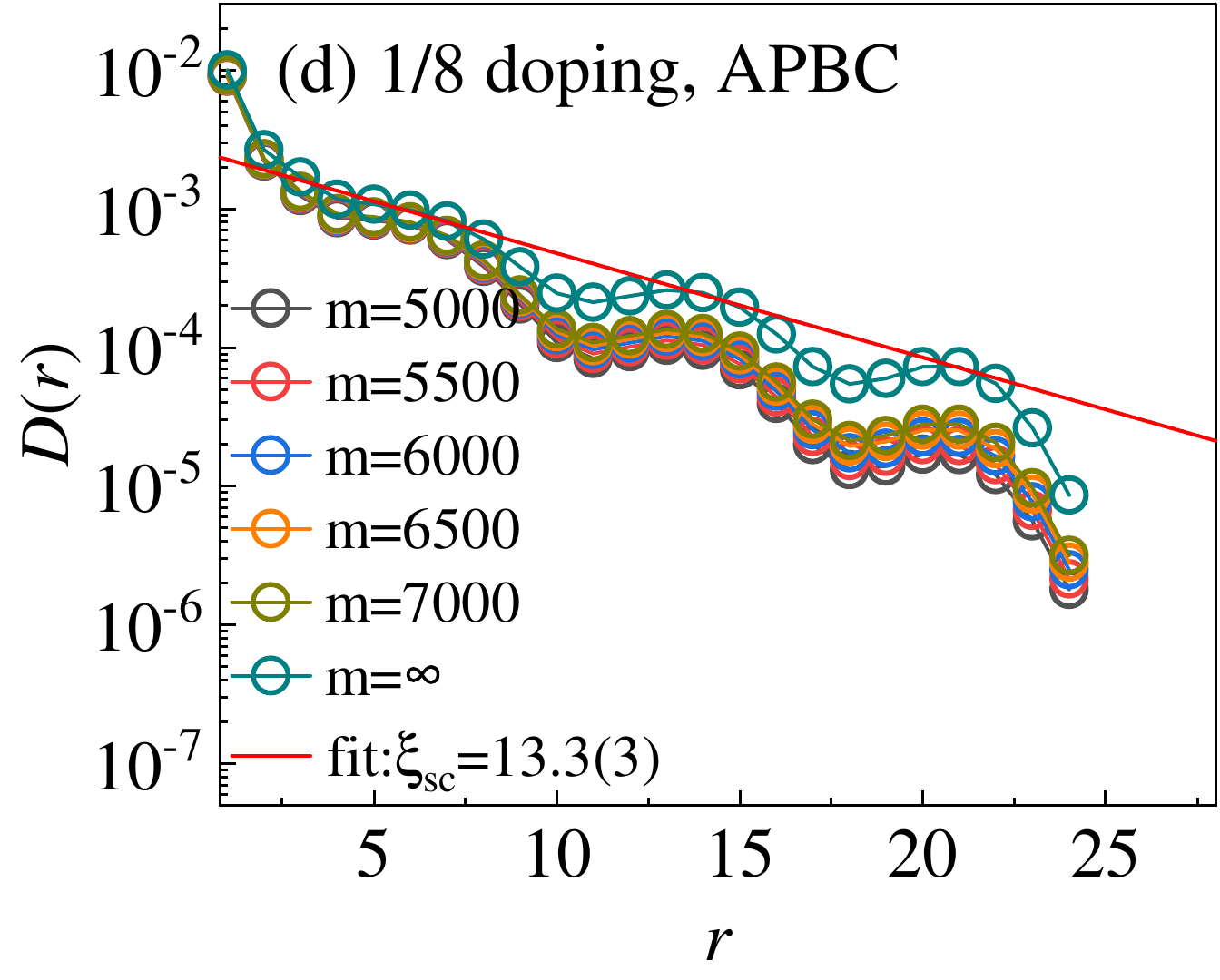}
	\includegraphics[width=0.23\textwidth]{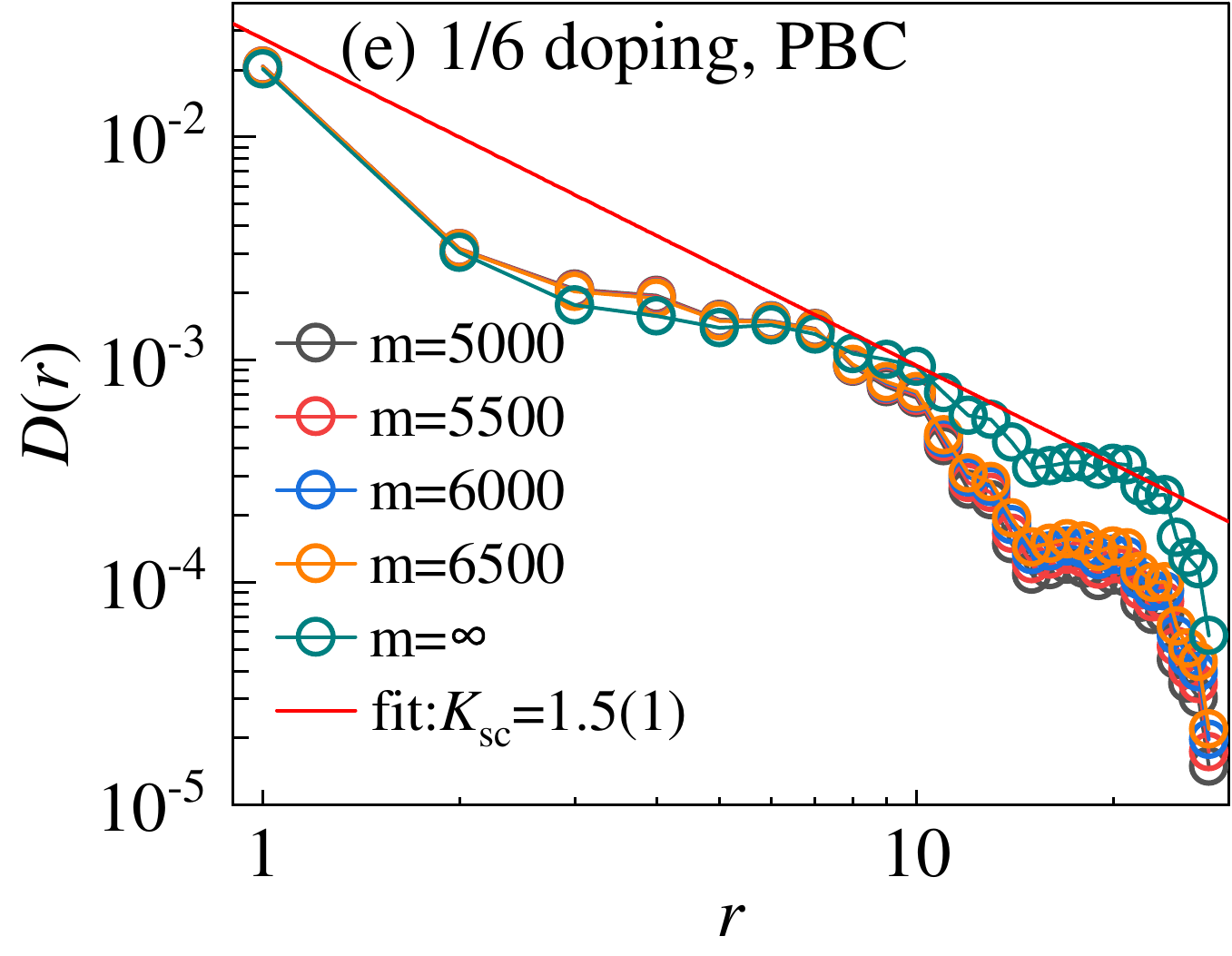}
	\includegraphics[width=0.23\textwidth]{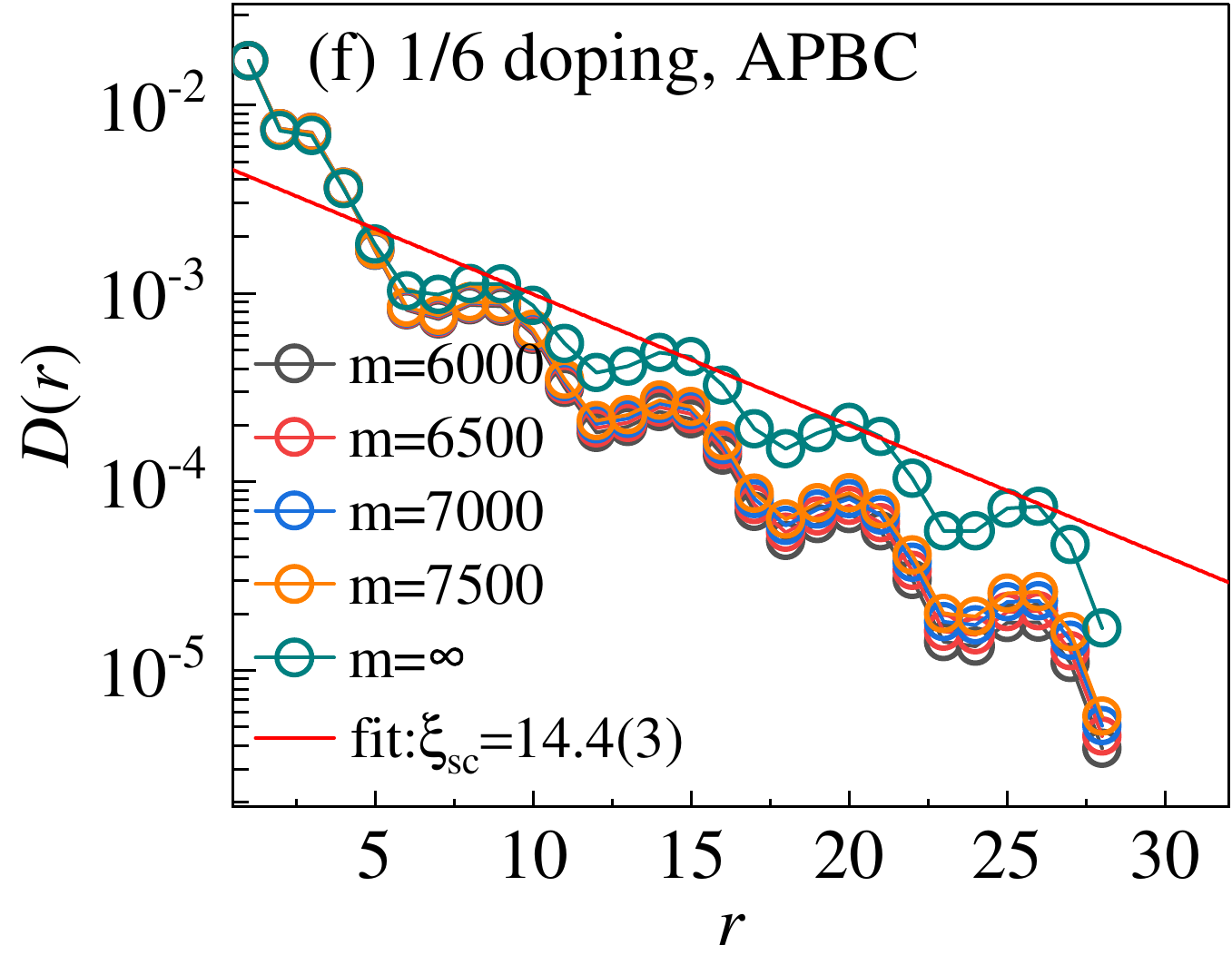}
	
	\caption{Singlet pair-pair correlation for $1/12$ (top), $1/8$ (middle) and $1/6$ (bottom) dopings on width-4 cylinders under PBC (left) and APBC (right). The reference bond is placed at the bond consisting of sites (8, 2) and (8, 3). Both results for finite kept states and the extrapolated with truncation error results are shown. Solid lines are algebraic (left) and exponential (right) fits. The correlation lengths or exponents are listed in the legend.}
	\label{width-4-pair}
\end{figure}

\begin{figure}[t]
	\includegraphics[width=0.23\textwidth]{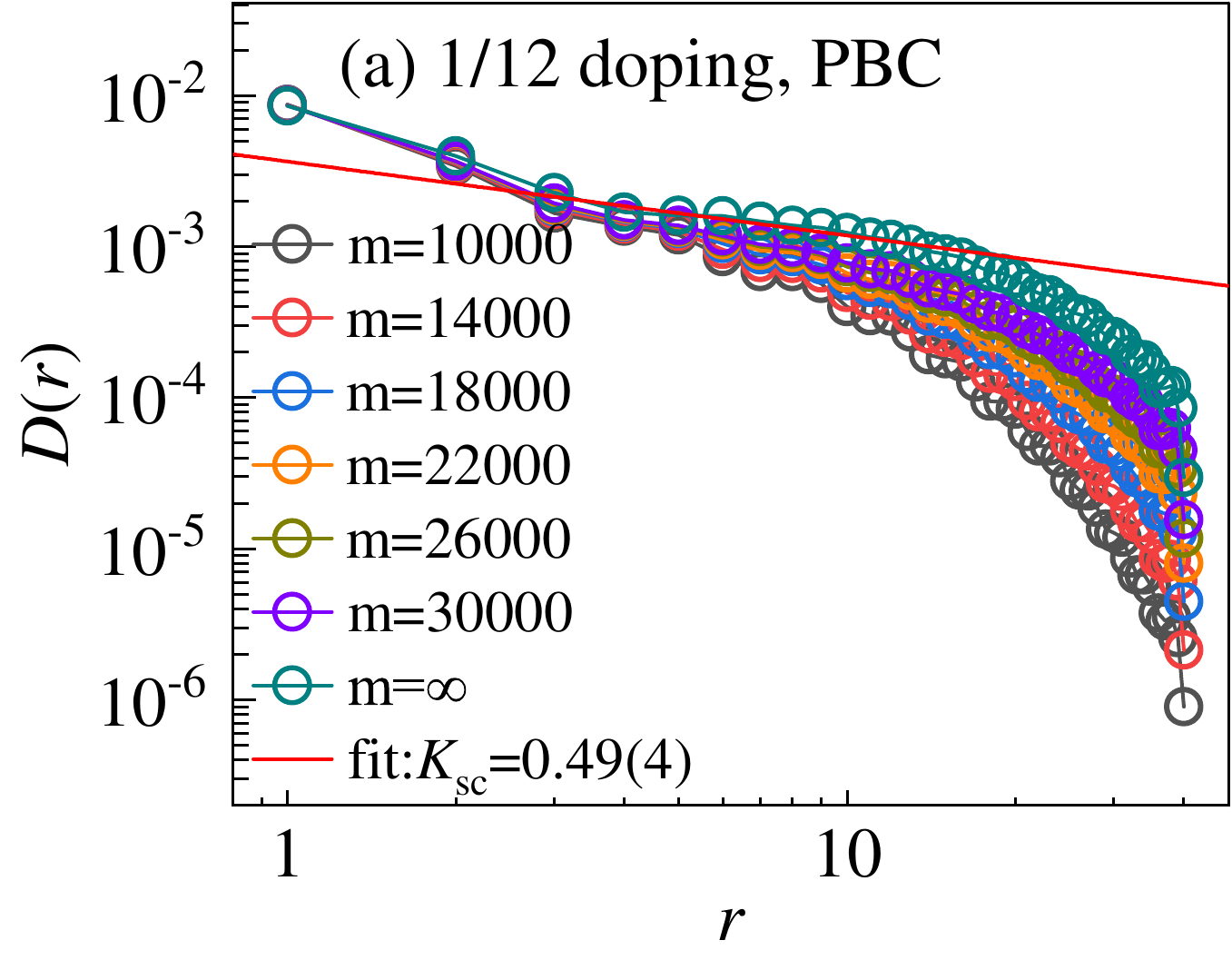}
	\includegraphics[width=0.23\textwidth]{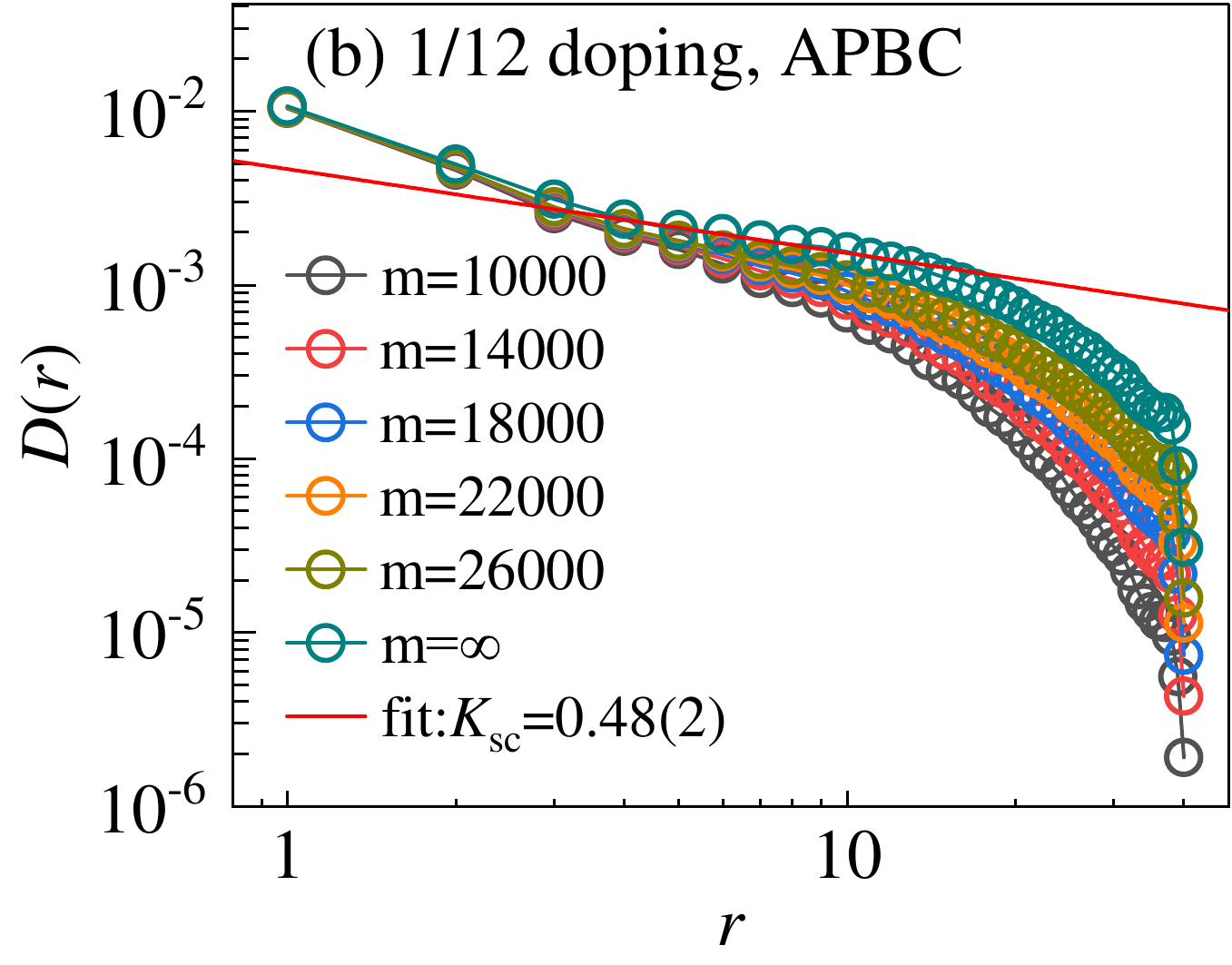}
	\includegraphics[width=0.23\textwidth]{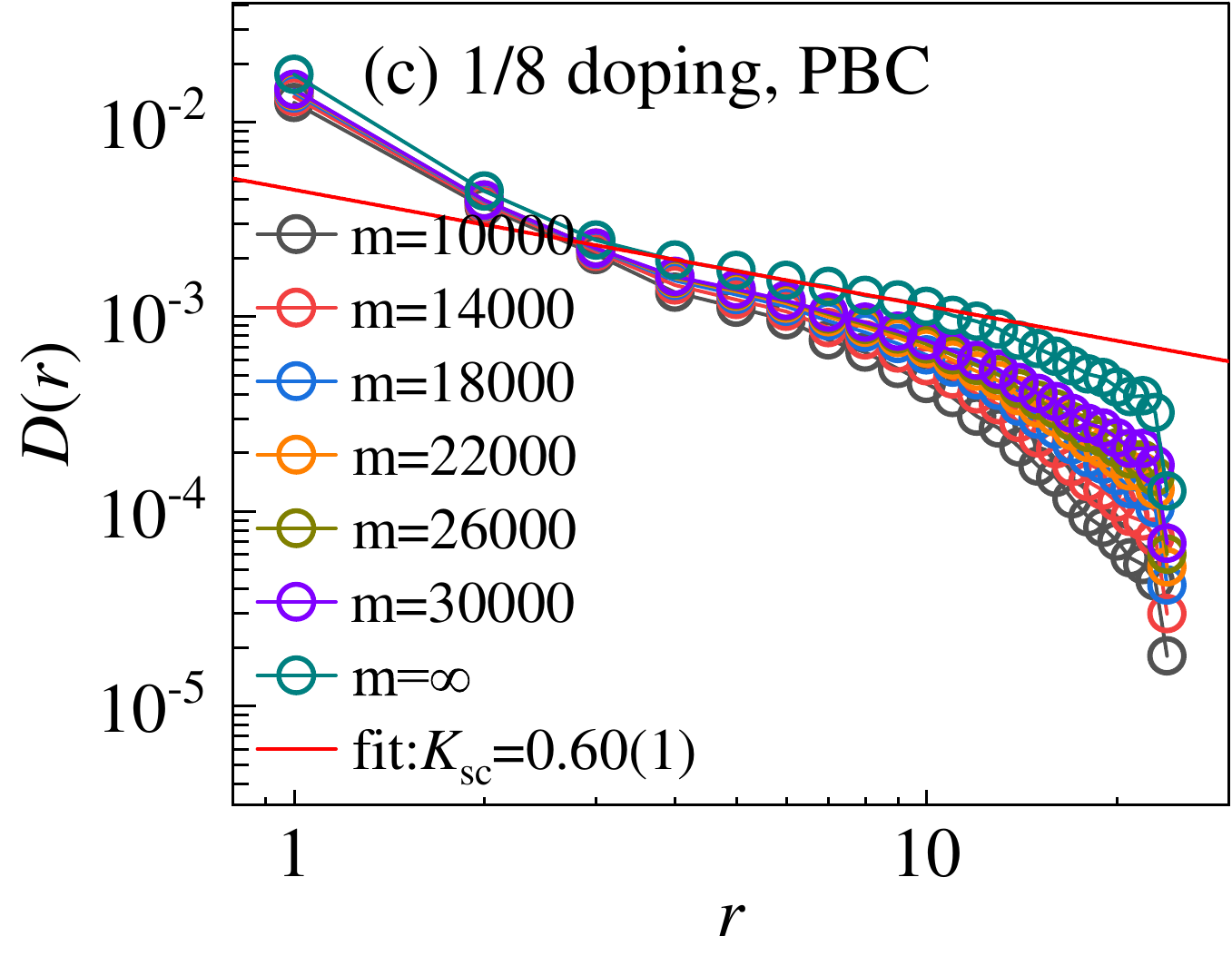}
	\includegraphics[width=0.23\textwidth]{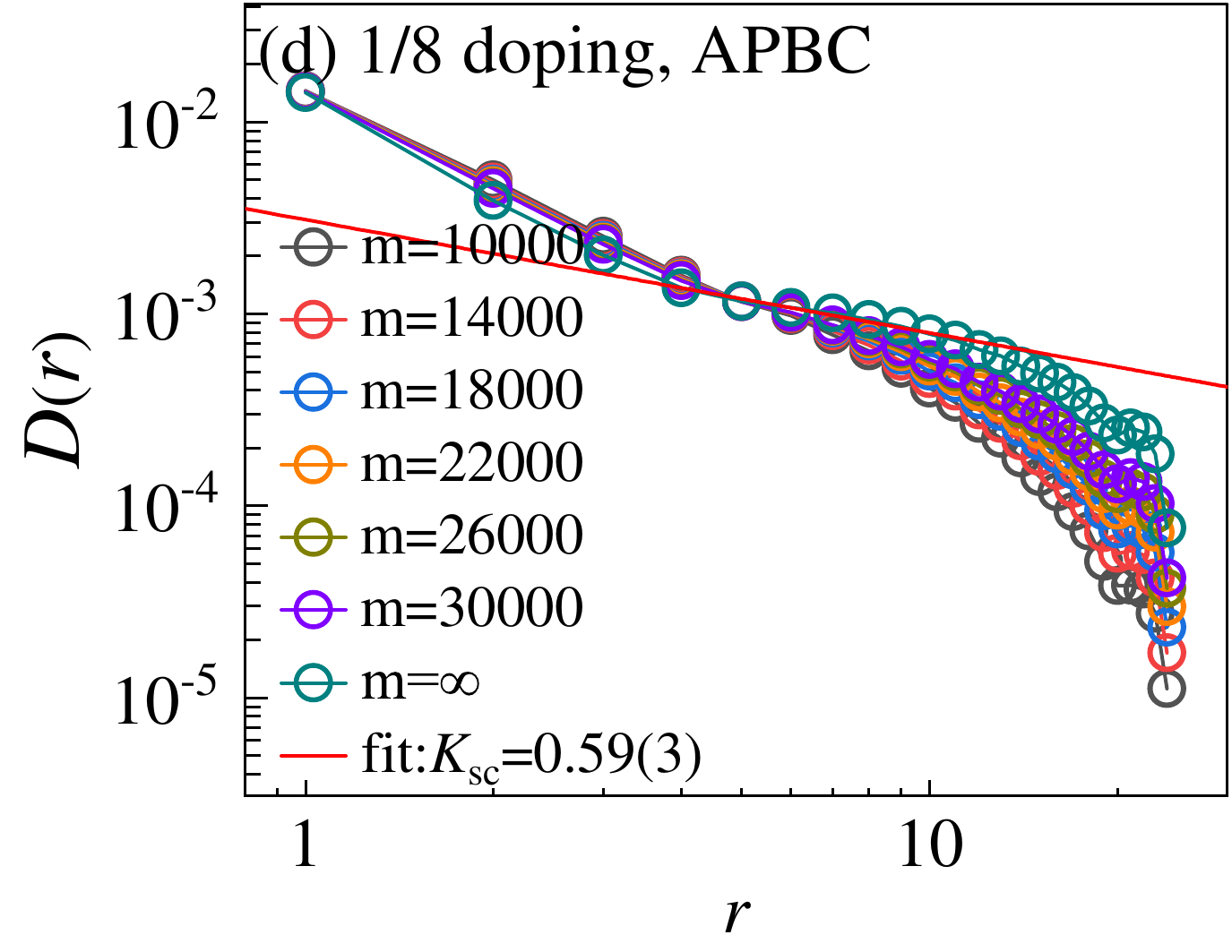}
	\includegraphics[width=0.23\textwidth]{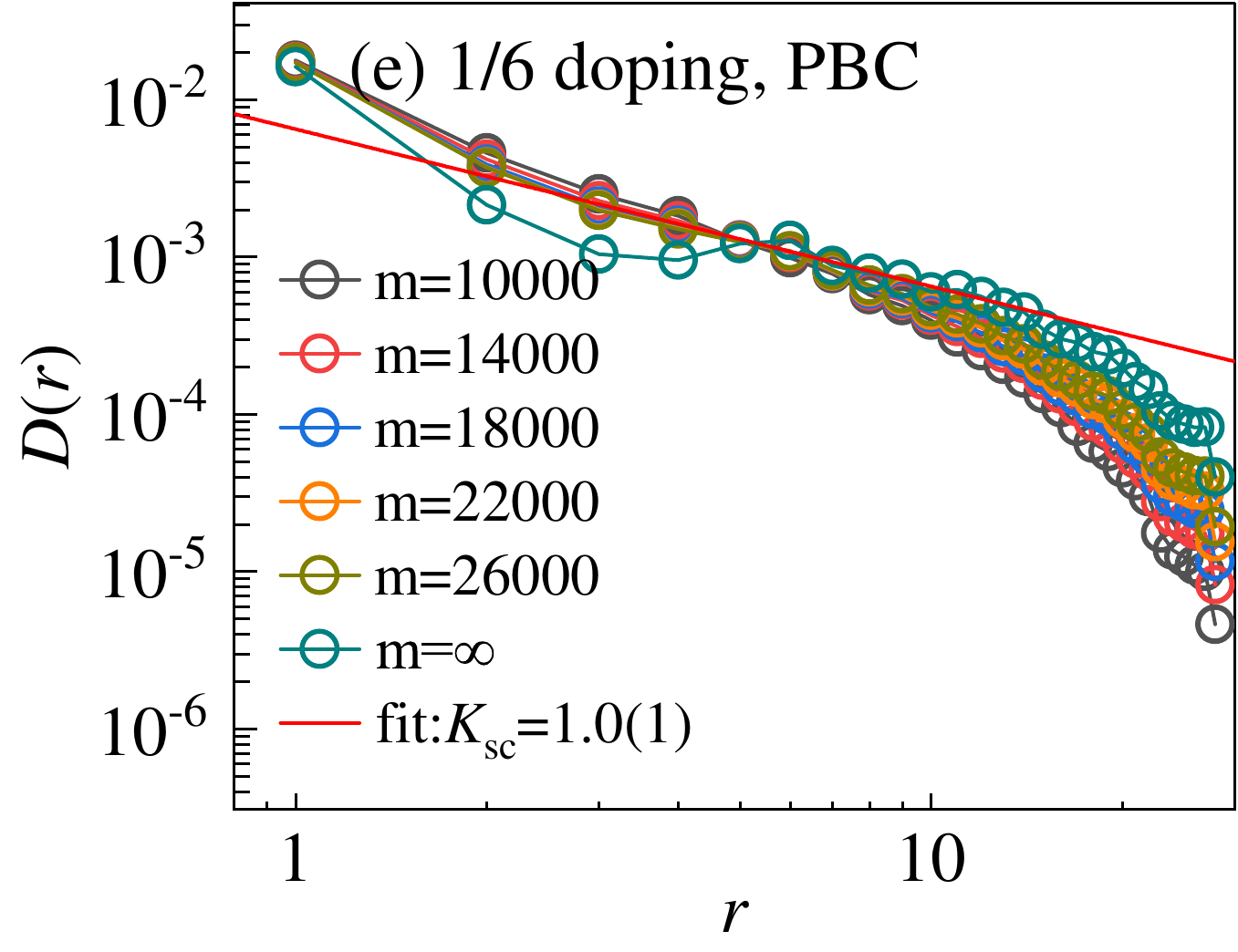}
	\includegraphics[width=0.23\textwidth]{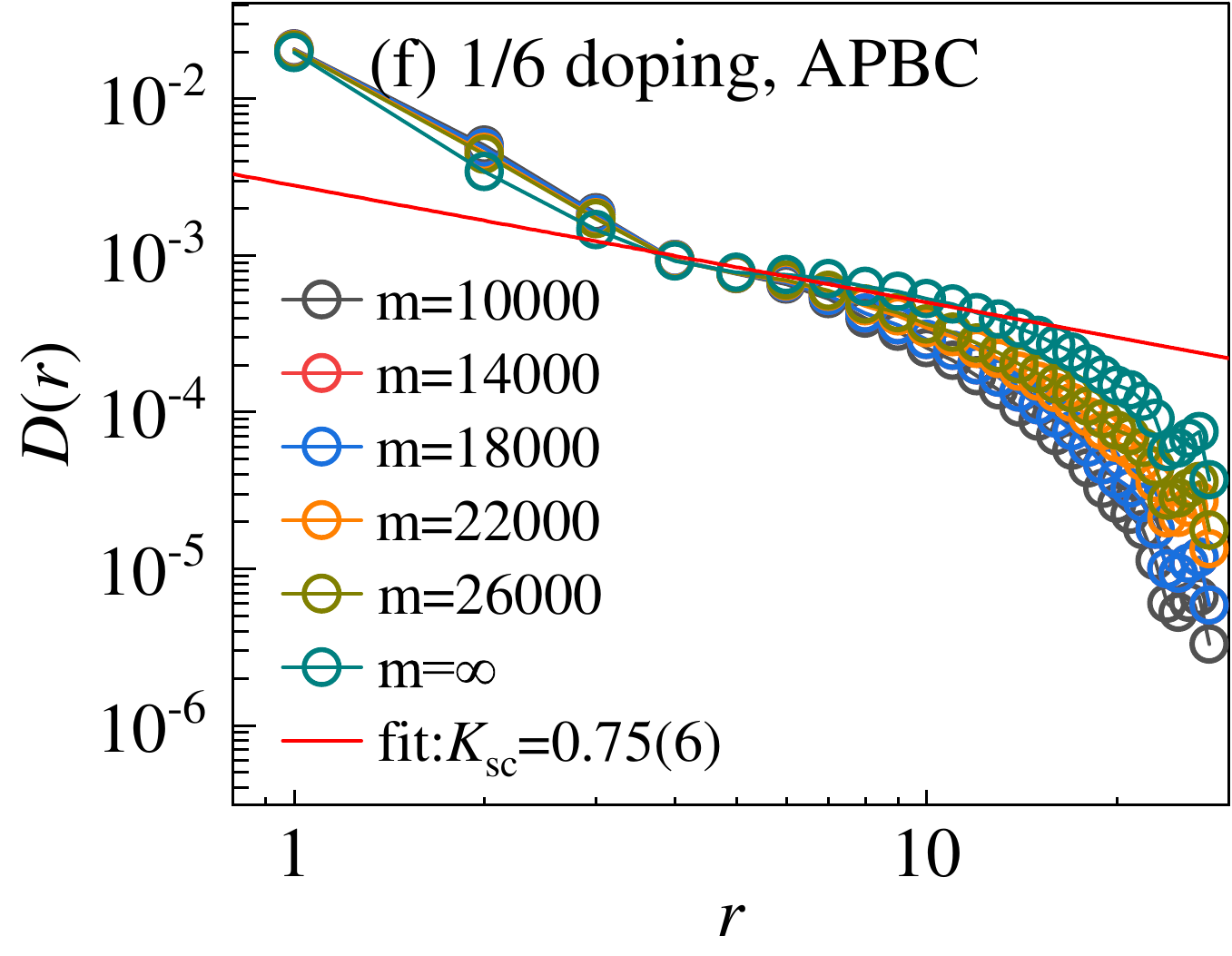}
	
	\caption{Similar as Fig.~\ref{width-4-pair} but for width 6 cylinders. Singlet pair-pair correlation for $1/12$ (top), $1/8$ (middle) and $1/6$ (bottom) dopings on width-6 cylinders with PBC (left) and APBC (right). The reference bond is placed at the bond consisting of sites (8, 2) and (8, 3). Both results for finite kept states and extrapolated with truncation error results are shown. Solid lines are algebraic fits and the corresponding exponents are listed in the legend.}
	\label{width-6-pair}
\end{figure}

\subsection{pairing correlation}
The singlet pair-pair correlation function between bond $i$ (the reference bond, consisting of sites ({$i_x$}, $i_y$) and ({$i_x$}, $i_y + 1$)) and bond $i+r$, defined as $D(r) = \langle \hat{\Delta}_i^{\dagger}\hat{\Delta}_{i+r}\rangle$ with $\hat{\Delta}_i^{\dagger}=1/\sqrt{2} (\hat{c}_{(i_x,i_y),\uparrow}^{\dagger}\hat{c}_{(i_x,i_y+1),\downarrow}^{\dagger}-\hat{c}_{(i_x,i_y),\downarrow}^{\dagger}\hat{c}_{(i_x,i_y+1),\uparrow}^{\dagger})$, are shown in Fig.~\ref{width-4-pair} (width-4) and Fig.~\ref{width-6-pair} (width-6). The pairing symmetry is found to be $d$-wave with opposite sign for vertical and horizontal bonds except the width-4 PBC system at $1/6$ doping, for which the pairing symmetry is plaquette $d$-wave \cite{PhysRevB.102.041106}. As mentioned above, the width-8 calculations are not well converged, we don't show the pair-pair correlation for width-8 systems.  In Figs.~\ref{width-4-pair} and ~\ref{width-6-pair}, both results for finite kept states and the extrapolated with truncation error results are shown. 

We only show results for $D(r)$ with reference bond as the 8th vertical bond (consisting of site(8, 2) and site(8, 3)) from the left open edge. It was known that the value of $D(r)$ depends on the position of reference bond \cite{PhysRevB.108.165113}. We also calculate $D(r)$ with reference bond as the 4th vertical bond (consisting of site(4, 2) and site(4, 3)). The results don't change qualitatively but there are indeed variations for the correlation lengths/exponents (see Table.~\ref{table_sum}).

\subsubsection{Width-4 results}
Fig.~\ref{width-4-pair} shows the pair-pair correlation function $D(r)$ for width-4 systems. We find that for all dopings, $D(r)$ decays algebraically with $D(r) \propto r^{-K_{sc}}$ (exponentially with $D(r) \propto e^{-r / \xi_{sc}}$ ) under PBC (APBC). The difference in ground state between systems under PBC and APBC is also observed for spin and charge densities in Fig.~\ref{width-4}. The fitted exponents $K_{sc}$ or correlation length $\xi_{sc}$ using the extrapolated with truncation error data are also shown in Fig.~\ref{width-4-pair}. We notice that for system under PBC, when the spin and charge density show a strong filled stripe state, $D(r)$ decays exponentially. This correlation indicates strong filled stripe state is harmful for pairing as seen in the pure Hubbard model \cite{PhysRevX.10.031016}.

\subsubsection{Width-6 results}
Fig.~\ref{width-6-pair} shows the pair-pair correlation function $D(r)$ for width-6 systems. For all the results in Fig.~\ref{width-6-pair}, $D(r)$ has a fast decaying tail which bends upward with the increase of kept states. We attribute this tail as the intrinsic consequence of finite number of kept states in DMRG calculations. From the data we find the extrapolation of $D(r)$ with truncation error at the tail is not as reliable as at the head. We also notice that the behavior of $D(r)$ for $r \le 4$ doesn't fit in the long-range behavior. Based on these observations, the small $r$ and very large $r$ data are not included when performing the fit of $D(r)$ to obtain the exponent $K_{sc}$.

For all doping and boundary conditions, $D(r)$ decays polynomially for width-6 systems as shown in Fig.~\ref{width-6-pair}. All the systems display quasi long-range pairing with $K_{sc} < 2$. For fixed doping, the difference of $K_{sc}$ between systems under PBC and APBC are small. For all the dopings, $K_{sc}$ is reduced comparing to the width-4 results in Fig.~\ref{width-4-pair}, indicating superconducting pairing is enhanced in wider systems and the existence of possible long-ranged superconductivity in the thermodynamic limit. We also notice the increase of $K_{sc}$ when the doping is reduced.

\section{discussions}
\label{discussions}

It was known that the filled stripe state is stable in the pure Hubbard model for different widths \cite{PhysRevX.10.031016}, where long-ranged pairing is absent. In the $t-t'-J$ model, we also notice that when strong filled stripe order is found in the ground state for the width-4 system under APBC, superconducting pairing is also short-ranged. These results indicate the strong filled stripe order is likely to be harmful for superconductivity. In other cases, short-ranged stripe states with fractional fillings is accompanied with quasi-long-ranged superconducting pairing. In general, the $t'$ term makes the electron more mobile, causing the paired electrons in the stripe state \cite{PhysRevX.10.031016} to gain coherence and display superconductivity \cite{xu2023coexistence}. The sensitivity of the ground state of $t-t'-J$ and $t'$-Hubbard models \cite{xu2023coexistence} with boundary conditions and widths can also be viewed as a sign of the fragility of stripe state when $t'$ term is included. These results underscore the necessity to check the finite size effect in the future study of systems on narrow cylinders. 

\section{conclusion and perspective}
\label{conclusion and perspective}
We study the finite size effect in the electron doped $t-t'-J$ model on cylinders. We find the sensitivity of ground state to both boundary conditions and system width, similar to what was found in the $t'$-Hubbard model \cite{xu2023coexistence}. For width-4 and 6 systems, the ground state switches between Neel and stripe states under different boundary conditions and with different system widths. For with-8 system and up to the number of kept states we can reach, only Neel (stripe) state can be stabilized in DMRG calculation for $1/12$ ($1/6$) doping, while both stripe and Neel states are stable in the DMRG sweep for $1/8$ doping, regardless of the boundary conditions. These results indicate that $1/8$ doping is likely to lie in the boundary of the phase transition between the Neel phase at lower doping and the stripe phase at higher doping, consistent with previous studies \cite{doi:10.1073/pnas.2109978118}. Our results underscore the necessity to check the finite size effect in the future study of systems on narrow cylinders. It will be interesting to also analyze the finite size effect in the hole doped region in the future, which is more complicated with different competing states in the ground state \cite{RevModPhys.84.1383}.

\begin{acknowledgments}
We thank Shiwei Zhang, Steven R. White, Ulrich Schollw\"{o}ck, Hao Xu, and Chia-Min Chung for useful discussions. We thank Kang Wang for useful suggestions on the manuscript. The calculation in this work is carried out with TensorKit \cite{foot}. The computation in this paper were run on the Siyuan-1 cluster supported by the Center for High Performance Computing at Shanghai Jiao Tong University. Y. Shen and M. P. Qin thank Weidong Luo for his generosity to provide computational resources. M. P. Qin acknowledges the
support from the National Key Research and Development Program of MOST of China (2022YFA1405400), the Innovation Program for Quantum Science and Technology (2021ZD0301902),
the National Natural Science Foundation of China (Grant
No. 12274290) and the sponsorship from Yangyang Development Fund.
\end{acknowledgments}

\bibliography{tj}

\begin{thebibliography}{34}
\expandafter\ifx\csname natexlab\endcsname\relax\def\natexlab#1{#1}\fi
\expandafter\ifx\csname bibnamefont\endcsname\relax
  \def\bibnamefont#1{#1}\fi
\expandafter\ifx\csname bibfnamefont\endcsname\relax
  \def\bibfnamefont#1{#1}\fi
\expandafter\ifx\csname citenamefont\endcsname\relax
  \def\citenamefont#1{#1}\fi
\expandafter\ifx\csname url\endcsname\relax
  \def\url#1{\texttt{#1}}\fi
\expandafter\ifx\csname urlprefix\endcsname\relax\def\urlprefix{URL }\fi
\providecommand{\bibinfo}[2]{#2}
\providecommand{\eprint}[2][]{\url{#2}}

\bibitem[{\citenamefont{Hubbard}(1963)}]{Hubbard}
\bibinfo{author}{\bibfnamefont{J.}~\bibnamefont{Hubbard}},
  \bibinfo{journal}{Proc. R. Soc. Lond. Ser. A} \textbf{\bibinfo{volume}{276}},
  \bibinfo{pages}{238} (\bibinfo{year}{1963}).

\bibitem[{\citenamefont{Zhang and Rice}(1988)}]{PhysRevB.37.3759}
\bibinfo{author}{\bibfnamefont{F.~C.} \bibnamefont{Zhang}} \bibnamefont{and}
  \bibinfo{author}{\bibfnamefont{T.~M.} \bibnamefont{Rice}},
  \bibinfo{journal}{Phys. Rev. B} \textbf{\bibinfo{volume}{37}},
  \bibinfo{pages}{3759} (\bibinfo{year}{1988}),
  \urlprefix\url{https://link.aps.org/doi/10.1103/PhysRevB.37.3759}.

\bibitem[{\citenamefont{Qin et~al.}(2022)\citenamefont{Qin, Sch\"{a}fer,
  Andergassen, Corboz, and Gull}}]{qin2022hubbard}
\bibinfo{author}{\bibfnamefont{M.}~\bibnamefont{Qin}},
  \bibinfo{author}{\bibfnamefont{T.}~\bibnamefont{Sch\"{a}fer}},
  \bibinfo{author}{\bibfnamefont{S.}~\bibnamefont{Andergassen}},
  \bibinfo{author}{\bibfnamefont{P.}~\bibnamefont{Corboz}}, \bibnamefont{and}
  \bibinfo{author}{\bibfnamefont{E.}~\bibnamefont{Gull}},
  \bibinfo{journal}{Annual Review of Condensed Matter Physics}
  \textbf{\bibinfo{volume}{13}}, \bibinfo{pages}{275} (\bibinfo{year}{2022}),
  \eprint{https://doi.org/10.1146/annurev-conmatphys-090921-033948},
  \urlprefix\url{https://doi.org/10.1146/annurev-conmatphys-090921-033948}.

\bibitem[{\citenamefont{Arovas et~al.}(2022)\citenamefont{Arovas, Berg,
  Kivelson, and Raghu}}]{annurev-conmatphys-031620-102024}
\bibinfo{author}{\bibfnamefont{D.~P.} \bibnamefont{Arovas}},
  \bibinfo{author}{\bibfnamefont{E.}~\bibnamefont{Berg}},
  \bibinfo{author}{\bibfnamefont{S.~A.} \bibnamefont{Kivelson}},
  \bibnamefont{and} \bibinfo{author}{\bibfnamefont{S.}~\bibnamefont{Raghu}},
  \bibinfo{journal}{Annual Review of Condensed Matter Physics}
  \textbf{\bibinfo{volume}{13}}, \bibinfo{pages}{239} (\bibinfo{year}{2022}),
  \eprint{https://doi.org/10.1146/annurev-conmatphys-031620-102024},
  \urlprefix\url{https://doi.org/10.1146/annurev-conmatphys-031620-102024}.

\bibitem[{\citenamefont{LeBlanc et~al.}(2015)\citenamefont{LeBlanc, Antipov,
  Becca, Bulik, Chan, Chung, Deng, Ferrero, Henderson, Jim\'enez-Hoyos
  et~al.}}]{PhysRevX.5.041041}
\bibinfo{author}{\bibfnamefont{J.~P.~F.} \bibnamefont{LeBlanc}},
  \bibinfo{author}{\bibfnamefont{A.~E.} \bibnamefont{Antipov}},
  \bibinfo{author}{\bibfnamefont{F.}~\bibnamefont{Becca}},
  \bibinfo{author}{\bibfnamefont{I.~W.} \bibnamefont{Bulik}},
  \bibinfo{author}{\bibfnamefont{G.~K.-L.} \bibnamefont{Chan}},
  \bibinfo{author}{\bibfnamefont{C.-M.} \bibnamefont{Chung}},
  \bibinfo{author}{\bibfnamefont{Y.}~\bibnamefont{Deng}},
  \bibinfo{author}{\bibfnamefont{M.}~\bibnamefont{Ferrero}},
  \bibinfo{author}{\bibfnamefont{T.~M.} \bibnamefont{Henderson}},
  \bibinfo{author}{\bibfnamefont{C.~A.} \bibnamefont{Jim\'enez-Hoyos}},
  \bibnamefont{et~al.} (\bibinfo{collaboration}{Simons Collaboration on the
  Many-Electron Problem}), \bibinfo{journal}{Phys. Rev. X}
  \textbf{\bibinfo{volume}{5}}, \bibinfo{pages}{041041} (\bibinfo{year}{2015}),
  \urlprefix\url{https://link.aps.org/doi/10.1103/PhysRevX.5.041041}.

\bibitem[{\citenamefont{Scalapino}(2012)}]{RevModPhys.84.1383}
\bibinfo{author}{\bibfnamefont{D.~J.} \bibnamefont{Scalapino}},
  \bibinfo{journal}{Rev. Mod. Phys.} \textbf{\bibinfo{volume}{84}},
  \bibinfo{pages}{1383} (\bibinfo{year}{2012}).

\bibitem[{\citenamefont{Dagotto}(1994)}]{RevModPhys.66.763}
\bibinfo{author}{\bibfnamefont{E.}~\bibnamefont{Dagotto}},
  \bibinfo{journal}{Rev. Mod. Phys.} \textbf{\bibinfo{volume}{66}},
  \bibinfo{pages}{763} (\bibinfo{year}{1994}),
  \urlprefix\url{https://link.aps.org/doi/10.1103/RevModPhys.66.763}.

\bibitem[{\citenamefont{Anderson}(1987)}]{doi:10.1126/science.235.4793.1196}
\bibinfo{author}{\bibfnamefont{P.~W.} \bibnamefont{Anderson}},
  \bibinfo{journal}{Science} \textbf{\bibinfo{volume}{235}},
  \bibinfo{pages}{1196} (\bibinfo{year}{1987}),
  \eprint{https://www.science.org/doi/pdf/10.1126/science.235.4793.1196},
  \urlprefix\url{https://www.science.org/doi/abs/10.1126/science.235.4793.1196}.

\bibitem[{\citenamefont{B.~Keimer}(2015)}]{nature14165}
\bibinfo{author}{\bibfnamefont{M.~N.} \bibnamefont{B.~Keimer},
  \bibfnamefont{S.Kivelson}}, \bibinfo{journal}{Nature}
  \textbf{\bibinfo{volume}{518}}, \bibinfo{pages}{179} (\bibinfo{year}{2015}),
  \urlprefix\url{https://doi.org/10.1038/nature14165}.

\bibitem[{\citenamefont{Zheng et~al.}(2017)\citenamefont{Zheng, Chung, Corboz,
  Ehlers, Qin, Noack, Shi, White, Zhang, and
  Chan}}]{doi:10.1126/science.aam7127}
\bibinfo{author}{\bibfnamefont{B.-X.} \bibnamefont{Zheng}},
  \bibinfo{author}{\bibfnamefont{C.-M.} \bibnamefont{Chung}},
  \bibinfo{author}{\bibfnamefont{P.}~\bibnamefont{Corboz}},
  \bibinfo{author}{\bibfnamefont{G.}~\bibnamefont{Ehlers}},
  \bibinfo{author}{\bibfnamefont{M.-P.} \bibnamefont{Qin}},
  \bibinfo{author}{\bibfnamefont{R.~M.} \bibnamefont{Noack}},
  \bibinfo{author}{\bibfnamefont{H.}~\bibnamefont{Shi}},
  \bibinfo{author}{\bibfnamefont{S.~R.} \bibnamefont{White}},
  \bibinfo{author}{\bibfnamefont{S.}~\bibnamefont{Zhang}}, \bibnamefont{and}
  \bibinfo{author}{\bibfnamefont{G.~K.-L.} \bibnamefont{Chan}},
  \bibinfo{journal}{Science} \textbf{\bibinfo{volume}{358}},
  \bibinfo{pages}{1155} (\bibinfo{year}{2017}),
  \eprint{https://www.science.org/doi/pdf/10.1126/science.aam7127},
  \urlprefix\url{https://www.science.org/doi/abs/10.1126/science.aam7127}.

\bibitem[{\citenamefont{Xu et~al.}(2022)\citenamefont{Xu, Shi, Vitali, Qin, and
  Zhang}}]{PhysRevResearch.4.013239}
\bibinfo{author}{\bibfnamefont{H.}~\bibnamefont{Xu}},
  \bibinfo{author}{\bibfnamefont{H.}~\bibnamefont{Shi}},
  \bibinfo{author}{\bibfnamefont{E.}~\bibnamefont{Vitali}},
  \bibinfo{author}{\bibfnamefont{M.}~\bibnamefont{Qin}}, \bibnamefont{and}
  \bibinfo{author}{\bibfnamefont{S.}~\bibnamefont{Zhang}},
  \bibinfo{journal}{Phys. Rev. Res.} \textbf{\bibinfo{volume}{4}},
  \bibinfo{pages}{013239} (\bibinfo{year}{2022}),
  \urlprefix\url{https://link.aps.org/doi/10.1103/PhysRevResearch.4.013239}.

\bibitem[{\citenamefont{Qin et~al.}(2020)\citenamefont{Qin, Chung, Shi, Vitali,
  Hubig, Schollw\"ock, White, and Zhang}}]{PhysRevX.10.031016}
\bibinfo{author}{\bibfnamefont{M.}~\bibnamefont{Qin}},
  \bibinfo{author}{\bibfnamefont{C.-M.} \bibnamefont{Chung}},
  \bibinfo{author}{\bibfnamefont{H.}~\bibnamefont{Shi}},
  \bibinfo{author}{\bibfnamefont{E.}~\bibnamefont{Vitali}},
  \bibinfo{author}{\bibfnamefont{C.}~\bibnamefont{Hubig}},
  \bibinfo{author}{\bibfnamefont{U.}~\bibnamefont{Schollw\"ock}},
  \bibinfo{author}{\bibfnamefont{S.~R.} \bibnamefont{White}}, \bibnamefont{and}
  \bibinfo{author}{\bibfnamefont{S.}~\bibnamefont{Zhang}}
  (\bibinfo{collaboration}{Simons Collaboration on the Many-Electron Problem}),
  \bibinfo{journal}{Phys. Rev. X} \textbf{\bibinfo{volume}{10}},
  \bibinfo{pages}{031016} (\bibinfo{year}{2020}),
  \urlprefix\url{https://link.aps.org/doi/10.1103/PhysRevX.10.031016}.

\bibitem[{\citenamefont{Andersen et~al.}(1995)\citenamefont{Andersen,
  Liechtenstein, Jepsen, and Paulsen}}]{ANDERSEN19951573}
\bibinfo{author}{\bibfnamefont{O.}~\bibnamefont{Andersen}},
  \bibinfo{author}{\bibfnamefont{A.}~\bibnamefont{Liechtenstein}},
  \bibinfo{author}{\bibfnamefont{O.}~\bibnamefont{Jepsen}}, \bibnamefont{and}
  \bibinfo{author}{\bibfnamefont{F.}~\bibnamefont{Paulsen}},
  \bibinfo{journal}{Journal of Physics and Chemistry of Solids}
  \textbf{\bibinfo{volume}{56}}, \bibinfo{pages}{1573 } (\bibinfo{year}{1995}),
  ISSN \bibinfo{issn}{0022-3697}, \bibinfo{note}{proceedings of the Conference
  on Spectroscopies in Novel Superconductors},
  \urlprefix\url{http://www.sciencedirect.com/science/article/pii/0022369795002693}.

\bibitem[{\citenamefont{Hirayama et~al.}(2018)\citenamefont{Hirayama, Yamaji,
  Misawa, and Imada}}]{PhysRevB.98.134501}
\bibinfo{author}{\bibfnamefont{M.}~\bibnamefont{Hirayama}},
  \bibinfo{author}{\bibfnamefont{Y.}~\bibnamefont{Yamaji}},
  \bibinfo{author}{\bibfnamefont{T.}~\bibnamefont{Misawa}}, \bibnamefont{and}
  \bibinfo{author}{\bibfnamefont{M.}~\bibnamefont{Imada}},
  \bibinfo{journal}{Phys. Rev. B} \textbf{\bibinfo{volume}{98}},
  \bibinfo{pages}{134501} (\bibinfo{year}{2018}),
  \urlprefix\url{https://link.aps.org/doi/10.1103/PhysRevB.98.134501}.

\bibitem[{\citenamefont{White and Scalapino}(1999)}]{PhysRevB.60.R753}
\bibinfo{author}{\bibfnamefont{S.~R.} \bibnamefont{White}} \bibnamefont{and}
  \bibinfo{author}{\bibfnamefont{D.~J.} \bibnamefont{Scalapino}},
  \bibinfo{journal}{Phys. Rev. B} \textbf{\bibinfo{volume}{60}},
  \bibinfo{pages}{R753} (\bibinfo{year}{1999}),
  \urlprefix\url{https://link.aps.org/doi/10.1103/PhysRevB.60.R753}.

\bibitem[{\citenamefont{Jiang et~al.}(2020)\citenamefont{Jiang, Zaanen,
  Devereaux, and Jiang}}]{PhysRevResearch.2.033073}
\bibinfo{author}{\bibfnamefont{Y.-F.} \bibnamefont{Jiang}},
  \bibinfo{author}{\bibfnamefont{J.}~\bibnamefont{Zaanen}},
  \bibinfo{author}{\bibfnamefont{T.~P.} \bibnamefont{Devereaux}},
  \bibnamefont{and} \bibinfo{author}{\bibfnamefont{H.-C.} \bibnamefont{Jiang}},
  \bibinfo{journal}{Phys. Rev. Res.} \textbf{\bibinfo{volume}{2}},
  \bibinfo{pages}{033073} (\bibinfo{year}{2020}),
  \urlprefix\url{https://link.aps.org/doi/10.1103/PhysRevResearch.2.033073}.

\bibitem[{\citenamefont{Dodaro et~al.}(2017)\citenamefont{Dodaro, Jiang, and
  Kivelson}}]{PhysRevB.95.155116}
\bibinfo{author}{\bibfnamefont{J.~F.} \bibnamefont{Dodaro}},
  \bibinfo{author}{\bibfnamefont{H.-C.} \bibnamefont{Jiang}}, \bibnamefont{and}
  \bibinfo{author}{\bibfnamefont{S.~A.} \bibnamefont{Kivelson}},
  \bibinfo{journal}{Phys. Rev. B} \textbf{\bibinfo{volume}{95}},
  \bibinfo{pages}{155116} (\bibinfo{year}{2017}),
  \urlprefix\url{https://link.aps.org/doi/10.1103/PhysRevB.95.155116}.

\bibitem[{\citenamefont{Chung et~al.}(2020)\citenamefont{Chung, Qin, Zhang,
  Schollw\"ock, and White}}]{PhysRevB.102.041106}
\bibinfo{author}{\bibfnamefont{C.-M.} \bibnamefont{Chung}},
  \bibinfo{author}{\bibfnamefont{M.}~\bibnamefont{Qin}},
  \bibinfo{author}{\bibfnamefont{S.}~\bibnamefont{Zhang}},
  \bibinfo{author}{\bibfnamefont{U.}~\bibnamefont{Schollw\"ock}},
  \bibnamefont{and} \bibinfo{author}{\bibfnamefont{S.~R.} \bibnamefont{White}}
  (\bibinfo{collaboration}{The Simons Collaboration on the Many-Electron
  Problem}), \bibinfo{journal}{Phys. Rev. B} \textbf{\bibinfo{volume}{102}},
  \bibinfo{pages}{041106} (\bibinfo{year}{2020}),
  \urlprefix\url{https://link.aps.org/doi/10.1103/PhysRevB.102.041106}.

\bibitem[{\citenamefont{{Huang} et~al.}(2018)\citenamefont{{Huang}, {Mendl},
  {Jiang}, {Moritz}, and {Devereaux}}}]{2018npjQM...3...22H}
\bibinfo{author}{\bibfnamefont{E.~W.} \bibnamefont{{Huang}}},
  \bibinfo{author}{\bibfnamefont{C.~B.} \bibnamefont{{Mendl}}},
  \bibinfo{author}{\bibfnamefont{H.-C.} \bibnamefont{{Jiang}}},
  \bibinfo{author}{\bibfnamefont{B.}~\bibnamefont{{Moritz}}}, \bibnamefont{and}
  \bibinfo{author}{\bibfnamefont{T.~P.} \bibnamefont{{Devereaux}}},
  \bibinfo{journal}{npj Quantum Materials} \textbf{\bibinfo{volume}{3}},
  \bibinfo{eid}{22} (\bibinfo{year}{2018}), \eprint{1709.02398}.

\bibitem[{\citenamefont{Gong et~al.}(2021)\citenamefont{Gong, Zhu, and
  Sheng}}]{PhysRevLett.127.097003}
\bibinfo{author}{\bibfnamefont{S.}~\bibnamefont{Gong}},
  \bibinfo{author}{\bibfnamefont{W.}~\bibnamefont{Zhu}}, \bibnamefont{and}
  \bibinfo{author}{\bibfnamefont{D.~N.} \bibnamefont{Sheng}},
  \bibinfo{journal}{Phys. Rev. Lett.} \textbf{\bibinfo{volume}{127}},
  \bibinfo{pages}{097003} (\bibinfo{year}{2021}),
  \urlprefix\url{https://link.aps.org/doi/10.1103/PhysRevLett.127.097003}.

\bibitem[{\citenamefont{Jiang and Kivelson}(2021)}]{PhysRevLett.127.097002}
\bibinfo{author}{\bibfnamefont{H.-C.} \bibnamefont{Jiang}} \bibnamefont{and}
  \bibinfo{author}{\bibfnamefont{S.~A.} \bibnamefont{Kivelson}},
  \bibinfo{journal}{Phys. Rev. Lett.} \textbf{\bibinfo{volume}{127}},
  \bibinfo{pages}{097002} (\bibinfo{year}{2021}),
  \urlprefix\url{https://link.aps.org/doi/10.1103/PhysRevLett.127.097002}.

\bibitem[{\citenamefont{Jiang et~al.}(2021)\citenamefont{Jiang, Scalapino, and
  White}}]{doi:10.1073/pnas.2109978118}
\bibinfo{author}{\bibfnamefont{S.}~\bibnamefont{Jiang}},
  \bibinfo{author}{\bibfnamefont{D.~J.} \bibnamefont{Scalapino}},
  \bibnamefont{and} \bibinfo{author}{\bibfnamefont{S.~R.} \bibnamefont{White}},
  \bibinfo{journal}{Proceedings of the National Academy of Sciences}
  \textbf{\bibinfo{volume}{118}}, \bibinfo{pages}{e2109978118}
  (\bibinfo{year}{2021}),
  \eprint{https://www.pnas.org/doi/pdf/10.1073/pnas.2109978118},
  \urlprefix\url{https://www.pnas.org/doi/abs/10.1073/pnas.2109978118}.

\bibitem[{\citenamefont{Lu et~al.}(2023)\citenamefont{Lu, Zhang, Gong, Sheng,
  and Weng}}]{lu2023sign}
\bibinfo{author}{\bibfnamefont{X.}~\bibnamefont{Lu}},
  \bibinfo{author}{\bibfnamefont{J.-X.} \bibnamefont{Zhang}},
  \bibinfo{author}{\bibfnamefont{S.-S.} \bibnamefont{Gong}},
  \bibinfo{author}{\bibfnamefont{D.}~\bibnamefont{Sheng}}, \bibnamefont{and}
  \bibinfo{author}{\bibfnamefont{Z.-Y.} \bibnamefont{Weng}},
  \bibinfo{journal}{arXiv preprint arXiv:2303.13498}  (\bibinfo{year}{2023}).

\bibitem[{\citenamefont{Lu et~al.}(2024)\citenamefont{Lu, Chen, Zhu, Sheng, and
  Gong}}]{PhysRevLett.132.066002}
\bibinfo{author}{\bibfnamefont{X.}~\bibnamefont{Lu}},
  \bibinfo{author}{\bibfnamefont{F.}~\bibnamefont{Chen}},
  \bibinfo{author}{\bibfnamefont{W.}~\bibnamefont{Zhu}},
  \bibinfo{author}{\bibfnamefont{D.~N.} \bibnamefont{Sheng}}, \bibnamefont{and}
  \bibinfo{author}{\bibfnamefont{S.-S.} \bibnamefont{Gong}},
  \bibinfo{journal}{Phys. Rev. Lett.} \textbf{\bibinfo{volume}{132}},
  \bibinfo{pages}{066002} (\bibinfo{year}{2024}),
  \urlprefix\url{https://link.aps.org/doi/10.1103/PhysRevLett.132.066002}.

\bibitem[{\citenamefont{Jiang et~al.}(2022)\citenamefont{Jiang, Scalapino, and
  White}}]{PhysRevB.106.174507}
\bibinfo{author}{\bibfnamefont{S.}~\bibnamefont{Jiang}},
  \bibinfo{author}{\bibfnamefont{D.~J.} \bibnamefont{Scalapino}},
  \bibnamefont{and} \bibinfo{author}{\bibfnamefont{S.~R.} \bibnamefont{White}},
  \bibinfo{journal}{Phys. Rev. B} \textbf{\bibinfo{volume}{106}},
  \bibinfo{pages}{174507} (\bibinfo{year}{2022}),
  \urlprefix\url{https://link.aps.org/doi/10.1103/PhysRevB.106.174507}.

\bibitem[{\citenamefont{Marino et~al.}(2022)\citenamefont{Marino, Becca, and
  Tocchio}}]{10.21468/SciPostPhys.12.6.180}
\bibinfo{author}{\bibfnamefont{V.}~\bibnamefont{Marino}},
  \bibinfo{author}{\bibfnamefont{F.}~\bibnamefont{Becca}}, \bibnamefont{and}
  \bibinfo{author}{\bibfnamefont{L.~F.} \bibnamefont{Tocchio}},
  \bibinfo{journal}{SciPost Phys.} \textbf{\bibinfo{volume}{12}},
  \bibinfo{pages}{180} (\bibinfo{year}{2022}),
  \urlprefix\url{https://scipost.org/10.21468/SciPostPhys.12.6.180}.

\bibitem[{\citenamefont{Ponsioen et~al.}(2019)\citenamefont{Ponsioen, Chung,
  and Corboz}}]{PhysRevB.100.195141}
\bibinfo{author}{\bibfnamefont{B.}~\bibnamefont{Ponsioen}},
  \bibinfo{author}{\bibfnamefont{S.~S.} \bibnamefont{Chung}}, \bibnamefont{and}
  \bibinfo{author}{\bibfnamefont{P.}~\bibnamefont{Corboz}},
  \bibinfo{journal}{Phys. Rev. B} \textbf{\bibinfo{volume}{100}},
  \bibinfo{pages}{195141} (\bibinfo{year}{2019}),
  \urlprefix\url{https://link.aps.org/doi/10.1103/PhysRevB.100.195141}.

\bibitem[{\citenamefont{{Chen} et~al.}(2023)\citenamefont{{Chen}, {Qiao},
  {Zhang}, and {Zhu}}}]{2023arXiv231205893C}
\bibinfo{author}{\bibfnamefont{Q.}~\bibnamefont{{Chen}}},
  \bibinfo{author}{\bibfnamefont{L.}~\bibnamefont{{Qiao}}},
  \bibinfo{author}{\bibfnamefont{F.}~\bibnamefont{{Zhang}}}, \bibnamefont{and}
  \bibinfo{author}{\bibfnamefont{Z.}~\bibnamefont{{Zhu}}},
  \bibinfo{journal}{arXiv e-prints} \bibinfo{eid}{arXiv:2312.05893}
  (\bibinfo{year}{2023}), \eprint{2312.05893}.

\bibitem[{\citenamefont{Xu et~al.}(2023)\citenamefont{Xu, Chung, Qin,
  Schollw{\"o}ck, White, and Zhang}}]{xu2023coexistence}
\bibinfo{author}{\bibfnamefont{H.}~\bibnamefont{Xu}},
  \bibinfo{author}{\bibfnamefont{C.-M.} \bibnamefont{Chung}},
  \bibinfo{author}{\bibfnamefont{M.}~\bibnamefont{Qin}},
  \bibinfo{author}{\bibfnamefont{U.}~\bibnamefont{Schollw{\"o}ck}},
  \bibinfo{author}{\bibfnamefont{S.~R.} \bibnamefont{White}}, \bibnamefont{and}
  \bibinfo{author}{\bibfnamefont{S.}~\bibnamefont{Zhang}},
  \bibinfo{journal}{arXiv preprint arXiv:2303.08376}  (\bibinfo{year}{2023}).

\bibitem[{\citenamefont{White}(1992)}]{PhysRevLett.69.2863}
\bibinfo{author}{\bibfnamefont{S.~R.} \bibnamefont{White}},
  \bibinfo{journal}{Phys. Rev. Lett.} \textbf{\bibinfo{volume}{69}},
  \bibinfo{pages}{2863} (\bibinfo{year}{1992}),
  \urlprefix\url{https://link.aps.org/doi/10.1103/PhysRevLett.69.2863}.

\bibitem[{\citenamefont{White}(1993)}]{PhysRevB.48.10345}
\bibinfo{author}{\bibfnamefont{S.~R.} \bibnamefont{White}},
  \bibinfo{journal}{Phys. Rev. B} \textbf{\bibinfo{volume}{48}},
  \bibinfo{pages}{10345} (\bibinfo{year}{1993}),
  \urlprefix\url{https://link.aps.org/doi/10.1103/PhysRevB.48.10345}.

\bibitem[{\citenamefont{Schollw\"ock}(2011)}]{SCHOLLWOCK201196}
\bibinfo{author}{\bibfnamefont{U.}~\bibnamefont{Schollw\"ock}},
  \bibinfo{journal}{Annals of Physics} \textbf{\bibinfo{volume}{326}},
  \bibinfo{pages}{96} (\bibinfo{year}{2011}), ISSN \bibinfo{issn}{0003-4916},
  \bibinfo{note}{january 2011 Special Issue},
  \urlprefix\url{https://www.sciencedirect.com/science/article/pii/S0003491610001752}.

\bibitem[{\citenamefont{Shen et~al.}(2023)\citenamefont{Shen, Zhang, and
  Qin}}]{PhysRevB.108.165113}
\bibinfo{author}{\bibfnamefont{Y.}~\bibnamefont{Shen}},
  \bibinfo{author}{\bibfnamefont{G.-M.} \bibnamefont{Zhang}}, \bibnamefont{and}
  \bibinfo{author}{\bibfnamefont{M.}~\bibnamefont{Qin}},
  \bibinfo{journal}{Phys. Rev. B} \textbf{\bibinfo{volume}{108}},
  \bibinfo{pages}{165113} (\bibinfo{year}{2023}),
  \urlprefix\url{https://link.aps.org/doi/10.1103/PhysRevB.108.165113}.

\bibitem[{foo()}]{foot}
\bibinfo{note}{{Our in-house DMRG code is developed with TensorKit package at
  https://github.com/Jutho/TensorKit.jl}}.

\end{thebibliography}

\end{document}